\documentclass[traditabstract]{aa} 
\usepackage{txfonts}
\usepackage{natbib}
\bibpunct{(}{)}{;}{a}{}{,}
\usepackage{graphicx}
\usepackage{color}
\usepackage{hyperref}
\usepackage{breakurl}
\usepackage{ulem}
\usepackage{makecell}
\usepackage{cases}

\newcommand{\redshiftmin}{0.6}
\newcommand{\redshiftmax}{1.0}

\begin{document}

\title{The SDSS-IV extended Baryon Oscillation Spectroscopic Survey: selecting emission line galaxies using the Fisher discriminant}
\author{
A. Raichoor\inst{\ref{CEA}}$^,$\thanks{e-mail: \texttt{\href{mailto:anand.raichoor@cea.fr}{anand.raichoor@cea.fr}}}, 
J. Comparat\inst{\ref{UAM1},\ref{UAM2}}, 
T. Delubac\inst{\ref{EPFL}}, 
J.-P. Kneib\inst{\ref{EPFL},\ref{LAM}}, 
C. Y\`{e}che\inst{\ref{CEA}},
H. Zou\inst{\ref{Beijing}},
F.B. Abdalla\inst{\ref{UCL},\ref{Grahamstown}},
K. Dawson\inst{\ref{SLC}},
A. de la Macorra\inst{\ref{Mexico}},
X. Fan\inst{\ref{Tucson}},
Z. Fan\inst{\ref{Beijing}},
Z. Jiang\inst{\ref{Beijing}},
Y. Jing\inst{\ref{JiaoTong}},
S. Jouvel\inst{\ref{UCL}},
D. Lang\inst{\ref{Carnegie}},
M. Lesser\inst{\ref{Tucson}},
C. Li\inst{\ref{Shanghai}},
J. Ma\inst{\ref{Beijing}},
J.A. Newman\inst{\ref{Pittsburgh}},
J. Nie\inst{\ref{Beijing}},
N. Palanque-Delabrouille\inst{\ref{CEA}},
W.J. Percival\inst{\ref{Portsmouth}},
F. Prada\inst{\ref{Granada},\ref{Cantoblanco},\ref{UAM2}},
S. Shen\inst{\ref{Shanghai}},
J. Wang\inst{\ref{Beijing}},
Z. Wu\inst{\ref{Beijing}},
T. Zhang\inst{\ref{Beijing}},
X. Zhou\inst{\ref{Beijing}},
Z. Zhou\inst{\ref{Beijing}}
}
\institute{
CEA, Centre de Saclay, IRFU/SPP, F-91191 Gif-sur-Yvette, France \label{CEA} 
\and
Departamento de Fisica Teorica, Universidad Autonoma de Madrid, Cantoblanco E-28049, Madrid, Spain \label{UAM1}
\and
Instituto de F\'{\i}sica Te\'orica, (UAM/CSIC), Universidad Aut\'onoma de Madrid,  Cantoblanco, E-28049 Madrid, Spain \label{UAM2}
\and
Laboratoire d'Astrophysique, Ecole Polytechnique F\'{e}d\'{e}rale de Lausanne (EPFL), Observatoire de Sauverny, CH-1290 Versoix, Switzerland \label{EPFL}
\and
Aix Marseille Universit\'{e}, CNRS, LAM (Laboratoire d'Astrophysique de Marseille) UMR 7326, F-13388, Marseille, France \label{LAM}
\and
Key Laboratory of Optical Astronomy, National Astronomical Observatories, Chinese Academy of Sciences, Beijing, 100012, China \label{Beijing}
\and
Department of Physics and Astronomy, University College London, Gower Street, London WC1E6BT, UK \label{UCL}
\and
Department of Physics and Electronics, Rhodes University, PO Box 94, Grahamstown, 6140, South Africa \label{Grahamstown}
\and
Department of Physics and Astronomy, University of Utah, Salt Lake City, UT 84112, USA \label{SLC}
\and
Instituto de F\'{i}sica, Universidad Nacional Aut\'{o}noma de M\'{e}xico, A.P. 20-364, 01000, M\'{e}xico, D. F. M\'{e}xico \label{Mexico}
\and
Steward Observatory, University of Arizona, Tucson, AZ 85721, USA \label{Tucson}
\and
Center for Astronomy and Astrophysics, Department of Physics and Astronomy, Shanghai Jiao Tong University, Shanghai 200240, China \label{JiaoTong}
\and
McWilliams Center for Cosmology, Department of Physics, Carnegie Mellon University, 5000 Forbes Ave, Pittsburgh, PA, 15213, USA \label{Carnegie}
\and
Shanghai Astronomical Observatory, Chinese Academy of Sciences, Shanghai 200030, China \label{Shanghai}
\and
Department of Physics and Astronomy and PITT PACC, University of Pittsburgh, Pittsburgh, PA 15260, USA \label{Pittsburgh}
\and
Institute of Cosmology \& Gravitation, Dennis Sciama Building, University of Portsmouth, Portsmouth, PO1 3FX, UK \label{Portsmouth}
\and
Institutode Astrof\'{i}sica de Andaluc\'{i}a (CSIC),Glorieta de la Astronom\'{i}a, E-18080 Granada, Spain \label{Granada}
\and
Campus of International Excellence UAM+CSIC, Cantoblanco, E-28049 Madrid, Spain \label{Cantoblanco}
}

\abstract
{
% Context
We present a new selection technique of producing spectroscopic target catalogues for massive spectroscopic surveys for cosmology.
% Aims
This work was conducted in the context of the extended Baryon Oscillation Spectroscopic Survey (eBOSS), which will use $\sim$200~000 emission line galaxies (ELGs) at $0.6 \le z_{\rm spec} \le 1.0$ to obtain a precise baryon acoustic oscillation measurement.
% Method
Our proposed selection technique is based on optical and near-infrared broad-band filter photometry.
We used a training sample to define a quantity, the Fisher discriminant (linear combination of colours), which correlates best with the desired properties of the target: redshift and [O\textsc{ii}] flux.
The proposed selections are simply done by applying a cut on magnitudes and this Fisher discriminant.
We used public data and dedicated SDSS spectroscopy to quantify the redshift distribution and [O\textsc{ii}] flux of our ELG target selections.
% Results
We demonstrate that two of our selections fulfil the initial eBOSS/ELG redshift requirements: for a target density of 180 deg$^{-2}$, $\sim$70\% of the selected objects have $0.6 \le z_{\rm spec} \le 1.0$ and only $\sim$1\% of those galaxies in the range $0.6 \le z_{\rm spec} \le 1.0$ are expected to have a catastrophic $z_{\rm spec}$ estimate.
Additionally, the stacked spectra and stacked deep images for those two selections show characteristic features of star-forming galaxies.
% Conclusion
The proposed approach using the Fisher discriminant could, however, be used to efficiently select other galaxy populations, based on multi-band photometry, providing that spectroscopic information is available.
This technique could thus be useful for other future massive spectroscopic surveys such as PFS, DESI, and 4MOST.
}


\titlerunning{Spectroscopic target selection with Fisher Discriminant}
\authorrunning{A. Raichoor et al.}
\maketitle

%============================================================
% INTRODUCTION
%============================================================

\section{Introduction}

% Large photometric surveys
Large optical imaging surveys in astronomy, such as the Sloan Digital Sky Survey \citep[SDSS,][]{york00} or the Canada-France-Hawaii Telescope Legacy Survey \citep[CFHTLS,][]{gwyn12}, have revolutionised the fields of galaxy evolution and cosmology.
Indeed, they enable the photometric selection of large, controlled galaxy populations over either a very wide area or up to faint magnitudes.
Such large, homogeneous galaxy samples are needed to define target catalogues for intensive spectroscopic surveys (e.g., the Main Galaxy Sample: \citealt{strauss02}; the VIMOS VLT Deep Survey, VVDS: \citealt{le-fevre05,le-fevre13}).
Those spectroscopic surveys are then used to measure galaxy properties \citep[e.g.,][]{kauffmann03a,ilbert05} or cosmological parameters \citep[e.g.,][]{eisenstein05,percival10,anderson12} with high statistical accuracy.
In addition, the photometric galaxy samples themselves can also put interesting constraints on galaxy evolution or cosmology \citep[e.g,][]{van-dokkum10,seo12}.

% BOSS/eBOSS - eBOSS/ELG requirements
In this context, a significant step was taken by the Baryon Oscillation Spectroscopic Survey \citep[BOSS,][]{dawson13}, which uses 1.5 million galaxies over 10~000 deg$^2$  selected in the SDSS, to precisely measure the scale of the baryon acoustic oscillations (BAO) to redshifts $z<0.6$, and 160~000 quasars to produce measurements at $z>2.1$ using the quasar Lyman-$\alpha$ forest \citep{delubac14,font-ribera14}.

The main goal of the BOSS survey is to put a cosmological constraint on dark energy through the measurement of the BAO, but its legacy for galaxy evolution will also be unique.
The BOSS observations ended in early 2014, and the data were released as part of the DR11 and DR12 \citep{alam15}.
Building on the success of the BOSS survey, the {\sl extended Baryon Oscillation Spectroscopic Survey} (eBOSS, Dawson et al., 2015) will use four different tracers of the underlying density field to expand the volume covered by BOSS focussing on the redshift range $0.6<z<2.2$.
The four eBOSS tracers are 
{\sl i)}  luminous red galaxies (LRGs) at $z \sim 0.7$, 
{\sl ii)} emission line galaxies (ELGs) at $z \sim 0.8$, 
{\sl iii)} quasars (at $0.9 < z < 2.2$),
and 
{\sl iv)} Lyman-$\alpha$ absorbers in the line of sight of high-redshift ($2.1 < z < 3.5$) quasars.

For the ELG tracers, 300 spectroscopic plates (BOSS plates have a $\sim$7 deg$^2$ circular area and 900 fibres) are to be dedicated to observing $270~000$ targets as potential ELGs in the South Galactic Cap (SGC).
The choice of targeting ELGs is motivated by the presence of the [O\textsc{ii}] emission line in the ELG spectrum, which permits an efficient redshift measurement with $\sim$1h of exposure time.
The requirement for ensuring a measurement of BAO parameters with a precision of 2\% with ELGs is to obtain the spectroscopic redshift measurements of 190~000 ELGs in the $\redshiftmin< z_{\rm spec}<\redshiftmax$ redshift range with a precision better than 300 km.s$^{-1}$, with $\lesssim 1\%$ of catastrophic failures (precision greater than 1000 km.s$^{-1}$).
To fulfil those requirements, the initial eBOSS/ELG settings are an area of 1500 deg$^2$ and a 180 deg$^{-2}$ target density with a minimal efficiency of 70\%, where we define the efficiency as the number of ELGs with a reliable $z_{\rm spec}$ measurement with $\redshiftmin< z_{\rm spec}<\redshiftmax$ divided by the number of targets.
We used those values for the baseline of this study.
The technique we propose here has the advantage of being flexible, thus could be adapted in the case where the final eBOSS/ELG requirements should differ from those values.
For instance, for the eBOSS ELG programme a fibre density of 170 deg$^{-2}$ is assumed in Dawson et al. (2015) -- hence shifting the minimum required efficiency to 74\% -- as 10 deg$^{-2}$ fibres are reserved for other targets.

% eBOSS/ELGs
This paper  is part of a series of papers analysing the properties of $z \sim 0.8$ ELG selection, paving the way for the final eBOSS ELG selection. This paper (Paper II) introduces a new method of selecting $z \sim 0.8$ ELG based on the SDSS detected objects and describes the redshift and [O\textsc{ii}] properties of the selected galaxies.
In Delubac et al. (2015: Paper III), we present the catalogue of the selected ELGs, along with various homogeneity and systematics tests.
\citet[][Paper I]{comparat15} study the [O\textsc{ii}], H$\beta$, and [O\textsc{iii}] emission lines measurement at $z \sim 0.8$ with the BOSS spectrograph \citep{smee13}, aiming to better understand the redshift estimation and the selected galaxy properties. It also details the spectroscopic observations dedicated to the preliminary study of ELG selection.
\citet[][Paper IV]{jouvel15} analyses the properties (redshift, homogeneity) of a $z \sim 0.9$ ELG selection based on the Dark Energy Survey \citep[DES\footnote{\href{http://www.darkenergysurvey.org}{http://www.darkenergysurvey.org}};][]{des05} photometry.

% This paper
In this paper, we present a novel method of select $z \sim 0.8$ ELGs.
Compared to the initial tests, which only used the optical bands (see Paper I), our analysis additionally includes one near-infrared band, hence adding one dimension to the colour-colour space.
The most common method of selecting galaxies for a spectroscopic survey is to apply cuts in magnitudes and colour-colour spaces. It has been used for the surveys targeting a given redshift range (e.g., DEEP2: \citealt{newman13a}; VIPERS: \citealt{guzzo14}) or for the surveys used for BAO measurements (e.g., SDSS/LRG: \citealt{eisenstein01}; WiggleZ: \citealt{drinkwater10}; the upcoming DESI/LRG-ELG\footnote{Dark Energy Spectroscopic Instrument: \href{http://desi.lbl.gov/cdr/}{http://desi.lbl.gov/cdr/}}).
However, such an approach has some limitation when using a large number of multi-wavelength observations: when dealing with three or more colour-colour diagrams, the selection-box definition starts to be subjective, unless using an automatic exploration of all the possibilities.
One possibility is to use neural networks (e.g., for the BOSS/QSOs: \citealt{dawson13}), which can bring efficient selections but at the cost of a less tractable selection.
We introduce an alternative approach, the Fisher discriminant, which is equivalent to a hyperplane cut in the full colour space, that is to say, a cut on a simple linear combination of the colours.
This hyperplane is automatically defined from a training sample and a list of criteria, which are here a redshift of $\sim$0.8 and significant [O\textsc{ii}] emission.
We note that this approach -- not used in astrophysics to our knowledge -- is automatic and can be used in other situations where one wants to select a given population from multi-wavelength photometry, given that a training sample is available.
We present in Section \ref{sec:fishermethod} the Fisher discriminant approach, then we introduce in Section \ref{sec:data} the photometric and spectroscopic data used in this study.
Section \ref{sec:elg_selection} is dedicated to the tested $z \sim 0.8$ ELG selection schemes: we first describe them and then analyse their global properties in terms of redshift and [O\textsc{ii}] emission.
For two of the selections, we present stacked spectra and structural properties in Section \ref{sec:stack}.
Finally, we conclude in Section \ref{sec:conclusions}.

% Paper conventions
In this paper, we adopt $H_0 = 70$ km s$^{-1}$ Mpc$^{-1}$,  $\Omega_m = 0.30$, and $\Omega_\Lambda = 0.70$.
All magnitudes are expressed in the AB system and corrected for the Galactic foreground extinction using the \citet{schlegel98} maps.

%============================================================
% FISHER DISCRIMINANT METHOD
%============================================================
\section{Fisher discriminant method \label{sec:fishermethod}}

%=========================
% FISHER/Principle
%=========================
\subsection{Principle}
The goal of the linear discriminant analysis \citep[LDA,][]{fisher36}, also known as the Fisher method, is to define a discriminant (the Fisher discriminant $X_{FI}$) that separates two known classes of a set of events best.
We assume we have a collection of events (${\bf y}$) where each event ${\bf y}$ is known to belong to one of the two classes, $({\bf y_1})$ and $({\bf y_2})$.
To each event are associated $N$ measurements: ${\bf y} = (x_1, x_2, ..., x_N)$, each $x_i$ being a real variable measuring a given property.
For instance, in the original taxonomic work of \citet{fisher36}, the two classes $({\bf y_1})$ and $({\bf y_2})$ are two different species of iris, \textit{Iris setosa} and \textit{Iris versicolor}; the events ${\bf y} = (x_1, x_2, x_3, x_4)$ are a sample of fifty plants, for each of which are available four measurements done on the sepals and the petals.
We then let $n_1$ and $n_2$ be the number of events in each class, $\bar{\bf y}_1$ and $\bar{\bf y}_2$ the means over each class, and $T$ the total variance-covariance matrix of the sample (${\bf y}$).

% Figure: Fisher illustration
\begin{figure}
        \includegraphics[width=0.95\columnwidth]{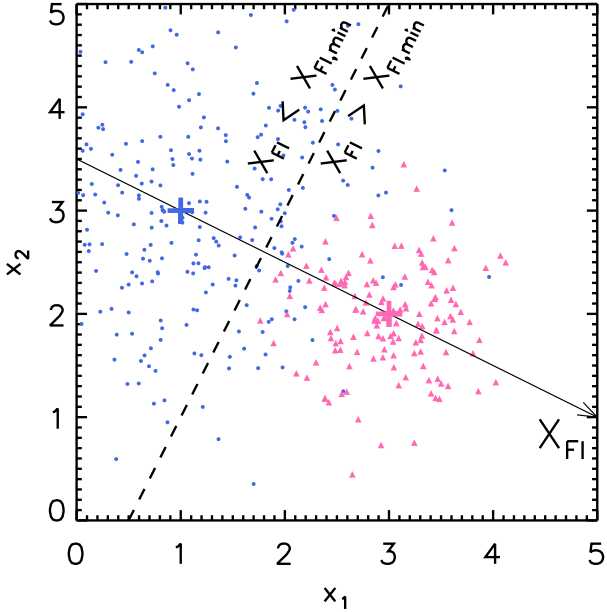}\\      
        \caption{Illustration of the Fisher discriminant method with $N=2$.
The $({\bf y_1})$ class is in blue dots, with the blue cross at $\bar{\bf y}_1 = (1,3)$, while the $({\bf y_2})$ class is in pink triangles, with the pink cross at $\bar{\bf y}_2 = (3,2)$.
For each event, the Fisher discriminant $X_{FI}$ corresponds to its orthogonal projection along the axis defined by $\bar{\bf y}_1$ and $\bar{\bf y}_2$.
The dashed line illustrates the hyperplane used to split the events in two classes.}
        \label{fig:fisher_illustration}
\end{figure}

The Fisher discriminant $X_{FI}$ is a linear combination of the $N$ variables $x_i$, aiming to provide the best separation between the two classes of events $({\bf y_1})$ and $({\bf y_2})$. In the $N$-dimension space of the measurement variables, it defines a hyperplane (dimension $N-1$).
This hyperplan is orthogonal to the axis defined by the line connecting $\bar{\bf y}_1$ and $\bar{\bf y}_2$, along which the distance between the projected points will naturally be a maximum.
In other terms, the Fisher discriminant $X_{FI}$ is the orthogonal projection on this axis, as illustrated in Figure \ref{fig:fisher_illustration} for $N=2$.
In his original work, \citet{fisher36} proposed to normalise the projected distance by the quadratic sum of the projected dispersion of each class:
\begin{equation}
        X_{FI} = \frac{\sqrt{n_{1}n_{2}}}{n_1+n_2}\,
                (\bar{{\bf y}}_1 - \bar{{\bf y}}_2)^T T^{-1}{\bf y}.
	\label{eq:fisher_eq}
\end{equation}
A threshold value $X_{FI,\rm min}$ is then used to associate the events with $X_{FI} < X_{FI,\rm min}$ to the $({\bf y_1})$ class and the events with $X_{FI} > X_{FI,\rm min}$ to the $({\bf y_2})$ class.

%=========================
% FISHER/Application to ELGs
%=========================
\subsection{Application to ELGs \label{sec:fisher_elg}}
We now describe how we apply the Fisher discriminant approach to ELGs.
The considered variable space is the galaxy colour space, and we compute the Fisher discriminant quantity $X_{FI}$ through the use of a spectroscopic sample of galaxies (presented in Section \ref{sec:spectrain}).
As for any learning method, the training and test sample should ideally be as representative as possible of the data we want to apply the method to. This is the case in our study, where the training and test samples have homogeneous photometry coming from the surveys planned to be eventually used in eBOSS/ELG (except for the DECaLS data that we mimic with degrading the CFHTLS data; see Section \ref{sec:photometry}).
Please note that, as we test several selection schemes defined with different Fisher discriminants $X_{FI}$, we rescale the Fisher discriminant $X_{FI}$ obtained with Eq.(\ref{eq:fisher_eq}) to a common scale, in order to facilitate the comparison between Fisher expressions.
More precisely, we multiply it by a coefficient and add a constant value, so that the $X_{FI}$ distribution for a complete subsample (the VVDS, see Section \ref{sec:spectrain}) has a mean value of 0 and a standard deviation of 1.

Our aim is to select ELGs at $\redshiftmin \le z_{\rm spec} \le \redshiftmax$: we need to choose how to define the two considered classes, the \texttt{Signal} and the \texttt{Background} classes.
A first possible approach to define the two considered classes is to use criteria on both the spectroscopic redshift $z_{\rm spec}$ and the total [O\textsc{ii}] flux $f_{[\rm O\textsc{ii}]}^{\rm tot}$, with defining our \texttt{Signal} class with galaxies in a high redshift range and with a significant total [O\textsc{ii}] flux.

However, $f_{[\rm O\textsc{ii}]}^{\rm tot}$ is available for relatively few galaxies compared to $z_{\rm spec}$, which is a quantity usually provided in public surveys.
To cover the maximum variable space, we can also take advantage of the fact that in our targeted redshift range, a large majority of galaxies are star-forming \citep[e.g.][]{ilbert13}, as the star-formation rate density of the Universe is about ten times higher at $z \sim 1$ than today \citep[e.g.][]{lilly96,madau98,hopkins06}.
A possible alternate approach is to define our \texttt{Signal} class only with galaxies in a high-redshift range.
Proceeding in this way would allow us to use a large spectroscopic training sample, representative of the main types of galaxies at all redshifts.

%============================================================
% DATA
%============================================================
\section{Data \label{sec:data}}
\begin{figure}
        \includegraphics[width=0.95\columnwidth,clip=true,trim=120 0 0 0]{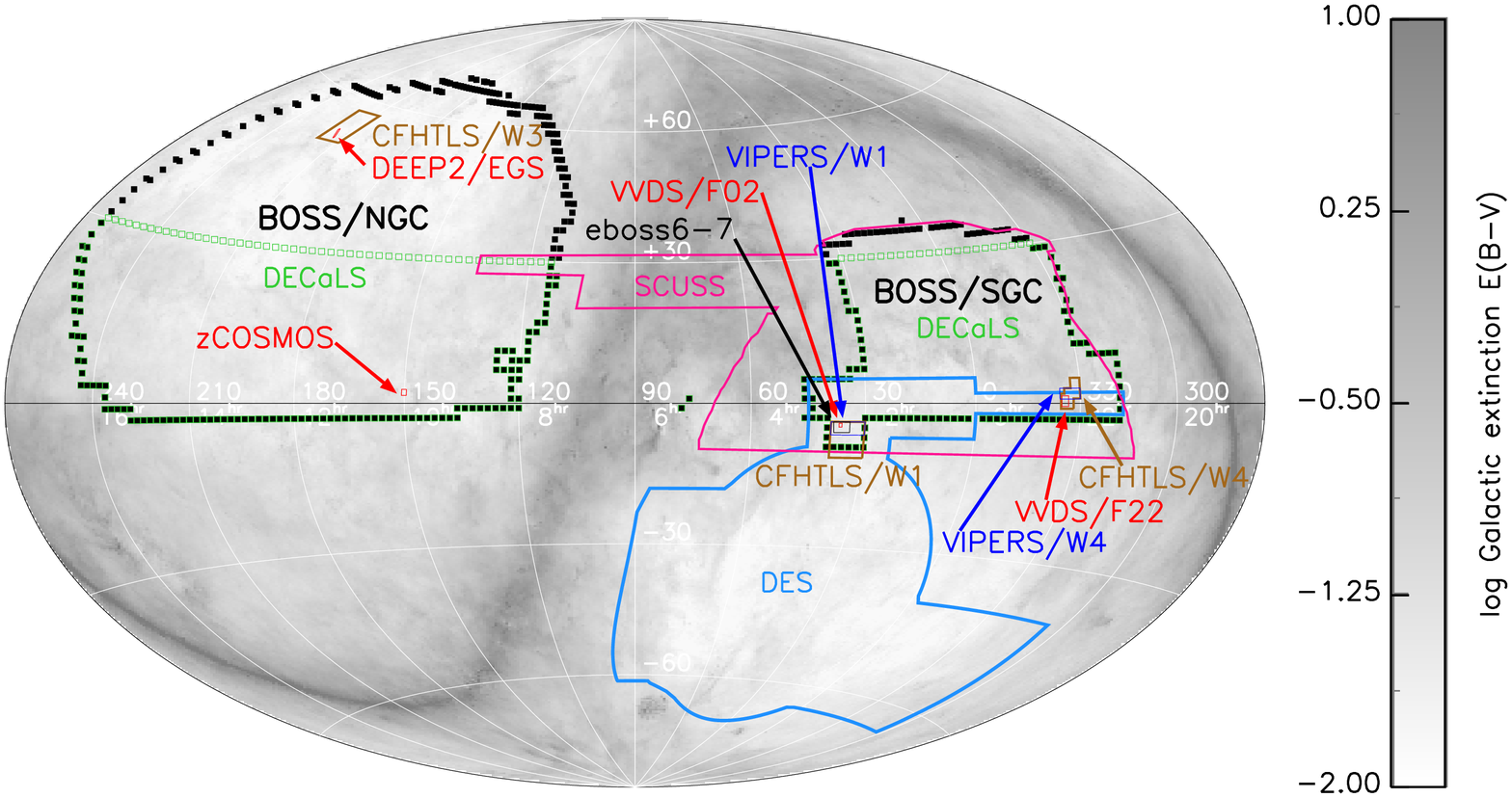}\\      
        \includegraphics[width=0.95\columnwidth]{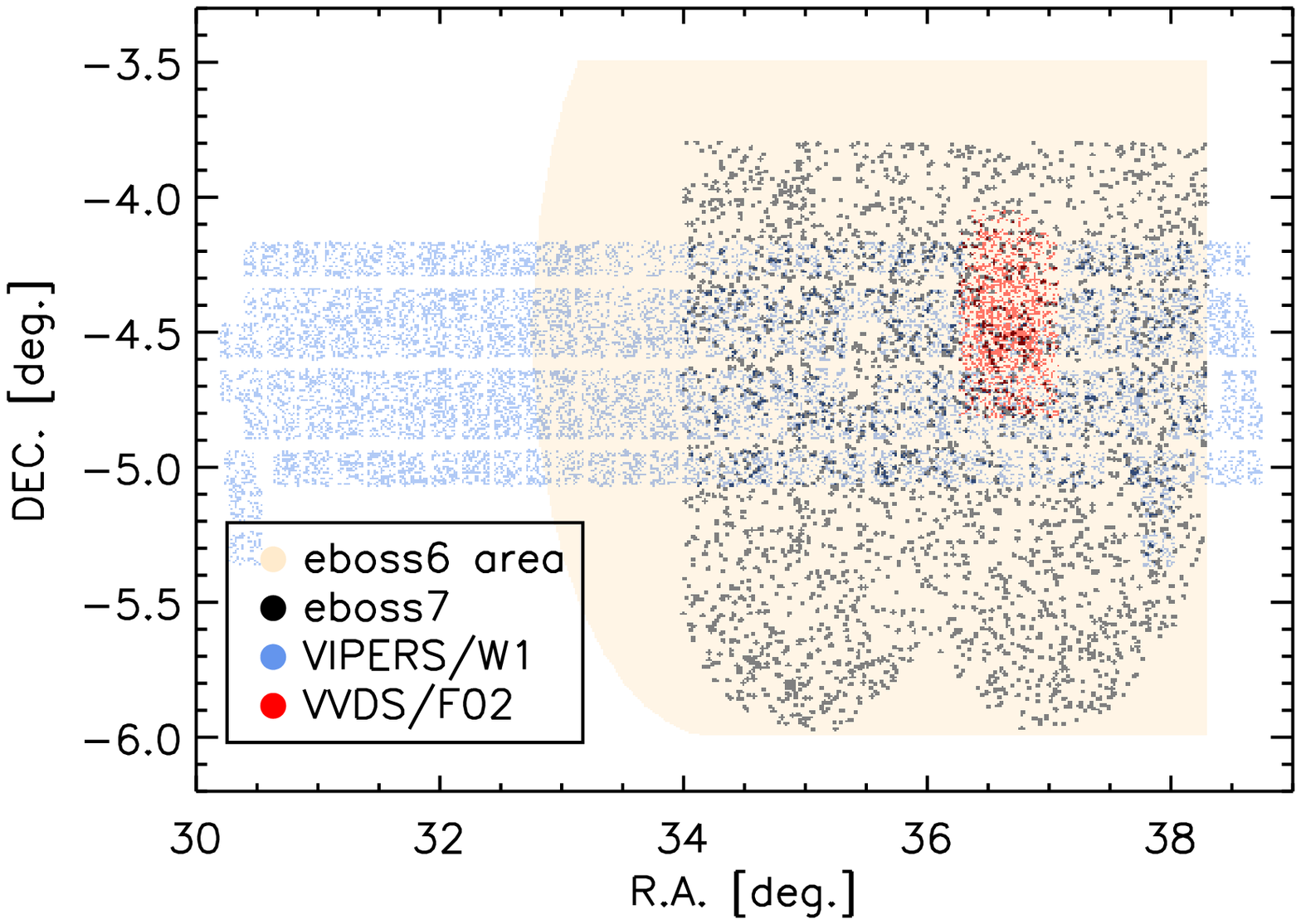}\\       
        \caption{Sky location of the different photometric and spectroscopic surveys used in this study.
\textit{Top panel}: overview of the whole sky in an Aitoff projection in J2000 equatorial coordinates; the WISE photometry covers the whole sky.
\textit{Bottom panel}: zoom on the location of \texttt{eboss6-7}, the ten eBOSS/ELG test plates observed in 2014; in this study, we do not use \texttt{eboss6} measurements outside of the \texttt{eboss7} area.
}
        \label{fig:surveys}
\end{figure}

In this section, we describe the data used in our study.
First, we present the photometric data with which the ELG selection is done, then we describe the Fisher training sample used to define the tested Fisher discriminants $X_{FI}$; finally, we introduce the spectroscopic data used to quantify the efficiencies of the tested Fisher selection algorithms.
We display in Figure \ref{fig:surveys} the sky locations of the different surveys of interest in this study.

% Figure: sky surveys

%=========================
% DATA/Photometry
%=========================
\subsection{Photometry \label{sec:photometry}}
We here present the different photometric surveys we use in this study to estimate object colours.
We test different schemes, based on objects detected in the SDSS.
The general properties of the used photometry are summed up in Table \ref{tab:phot}.
We note that the issues related to colours computed with magnitudes measured through different surveys are significantly mitigated for our SDSS-SCUSS-WISE colours, because our SCUSS and WISE photometry is done consistently with the SDSS \citep[forced photometry on SDSS objects, using SDSS structural information: ][ Hu Zou et al., 2015, in preparation]{lang14a}.\\

% SDSS
\textbf{\textit{SDSS.}}
The SDSS \citep[][; DR12]{york00, alam15} provides photometry in the optical $ugriz$ broad-bands to a depth of $r \sim 22.5$ mag over $\sim$15~000 deg$^2$ of high-latitude sky, split into two regions of $\sim$7~500 deg$^2$, namely the Northern Galactic Cap (NGC) and the Southern Galactic Cap (NGC).

Since we are interested in galaxy colours, we use the \texttt{modelMag} magnitudes\footnote{\href{http://www.sdss.org/dr12/algorithms/magnitudes/\#mag\_model}{http://www.sdss.org/dr12/algorithms/magnitudes/\#mag\_model}}: those are computed through a luminosity profile -- fitted to the $r$-band data -- convolved with the PSF in each band, and this permits unbiased colour measurements in the absence of colour gradients.
Our photometric object detection list is constituted of objects from the \texttt{PhotoPrimary} list\footnote{\href{http://skyserver.sdss.org/dr12/en/help/browser/browser.aspx}{http://skyserver.sdss.org/dr12/en/help/browser/browser.aspx}}.

% WISE
\textbf{\textit{WISE.}}
The Wide-Field Infrared Survey Explorer \citep[WISE;][]{wright10} measured the full sky in four mid-infrared bands centred on 3.4 $\mu$m, 4.6 $\mu$m, 12 $\mu$m, and 22 $\mu$m, known as $W1$ through $W4$.
We make use of the "forced photometry" done on the SDSS objects here \citep{lang14a}: this photometry uses measured SDSS source positions, star-galaxy separation, and galaxy profiles to define the sources whose fluxes are to be measured in the WISE images.
In this work, we use only the $W1$ channel, because it appeared during our preliminary studies that also using $W2$ creates spatial inhomogeneities in our selections, thus reflecting the underlying variations in the signal-to-noise ratio in $W2$.

% SCUSS
\textbf{\textit{SCUSS.}}
The South Galactic Cap $u$-band Sky Survey (SCUSS; Xu Zhou et al., 2015, in preparation; \citealt{zou15}) is an international cooperative project, which is undertaken by the National Astronomical Observatories of China (Chinese Academy of Sciences) and Steward Observatory (University of Arizona, USA).
It is a $u$-band (effective wavelength $\sim$ 3538 \AA) imaging survey programme with the 90-inch (2.3 m) Bok telescope located on Kitt Peak.
This survey has imaged $\sim$5~000 deg$^2$ in the SGC with 80\% of the area overlapped with the southern SDSS data, but 1-1.5 magnitudes deeper than the SDSS $u$-band photometry.
The $u$-band filter used in the SCUSS project is similar to the SDSS $u$-band filter but slightly bluer.
For our aims, this deep $u$-band imaging brings valuable improvement with respect to the SDSS $u$-band photometry, given that the $u$-band magnitude tightly correlates with the [\textsc{Oii}] flux \citep{comparat15a} and that the typical ELGs at $\redshiftmin \le z_{\rm spec} \le \redshiftmax$ are too faint to have a robust measurement in the SDSS $u$ band.

 For SCUSS we also use a forced \textit{model} photometry on SDSS objects (Hu Zou et al., 2015, in preparation).
This photometry constructs 2D models (de Vaucouleurs and exponential) based on SDSS $r$-band galaxy profiles and star-galaxy separation, and estimates object fluxes through comparing the models with the object images of SCUSS.
The \texttt{modelMag} magnitudes in SCUSS is derived from the object flux with higher likelihood in the de Vaucouleurs and exponential model fitting.

% CFHTLS degraded
\textbf{\textit{CFHTLS degraded to DECaLS.}}
We also test a scheme where the SDSS $z$ band is replaced by the DECaLS $z$ band.
The DECam Legacy Survey\footnote{\href{http://portal.nersc.gov/decals/}{http://portal.nersc.gov/decals/}} (DECaLS) will conduct a three-band imaging survey of the SDSS extragalactic footprint.
The Dark Energy Camera (DECam) will be used to image the 6700 deg$^2$ footprint overlapping SDSS in the region $-20<\textnormal{Dec. [deg]}<30$, to depths of $g=24.7$ mag, $r=23.9$ mag, and $z=23.0$ mag.
The survey will be conducted from 2014 through 2017, with periodic data releases beginning in March 2015.
As of January 2015,  $>$1000 deg$^2$ have already been observed in the SGC in the $z$ band to a depth of $z=22.7$ mag (whereas the $g$- and $r$-band observations have a significant lower coverage).
To simulate the DECaLS photometry, which is not public yet, we degrade the CFHTLS-Wide W1\footnote{\href{http://terapix.iap.fr/cplt/T0007/doc/T0007-doc.html}{http://terapix.iap.fr/cplt/T0007/doc/T0007-doc.html}} \citep{gwyn12} $z$-band photometry to the depth of the DECaLS $z$ band photometry by adding random Gaussian noise.
For both surveys we do a linear fit to the data: log$_{10}(em) =  zp + s \times m$, where $m$ is the magnitude and $em$ the magnitude error.
Then, if an object has $m_{\rm CFHTLS} = m_0$, we degrade it to $m_{\rm 0,deg} = m_0 + r \times \sqrt{10^{2 (zp_{\rm DECaLS}+s_{\rm DECaLS} \times m_0)} - 10^{2 (zp_{\rm CFHTLS}+s_{\rm CFHTLS} \times m_0)}}$, where $r$ is randomly drawn from a Gaussian distribution centred on 0 with a width of 1 and associate to this object a magnitude error of $10^{zp_{\rm DECaLS}+s_{\rm DECaLS} \times m_{\rm 0,deg}}$.
We note that we did not model the scatter in the magnitude error, since this feature has no influence on our selection process.

In addition to the DECaLS, the DES -- started in Autumn 2013 -- also uses the DECam instrument to image 5000 deg$^2$ in the $grizY$ bands to about two magnitudes deeper than the DECaLS.
Therefore, the DES $z$-band photometry for the $\sim$500 deg$^2$ overlapping region between the SDSS and DES could also be considered.\\

% Table: photometric surveys
\begin{table*}
        \begin{tabular}{lccc}
                \hline
                \hline
                Survey & BOSS/SGC ($\sim$3,100 deg$^2$) & BOSS/NGC ($\sim$7,500 deg$^2$) & Band depth\\
                & coverage & coverage & (mag)\\
                \hline
                SDSS & 100\% & 100\% & $g=23.1$, $r=22.7$, $i=22.2$, $z=20.7$\\
                WISE (forced photometry) & 100\% & 100\% & $W1=20.3$\\
                SCUSS &100\% & - & $u \sim 23$\\
                DECaLS/DR1 (CFHTLS-Wide degraded) & $\sim$30\% & $\sim$45\% & $z = 22.7$\\
                \hline
        \end{tabular}
        \caption{Photometric data properties summary (Section \ref{sec:photometry}).
Reported depths correspond to a 5$\sigma$ point-source detection.}
        \label{tab:phot}
\end{table*}

%=========================
% DATA/Fisher training sample
%=========================
\subsection{Fisher training sample \label{sec:spectrain}}

For each selection scheme, we need to define the Fisher discriminant quantity $X_{FI}$, i.e. the linear colour combination.
This is done through the use of a spectroscopic training sample, for which the used photometry and the quantity used to define the classes, the spectroscopic redshift $z_{\rm spec}$ (and eventually the total [O\textsc{ii}] flux $f_{[\rm O\textsc{ii}]}^{\rm tot}$), are known.
As explained in Section \ref{sec:fisher_elg}, our approach is to use a large, composite sample of galaxies, in order to cover the loci as much as possible in the colour space occupied by the different type of galaxies at all redshifts.
We list below the surveys we use to define our Fisher training sample.\\

% VIPERS
\textbf{\textit{VIPERS.}}
The VIMOS Public Extragalactic Redshift Survey \citep[VIPERS, ][]{guzzo14} is an on-going large programme that builds a spectroscopic sample of $10^5$ galaxies with $17.5 \le i \le 22.5$ and $0.5<z_{\rm spec}<1.5$ over a total area of 24 deg$^2$ within the CFHTLS-Wide W1 and W4 fields.
The observations are done with the VIMOS instrument \citep{le-fevre03} with the LR-RED grism (wavelength coverage: $\sim$5500-9500 \AA; spectral resolution: $R \sim 250$; 0.75h exposure time).
The low-redshift ($z_{\rm spec}<0.5$) galaxies are efficiently removed from the target section through a colour cut, resulting in a completeness $>95\%$ in the $0.6<z_{\rm spec}<1.2$ range \citep{guzzo14}.
This sample is crucial for our study, because it covers our targeted redshift range and is flux-selected with an $i$-band flux fainter than the SDSS $i$-band depth.
Among the 57~204 spectra of the First Data Release \citep{garilli14,franzetti14}, we restricted ourselves to those that have a secure redshift flag ($2 \le \textnormal{Flag} < 5$) and that are detected in the SDSS.

% Other public surveys
\textbf{\textit{Other public surveys.}}
In our Fisher training sample, we include SDSS galaxies belonging to the following public surveys:
the F02 and F22 fields of the VIMOS VLT Deep Survey (VVDS, \citealt{le-fevre05,le-fevre13}; VIMOS LR-RED grism: $\sim$5500-9500 \AA; $R \sim 250$; 0.75h and 4.5h),
the zCOSMOS 10k-Bright Spectroscopic Sample (\citealt{lilly09}; VIMOS MR grism: 5500-9650 \AA; $R \sim 600$; 1h exposure time),
and the EGS field of DEEP2 (\citealt{newman13a}; DEIMOS spectrograph: 6500-9300 \AA; $R \sim 5900$; 1h exposure time).
We consider only objects having reliable spectroscopic redshifts.
Those surveys, including VIPERS, are magnitude-limited, with a magnitude limit fainter than the SDSS depth (VVDS/F02: $17.5<i<24$, VVDS/F22: $17.5<i<22.5$, zCOSMOS/Bright: $i<22.5$, DEEP2/EGS: $R<24.1$), so that they include all possible types of galaxies detected in the SDSS (lying in the observed sky region).
We also add all the SDSS DR12 public spectroscopic redshifts (DR12, \citealt{alam15}; SDSS and BOSS spectrographs: 3800-9200 \AA~and  3650-10,400\AA; $R \sim 1500$-2500; 0.75-1h exposure time) covering those survey regions.\\

% Comparat+15
\textbf{\textit{Comparat et al. (2015) ELGs.}}
Futhermore, we enlarge our Fisher training sample by adding $\sim$10$^{4}$ ELGs observed as pilot programmes for eBOSS and DESI (Paper I).\\

% OII flux
\textbf{\textit{Total [O\textsc{ii}] flux.}}
In addition, we extracted the total [O\textsc{ii}] fluxes for the VIPERS, VVDS, and BOSS/eBOSS surveys in a consistent way from the spectra $f_{[\rm O\textsc{ii}]}^{\rm
tot}$ (see Paper I for details).

%=========================
% DATA/Spectroscopic verification sample
%=========================
\subsection{Spectroscopic verification sample \label{sec:speceboss}}
\begin{figure}
        \includegraphics[width=0.95\columnwidth]{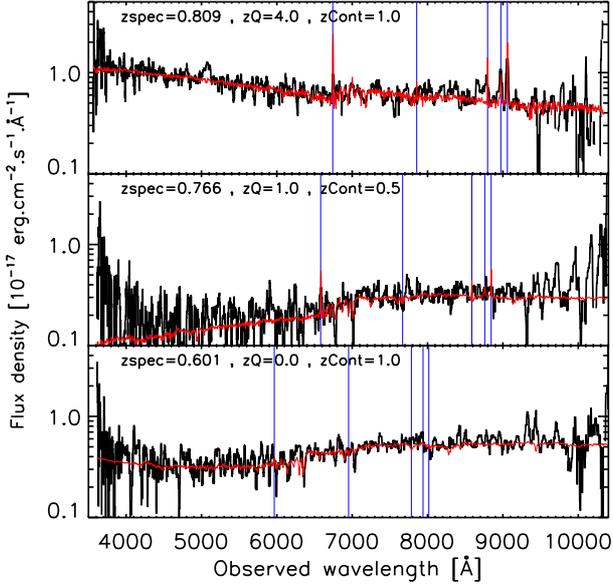}
        \caption{
        Example of three \texttt{eboss6-7} spectra, illustrating the confidence flags (zQ, zCont).
The observed spectra (black) is smoothed: each pixel is replaced by the median value of the 20 nearest pixels.
The best-fit model is in red.
Vertical blue lines illustrate the location of some expected emission lines in the observed frame: [O\textsc{ii}]$_{3725-3727}$, H$\gamma_{4342}$, H$\beta_{4862}$, [O\textsc{iii}]$_{4959}$, [O\textsc{iii}]$_{5007}$ (in increasing wavelength).}
        \label{fig:exspectra}
\end{figure}

We tested the efficiency of our ELG Fisher selection algorithms in a 8.82 deg$^2$ area centred approximately at (R.A., DEC.) $\sim$ (36.0,-4.8) (see bottom panel of Figure \ref{fig:surveys}), thus using a verification sample that is independent of our training sample ($\sim$6\% overlap).
This part of the sky has been extensively observed in 2014 with ten eBOSS/ELG test plates (\texttt{eboss6}: plates 8123-8130; \texttt{eboss7}: plates 8355, 8356), with $4\times$15 minutes exposures using the BOSS spectrograph (3650-10,400\AA; $R \sim 1500$-2500).
The \texttt{eboss7} plates have been specifically dedicated to two of our tested photometric Fisher discriminant selections (Fisher\_UGRIZW1 and  Fisher\_GRIW1, see Section \ref{sec:elg_selection}), which thus have a spectroscopic coverage of the target selection of $\sim$93\%, the remaining $\sim$7\% having not been targeted for tiling reasons.
The observations and their reduction are described in Paper I, to which we refer the reader.
In short, the observations were done with the BOSS spectrograph \citep{smee13} of the 2.5m telescope located at Apache Point Observatory (New Mexico, USA), using 2\arcsec~diameter fibers and an exposure time of $\sim$1h.
The reduction provides various information, with a confidence flag based on the continuum (zCont) and emission lines (zQ).
The zCont flag quantifies the degree to which the continuum is detected, and the zQ flag quantifies the number of detected emission lines, along with the signal-to-noise ratio ($S/N$) thereof.
We refer to Paper I for further details on the definition of zCont and zQ.
We have restricted this study to galaxies having a secure confidence flag; that is,\begin{subequations}
\begin{eqnarray}
        &\textnormal{zQ} \ge 1.5 \label{eq:spAll_flags_a}\\
        \textnormal{or} &  \textnormal{(zQ=1 and zCont} \ge 0.5) \label{eq:spAll_flags_b}\\
        \textnormal{or} &  \textnormal{(zQ=0 and zCont} \ge 1.0), \label{eq:spAll_flags_c}
\end{eqnarray}
\end{subequations}
meaning that the galaxy has either undoubted emission features -- one line at $S/N \ge 5$ or two or more lines at $S/N \ge 3$ (Eq.(\ref{eq:spAll_flags_a})) --
or a trustable combination of a detection of emission features and of the continuum:
one line at $S/N \ge 3$ and a continuum detected at $S/N \ge 8$ with at least three emission lines (Eq.(\ref{eq:spAll_flags_b})) or a continuum detected at $S/N \ge 10$ with at least three emission lines (Eq.(\ref{eq:spAll_flags_c})).
We illustrate in Figure \ref{fig:exspectra} those flags with three \texttt{eboss6-7} spectra.
The expected catastrophic $z_{\rm spec}$ estimate rate, estimated through the visual inspection of more than 10~000 BOSS spectra, can be estimated for each \{zCont, zQ\}.
We note that in the $\redshiftmin \le z \le \redshiftmax$ range, the galaxies selected with Eqs.(\ref{eq:spAll_flags_b},\ref{eq:spAll_flags_c}) have, on average. lower precision in the $z_{\rm spec}$ estimate (median value of $z_{\rm spec, err}/(1+z_{\rm spec})$ of 1.5-1.7$\times$10$^{-4}$ vs. 0.5$\times$10$^{-4}$) and a slightly higher catastrophic $z_{\rm spec}$ estimate rate when compared to the galaxies selected with Eq.(\ref{eq:spAll_flags_a}).

% Figure: example spectra

The \texttt{eboss6-7} data make it possible an unbiased and complete analysis for two of our tested selections (Fisher\_UGRIZW1 and  Fisher\_GRIW1).
For the remaining tested selections, those \texttt{eboss6-7} data represent a biased and incomplete subsample of the selections, so they cannot be used to reliably infer the selection properties.
To overcome this, we duplicate the analysis using complementary data for all the selections: the CFHTLS-Wide photometric redshifts for the redshift, and the VIPERS total [O\textsc{ii}] fluxes for the [O\textsc{ii}] diagnosis.
The CFHTLS-Wide photometric redshifts \citep[T0007 release\footnote{\href{http://terapix.iap.fr/rubrique.php?id\_rubrique=267}{http://terapix.iap.fr/rubrique.php?id\_rubrique=267}};][]{ilbert06a,coupon09} are of very good quality up to $i<22.5$ (bias below 1\%, scatter of $\sim$0.04, and less than 4\% outliers).
Using the \texttt{eboss6-7} test plates, we demonstrate the reliability of those photometric redshifts for ELGs up to redshift $\sim$1 in Appendix \ref{sec:zphot_properties}.
As explained in the previous section, the VIPERS sample is magnitude-complete down to $i = 22.5$ mag in the redshift range $0.6 \le z_{\rm spec} \le 1.2$.

Objects with a signal-to-noise ratio ($S/N$) of 3 in the  [O\textsc{ii}] flux measurements have on average $f_{[\rm O\textsc{ii}]}^{\rm tot} \sim 10^{-16.6}$ erg.s$^{-1}$.cm$^{-2}$ in the VIPERS observations and $f_{[\rm O\textsc{ii}]}^{\rm tot} \sim 10^{-16.4}$ erg.s$^{-1}$.cm$^{-2}$ in the \texttt{eboss6-7} observations.

%============================================================
% Improvement with using the WISE/W1 data
%============================================================
\section{Improvement with using the WISE/W1 data \label{sec:w1_improvement}}

\begin{figure*}
        \begin{tabular}{lcr}
        \includegraphics[width=0.31\linewidth,clip=true,trim=20 0 20 0]{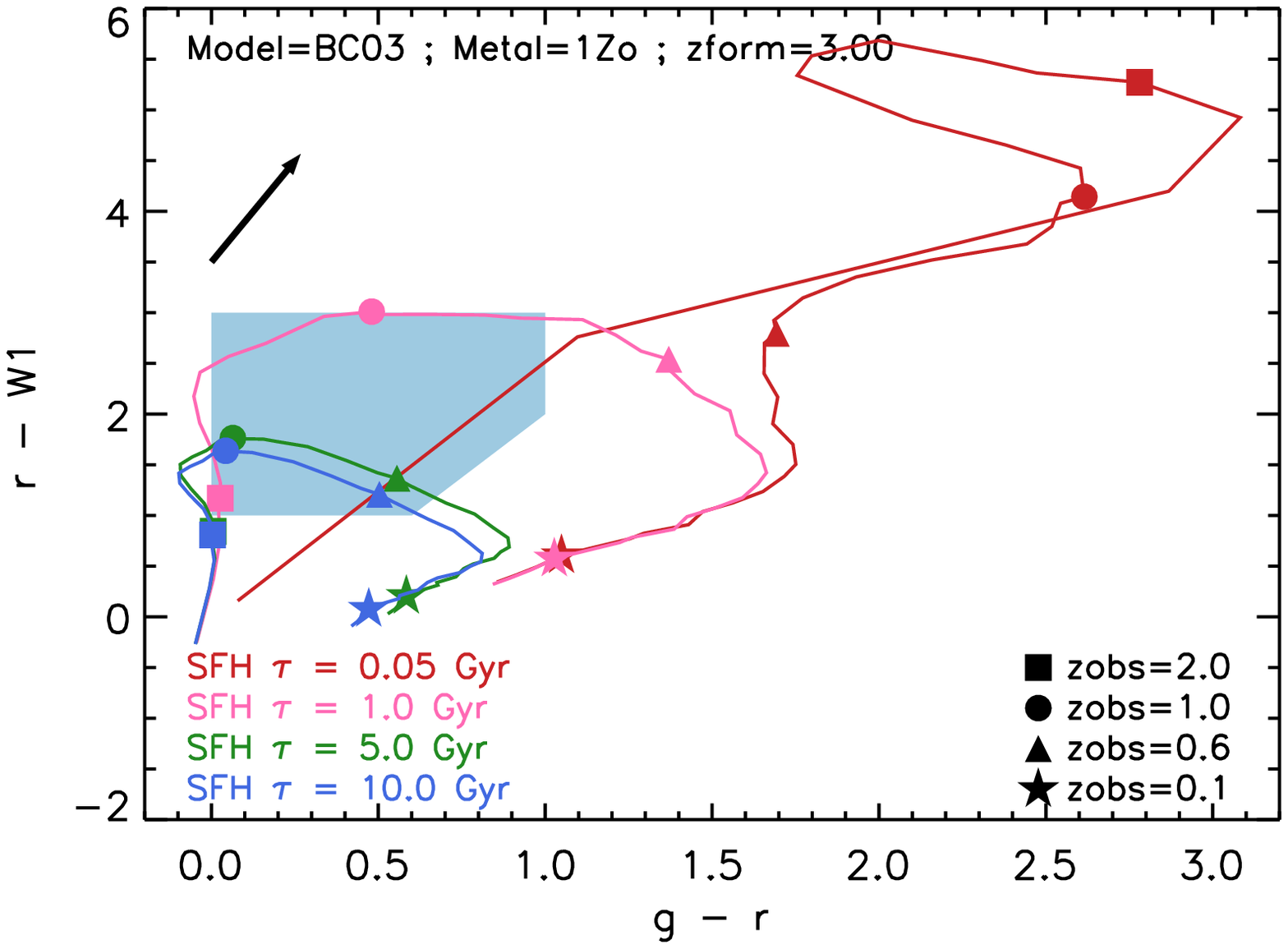} &
        \includegraphics[width=0.31\linewidth,clip=true,trim=20 0 20 0]{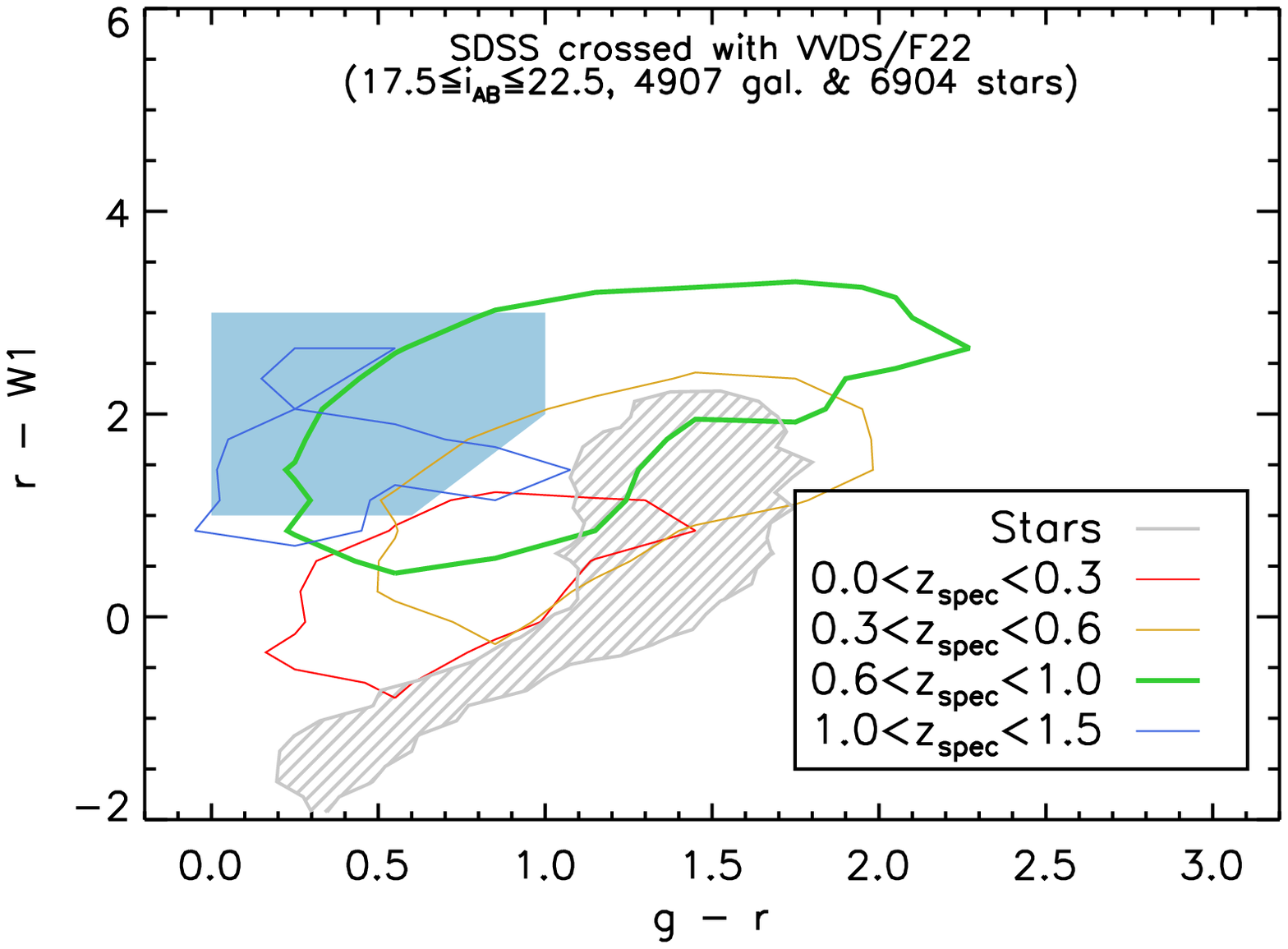} &
        \includegraphics[width=0.31\linewidth,clip=true,trim=20 0 20 0]{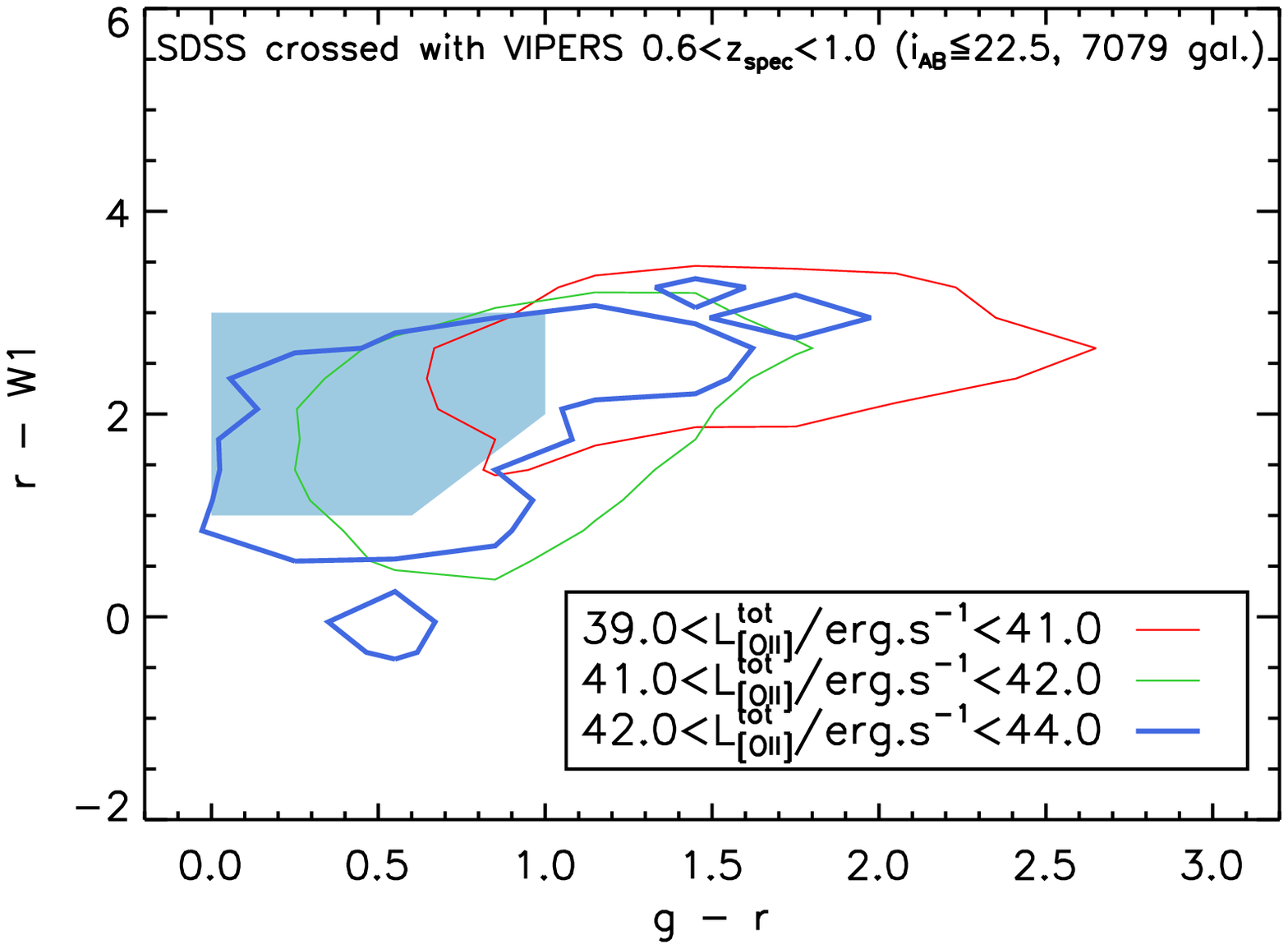}\\
        \end{tabular}
        \caption{
$r-W1$ vs. $g-r$ colour-colour plots illustrating how WISE helps in identifying $\redshiftmin \le z_{\rm spec} \le \redshiftmax$ ELGs.
\textit{Left panel}: BC03 stellar population models; models with SFH $\tau = 5, 10$ Gyr (SFH $\tau = 0.05, 1.0$ Gyr, respectively) are representative of ELGs (LRGs, respectively); all models are dust-free ($E(B-V)=0$); and the reddening due to the dust ($E(B-V) = 0.2$) is illustrated with the arrow in the top left corner.
To guide the eye, we indicate with blue shading the approximate locus where $\redshiftmin \le z_{\rm spec} \le \redshiftmax$ ELGs are expected to lie according to those models.
\textit{Middle panel}: SDSS objects in the VVDS/F22 survey ($17.5$ mag $< i < 22.5$ mag), binned in $z_{\rm spec}$; the hatched region is for stars.
\textit{Right panel}: SDSS $0.6 \le z_{\rm spec} \le 1.0$ galaxies in the VIPERS survey ($i < 22.5$ mag), binned in total [O\textsc{ii}] luminosity.
In the middle and right panels, we plot the region including 68\% of the data for each bin.}
        \label{fig:grW1}
\end{figure*}

The WISE/$W1$ near-infrared data bring crucial information for identifying the galaxy redshift, and the combination with other colours permits isolating $\redshiftmin \le z_{\rm spec} \le \redshiftmax$ ELGs.
We illustrate this point with the $r-W1$ vs. $g-r$ diagram in Figure \ref{fig:grW1}, using both model predictions and data.

In the left-hand panel, we plot the tracks predicted by the \citet[][]{bruzual03} stellar population models with standard settings ($z_{\rm form}=3$, solar metallicity) for four different exponentially declining star formation histories (SFH $\tau = 0.05, 1, 5, 10$ Gyr) with no dust.
Models with SFH $\tau = 5, 10$ Gyr are representative of ELGs, while models with SFH $\tau = 0.05, 1$ Gyr are representative of LRGs.
Besides, it is known that galaxies at $\redshiftmin \le z_{\rm spec} \le \redshiftmax$ with star formation can be dusty. We represent the effect of a $E(B-V) = 0.2$ reddening using the \citet{calzetti00} law, and this value corresponds to the median value for $\redshiftmin \le z \le \redshiftmax$ star-forming galaxies with $i < 22.5$ mag in the COSMOS catalogue \citep{ilbert09}.
To guide the eye, we shade the approximate locus where $\redshiftmin \le z_{\rm spec} \le \redshiftmax$ ELGs are expected to lie according to those models.

The assumptions made to compute the models (formation redshift, metallicity, SFH, dust) are simple and generic: to verify that the model predictions agree with observed galaxy properties, we look at the same colour-colour diagram, but with observed data.
We plot the loci of the SDSS objects belonging to complete spectroscopic surveys (see Section \ref{sec:spectrain}) with $i < 22.5$ mag.
In the middle panel, we gather the VVDS/F22 galaxies in bins of spectroscopic redshift: the colour evolution with the redshift agrees with the model predictions, with $\redshiftmin \le z_{\rm spec} \le \redshiftmax$ galaxies lying at $1 \lesssim r-W1 \lesssim 3$ and spanning a wide range of $g-r$ colours.
Then we plot in the right-hand panel the $0.6 \le z_{\rm spec} \le 1$ VIPERS galaxies binned by [O\textsc{ii}] luminosity. Again, the data and the model predictions agree, with $\redshiftmin \le z_{\rm spec} \le \redshiftmax$ ELGs having blue $g-r \lesssim 1$ colours.

Thus, we see that the $\redshiftmin \le z_{\rm spec} \le \redshiftmax$ ELGs can be isolated using colours that include the WISE/$W1$ data.
The $z_{\rm spec}<0.6$ galaxies should lie in a different locus at bluer $r-W1$ colours; stars are also expected to minimally contaminate the ELG selection (Figure \ref{fig:grW1}, middle panel).
The $z_{\rm spec}>1$ galaxies lie at a locus overlapping the blue-shaded area; however, we note that our requirement that the eBOSS/ELG galaxies are detected in the SDSS significantly reduces the contamination from them, because the SDSS is too shallow to detect a large number of $z_{\rm spec}>1$ galaxies.

% Figure: cb07 grz tracks

%============================================================
% ELG selection with Fisher Discriminant
%============================================================
\section{ELG selection with Fisher discriminant \label{sec:elg_selection}}
We present in this section different target selections to account for different possible strategies.
The eBOSS/ELG observations are planned to be done in the SGC and to begin in autumn 2016. A key point is the availability of the photometric data ahead of observations.
The SDSS, SCUSS, and WISE data are already available, while the DECaLS data are in the process of acquisition and reduction.
A first possibility is to make use of the maximum photometric information available today on the SGC, i.e. to combine the SDSS, SCUSS, and WISE data.
A second possibility is to minimise the number of combined surveys, i.e. combining the SDSS and WISE data, to minimise possible systematics.
A last possibility is to take advantage of the near availability of the DECam/$z$ data, which will be two magnitudes deeper than the SDSS/$z$ data. Indeed, the DECaLS/$z$-band imaging already covers more than 1000 deg$^2$ over the SGC and should be made public through annual releases starting in March 2015, and the DES/$z$-band data over the ``fat''-Stripe-82 region should be made public in 2015.\\

We first present in this section the tested ELG selection algorithms, which were built on the Fisher discriminant (Section \ref{sec:sel_schemes}).
We present in Section \ref{sec:sel_colcol} selections based on colour-colour cuts.
The aim is two-fold: (1) We present the initially tested selection using $ugri$-bands to illustrate the improvement due to the addition of the WISE/$W1$ band; (2) the colour-colour cuts using WISE/$W1$ allow the comparison of the Fisher selections with classical methods.
We then analyse the selection's redshift and [O\textsc{ii}] emission properties, along with their efficiencies (Section \ref{sec:sel_properties}).
This analysis is presented in two complementary manners to overcome the fact that the \texttt{eboss6-7} test plates cover a biased subsample of some selections.
%in fact, the two \texttt{eboss7} test plates were designed to have a $\sim$93\% coverage of the Fisher\_UGRIZW1 and Fisher\_GRIW1 selections; however none of the ten \texttt{eboss6-7} test plates has been designed specifically for the other Fisher selections.
%For the redshift, we present results for spectroscopic redshifts (Fisher\_UGRIZW1 and Fisher\_GRIW1 selections) and for photometric redshifts (all selections).
%For the [O\textsc{ii}] properties, we present results for the \texttt{eboss6-7} test plates (Fisher\_UGRIZW1 and Fisher\_GRIW1 selections) and for the VIPERS objects (all selections).
In fact, the \texttt{eboss6-7} test plates were designed to sample some selections, amongst which the Fisher\_UGRIZW1, Fisher\_GRIW1, and CC\_UGRI selections (defined in Sections \ref{sec:sel_schemes} and \ref{sec:sel_colcol}). Those three selections thus have a $\sim$95\% coverage with \texttt{eboss6-7}, since the remaining $\sim$5\% were not targeted for tiling reasons and are thus an unbiased subsample.
However, none of the ten \texttt{eboss6-7} test plates have been specifically designed for the other tested selections: even if up to $\sim$90\% of a selection is observed with \texttt{eboss6-7}, the unobserved $\sim$10\% is a biased subsample, thus preventing reliable statistics to be computed.
For the redshift, we present results for \texttt{eboss6-7} spectroscopic redshifts (Fisher\_UGRIZW1, Fisher\_GRIW1, and CC\_UGRI selections) and for photometric redshifts (all selections).
For the [O\textsc{ii}] properties, we present results for the \texttt{eboss6-7} test plates (Fisher\_UGRIZW1, Fisher\_GRIW1, and CC\_UGRI) and for the VIPERS objects (all selections).
Lastly, we present a brief illustration of the flexibility in terms of target density of our approach (Section \ref{sec:sel_sgc}).

%=========================
% ELGsel/Why do we need a Fisher discriminant?
%=========================
\subsection{Why use a Fisher discriminant approach? \label{sec:why_fisher}}

The initially tested selection based on an SDSS-detection for eBOSS/ELG uses cuts in the $uri$ and $gri$ colour-colour diagrams.
This selection, which we label CC\_UGRI ("CC" for colour-colour), is explained in Paper I and has been tested with the \texttt{eboss6} observations.
Our analysis deals with the five $ugriz$ optical bands and the WISE/$W1$ near-infrared band, hence five independent colours.
Though it is possible to define boxes in some colour-colour diagrams using those five colours, this task is complex and subjective, and this motivated us to use this alternative Fisher discriminant approach, which automatically defines the cut in the full colour-colour space.
The only requirement is the availability of a training sample, which we have in hand thanks to the numerous spectroscopic data covering the SDSS footprint (see Section \ref{sec:spectrain}).

We thus present below the tested Fisher selections, and to compare their performance with classical colour-colour cuts, we also present two colour-colour cuts using the WISE/$W1$ band.
In addition, we present the CC\_UGRI selection performance to illustrate the improvement due to the addition of the WISE/$W1$ band.

%=========================
% ELGsel/Selection schemes
%=========================
\subsection{Fisher selection schemes \label{sec:sel_schemes}}

The strength of this Fisher selection scheme is its simplicity: our selections are solely based on the Fisher discriminant quantity ($X_{FI, \rm min} < X_{FI}$) and on cuts in magnitudes and magnitude errors.
$X_{FI, \rm min}$ is tuned so that the selection has the desired object density.
The tested selection schemes (including the Fisher discriminant definition) are described in Table \ref{tab:selection}.\\

% Table: Fisher selections
\begin{table*}
        \begin{tabular}{lcccc}
                \hline
                \hline
                & Fisher\_UGRIZW1 & Fisher\_GRIW1 & Fisher\_GRIW1$^{OII}$ & Fisher\_GRZW1$_{180}$, Fisher\_GRZW1$_{300}$\\
                \hline
                Photometry & SCUSS/$u$& \multicolumn{2}{c}{SDSS/$gri$} & SDSS/$gr$\\
                &SDSS/$griz$& \multicolumn{2}{c}{WISE/$W1$} & DECaLS-like/$z$\\
                &WISE/$W1$ & \multicolumn{2}{c}{} & WISE/$W1$\\
                &&&&\\
                $X_{FI}$ training & 
                        $z_{\rm spec}$ only (Eq. \ref{eq:class_z}) &
                        $z_{\rm spec}$ only (Eq. \ref{eq:class_z}) &
                        $z_{\rm spec}$ and $f_{[\rm O\textsc{ii}]}$ (Eq. \ref{eq:class_zOII}) &
                        $z_{\rm spec}$ only (Eq. \ref{eq:class_z})\\
                &&&&\\
                $X_{FI}$ definition&
                        \multicolumn{4}{c}{$X_{FI} = \alpha_0 + \alpha_{ur} \times (u-r) + \alpha_{gr} \times (g-r) + \alpha_{ri} \times (r-i) + \alpha_{rz}  \times (r-z)+ \alpha_{rW1} \times (r-W1)$}\\
                &\multicolumn{4}{c}{}\\
                &       
                        % Fisher_UGRIZW1
                        \makecell{$\left\{\begin{array}{l}
                                \alpha_0=+0.956\\
                                \alpha_{ur}=-0.650\\
                                \alpha_{gr}=-0.781\\
                                \alpha_{ri}=+0.065\\
                                \alpha_{rz}=+0.229\\
                                \alpha_{rW1}=+0.739\\
                                \end{array}\right.$} &
                        % Fisher_GRIW1
                        \makecell{$\left\{\begin{array}{l}
                                \alpha_0=+0.104\\
                                \alpha_{ur}=0\\
                                \alpha_{gr}=-1.308\\
                                \alpha_{ri}=+0.870\\
                                \alpha_{rz}=0\\
                                \alpha_{rW1}=+0.782\\
                                \end{array}\right.$} &
                        % Fisher_GRIW1_OII
                        \makecell{$\left\{\begin{array}{l}
                                \alpha_0=+1.103\\
                                \alpha_{ur}=0\\
                                \alpha_{gr}=-1.982\\
                                \alpha_{ri}=+0.298\\
                                \alpha_{rz}=0\\
                                \alpha_{rW1}=+0.701\\
                                \end{array}\right.$} &
                        % Fisher_GRZW1
                        \makecell{$\left\{\begin{array}{l}
                                \alpha_0=+0.519\\
                                \alpha_{ur}=0\\
                                \alpha_{gr}= -1.483\\
                                \alpha_{ri}=0\\
                                \alpha_{rz}=-0.120\\
                                \alpha_{rW1}=+0.967\\
                                \end{array}\right.$}\\
                &&&&\\
                $X_{FI}$ cuts & 
                        $1.321 < X_{FI}$ & 
                        $1.492 < X_{FI}$ & 
                        $1.544 < X_{FI}$ & 
                        $1.475, 1.141^\dagger < X_{FI}  $\\
                &&&&\\
                Magnitude cuts & 
                        % Fisher_UGRIZW
                        \makecell{$\left\{\begin{array}{l}
                                20.0<u<23.5 \; , \; uerr<1.0\\
                                20.0<g<22.5 \; , \; gerr<0.5\\
                                19.0<r<22.5 \; , \; rerr<0.5\\
                                19.0<i<21.5 \; , \; ierr<0.5\\
                                17.0<W1<21.0 \; , \; W1err<0.5\\
                                \end{array}\right.$} &
                        % Fisher_GRIW
                        \multicolumn{2}{c}{
                                \makecell{$\left\{\begin{array}{l}
                                        20.0<g<22.5 \; , \; gerr<0.5\\
                                        19.0<r<22.5 \; , \; rerr<0.5\\
                                        19.0<i<21.5 \; , \; ierr<0.5\\
                                        17.0<W1<21.0 \; , \; W1err<0.5\\
                                \end{array}\right.$}} &
                        % Fisher_GRZW
                        \makecell{$\left\{\begin{array}{l}
                                20.0<g<22.5 \; , \; gerr<0.5\\
                                19.0<r<22.5 \; , \; rerr<0.5\\
                                18.0<z<21.5 \; , \; zerr<0.5\\
                                17.0<W1<21.0 \; , \; W1err<0.5\\
                                \end{array}\right.$}\\
                Other cuts & \multicolumn{4}{c}{$\texttt{BINNED2}=0$}\\
                (for all selections) & \multicolumn{4}{c}{$\texttt{OBJC\_TYPE}=3$ or $r>22$}\\
                        & \multicolumn{4}{c}{SDSS photometric masks (\texttt{bright\_star}, \texttt{bad\_field}, \texttt{bright\_object\_rykoff}) and custom $W1$ mask}\\
                \hline
        \end{tabular}
        \caption{Criteria used to define our Fisher selections.
For the Fisher\_GRZW1 selection, the $^\dagger$ denotes quantities related to the deep (300 objects per deg$^2$) selection.}
        \label{tab:selection}
\end{table*}

As explained in Section \ref{sec:fisher_elg}, there are two possible approaches to defining the \texttt{Signal} and \texttt{Background} classes for the Fisher discriminant training.
The first possible approach is to use criteria on both spectroscopic redshift $z_{\rm spec}$ and the total [O\textsc{ii}] flux:

\begin{equation}
        \begin{array}{ll}
                \texttt{Signal}:& (0.6<z_{\rm spec}<1.2 \; \textnormal{and} \; f_{[\rm O\textsc{ii}]}^{\rm tot} > 10^{-16.1} \textnormal{erg.s}^{-1}.\textnormal{cm}^{-2})\\
                & \textnormal{or} \; (0.9<z_{\rm spec}<1.2 \; \textnormal{and} \; g<22 \; \textnormal{mag})\\
                \texttt{Background}: &z_{\rm spec}<0.5.\\
        \label{eq:class_zOII}
        \end{array}
\end{equation}
With this, we account for the possibility that in our training sample, we miss some [O\textsc{ii}] line measurement due to sky lines at $z_{\rm spec}>0.9$, while keeping objects at $0.9<z_{\rm spec}<1.2$  in our \texttt{Signal} class with significant flux in the $g$ band, since it is correlated with the [O\textsc{ii}] flux \citep[see][]{comparat15a}.
Because we are working with objects detected in the SDSS, the vast majority of the galaxies we want to exclude will be at low redshift, so we
only include those in our \texttt{Background} class (regardless of the [O\textsc{ii}] flux).
The second possible approach is to define our $Signal$ class using only the $z_{\rm spec}$ quantity:
\begin{equation}
        \begin{array}{ll}
                \texttt{Signal}:& 0.75<z_{\rm spec}<1.3\\
                \texttt{Background}: &z_{\rm spec}<0.5.\\
        \end{array}
        \label{eq:class_z}
\end{equation}

We tested five selections, based on three different survey combinations:
\begin{itemize}
\item Fisher\_UGRIZW1: SCUSS/$u$ + SDSS/$griz$ + WISE/$W1$;
\item Fisher\_GRIW1: SDSS/$gri$ + WISE/$W1$;
\item Fisher\_GRIW1$^{OII}$: SDSS/$gri$ + WISE/$W1$, with a Fisher training using Eq.(\ref{eq:class_zOII});
\item Fisher\_GRZW1$_{180}$: SDSS/$gr$ + CFHTLS-W/$z$ (degraded to the DECaLS depth) + WISE/$W1$;
\item Fisher\_GRZW1$_{300}$: same as Fisher\_GRZW1$_{180}$, but with higher object density.
\end{itemize}

Except for the Fisher\_GRIW1$^{OII}$ selection, all the Fisher trainings were done with Eq.(\ref{eq:class_z}).
Our choice to define the tested selections were guided by the eBOSS/ELG experiment requirements (number of targets, available imaging at the start of spectroscopic observations).
On the one hand, the Fisher\_UGRIZW1 selection is based on a broad wavelength coverage (from the $u$ band to the $W$) and is limited to the SGC: as mentioned in Section \ref{sec:photometry}, the SCUSS deep $u$-band photometry provides us with a measurement of the ultra-violet emission at redshifts \redshiftmin-\redshiftmax.
On the other hand, the Fisher\_GRIW1 selection has a narrower wavelength coverage, but has the advantages of being available on both the SGC and the NGC and of minimising the number of surveys used.
Eventually, the Fisher\_GRZW1 selections use a deeper -- hence less scattered -- $z$-band photometry. We tested two scenarios (target densities of 180 deg$^{-2}$ and  300 deg$^{-2}$) to see to what extent the target density can be increased, in case the available DECaLS $z$-band photometry is available for an area smaller than 1500 deg$^2$.

% Figure: Fisher vs. OII-zspec
\begin{figure}
        \includegraphics[width=0.95\columnwidth]{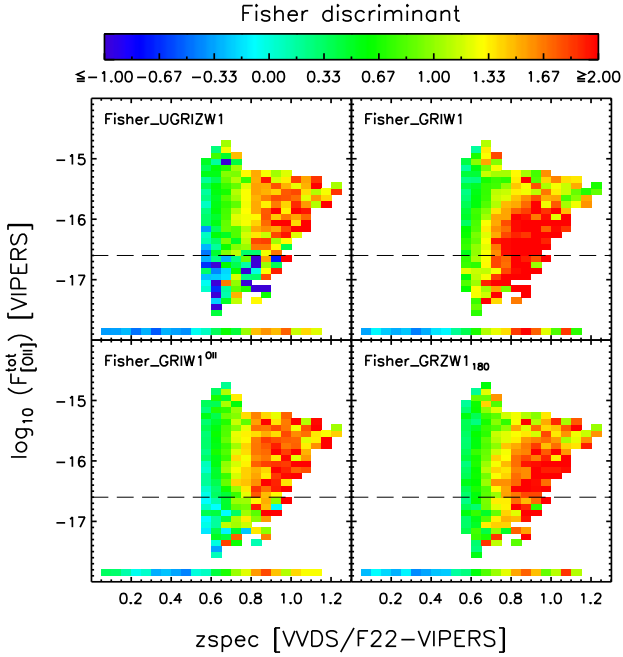}\\        
        \caption{Dependence of the four defined Fisher discriminants on $z_{\rm spec}$ and $F_{[\rm O\textsc{ii}]}^{\rm tot}$.
For each Fisher discriminant, we present the VIPERS data (main patch), while the VVDS/F22 data are displayed at log$_{10}$($F_{[\rm O\textsc{ii}]}^{\rm tot})=-17.8$ in $z_{\rm spec}$ bin only to illustrate the redshift dependence over $0<z_{\rm spec}<1.2$.
The horizontal dashed line represents the approximate $F_{[\rm O\textsc{ii}]}^{\rm tot}$ of objects with $S/N = 3$.
We recall that we only consider galaxies detected in the SDSS.
We only display bins with at least ten galaxies.}
        \label{fig:Fisher_OIIzspec}
\end{figure}

Figure \ref{fig:Fisher_OIIzspec} illustrates how the Fisher correlates with redshift and [O\textsc{ii}] flux.
The VIPERS sample, with [O\textsc{ii}] flux measurement in $0.6 \lesssim z_{\rm spec} \lesssim 1.2,$ allows us to see the simultaneous dependence on $z_{\rm spec}$ and $F_{[\rm O\textsc{ii}]}^{\rm tot}$, whereas we use the VVDS/F22 sample to probe the dependence only on $z_{\rm spec}$, but over a wide redshift range ($0 < z_{\rm spec} < 1.2$).
We recall that the data plotted here are the subsamples of the VIPERS and VVDS/F22 detected in the SDSS and that, due to the SDSS depth, the number of galaxies per redshift bin decreases for $z_{\rm spec} \gtrsim 0.6$.
The Fisher\_UGRIZW1 quantity correlates with both redshift and [O\textsc{ii}] flux, which means that selecting objects with high values of Fisher\_UGRIZW1 should be efficient in selecting ELGs in $\redshiftmin \le z \le \redshiftmax$.
We note that the Fisher training here is done on $z_{\rm spec}$ only (Eq. (\ref{eq:class_z})): the efficiency in selecting [O\textsc{ii}] emitters is a byproduct of the presence of the deep SCUSS $u$ band.
Indeed, the faint magnitude cut on the SCUSS photometry favours $z \sim 0.8$ star-forming galaxies against $z \sim 0.8$ passive galaxies, which have faint emission in the $u$ band \citep[see for instance][]{comparat15a}.
In contrast, we observe that, though having a strong correlation with redshift by training, the Fisher\_GRIW1 is inefficient at distinguishing [O\textsc{ii}] emitters.
Adding [O\textsc{ii}] in the training (Fisher\_GRIW1$^{OII}$, trained with Eq.(\ref{eq:class_zOII})) improves the efficiency in selecting [O\textsc{ii}] emitters, but at the cost of slightly reducing the correlation with redshift.
Lastly, the Fisher\_GRZW1 efficiently distinguishes the redshift, but favours low [O\textsc{ii}] emitters at $z_{\rm spec} \sim 1$.\\

Before applying the cut on the Fisher discriminant, we apply cuts on the magnitudes and their errors (see Table \ref{tab:selection}).
The aim is twofold: the cut on the bright magnitudes removes objects that surely are at low redshift ($z \le \redshiftmin$) from the samples. The cut on the faint magnitudes and on the magnitude errors ensures that we use reasonable photometry and mitigates inhomogeneous spatial distribution.
We note that the Fisher discriminant's dependence on redshift and [O\textsc{ii}] flux illustrated in Figure \ref{fig:Fisher_OIIzspec} does not change when those cuts are applied, though the sample is significantly smaller.

We also apply the following cuts:
1) we reject potential stars,
2) we reject objects lying in regions where the photometry is uncertain,
3) we reject objects with $\texttt{BINNED2} \neq 0$.
To reject potential stars, we reject objects with $\texttt{OBJC\_TYPE} \neq 3$ and $r<22$, where \texttt{OBJC\_TYPE} is the SDSS star/galaxy separator\footnote{\href{http://www.sdss.org/dr12/algorithms/classify/\#photo\_class}{http://www.sdss.org/dr12/algorithms/classify/\#photo\_class}}).
Regarding the regions where the photometry is uncertain, we use the SDSS masks (\texttt{bright\_star}, \texttt{bad\_field}, \texttt{bright\_object\_rykoff}; see \href{http://data.sdss3.org/sas/dr10/boss/lss/}{http://data.sdss3.org/sas/dr10/boss/lss/}) and a custom mask for the $W1$ bright-object neighbourhood.
In fact, as explained in \citet[][see their Figure 10]{lang14a}, SDSS objects falling in $W1$ halo outskirts are not masked and have a significantly overestimated $W1$ flux -- so are very red in $r-W1$.

To build our custom $W1$ mask, we detect spatial 5$\sigma$ overdensities of SDSS $r<22.5$ mag objects with $3 \; \textnormal{mag} < r-W1 < 5 \; \textnormal{mag}$ and mask objects within a radius of $0.4^{\circ}$ around those overdensities\footnote{To detect overdensities, we first create a density map with a $0.1^{\circ} \times 0.1^{\circ}$ binning and then run \textsc{SExtractor} \citep{bertin96} with $\texttt{DETECT\_MINAREA}=10$  and $\texttt{DETECT\_THRESH}=5$.}.
Our custom $W1$ mask includes $\sim$200 overdensities in the SGC (and $\sim$350 in the NGC).
We note that the masking is independent of any ELG selection.
We also notice that, though the CFHTLS/W1 region used in this study does not include bright $W1$ objects, such a $W1$ mask is nevertheless required when using our selections over larger regions, as for instance in Paper III, where the homogeneity and systematics analysis of our selections over the SGC is done.
Eventually we only keep objects with $\texttt{BINNED2}=0$, because this (slightly) reduces the number of objects with unexploitable spectra. (\texttt{BINNED2} is one of the SDSS photometric flags\footnote{\href{http://skyserver.sdss.org/dr9/en/help/browser/enum.asp?n=PhotoFlags}{http://skyserver.sdss.org/dr9/en/help/browser/enum.asp?n=PhotoFlags}}.)

%=========================
% ELGsel/Color-color selections
%=========================
\subsection{Colour-colour cut selections} \label{sec:sel_colcol}

We also present some selections using colour-colour cuts in the following analysis.
First, we present the CC\_UGRI selection, which uses cuts in the $uri$ and $gri$ colour-colour diagrams.
This selection is detailed in Paper I and has been tested with the \texttt{eboss6} observations.
It allows the improvement in the selection to be quantified thanks to the addition of the WISE/$W1$ band.

Additionally, we present two selections based on colour-colour cuts using the $ugrizW1$ bandset and the $griW1$ bandset, which we label CC\_UGRIZW1 and CC\_GRIW1.
Those allow a comparison of the Fisher discriminant selection performances with classical colour-colour cuts.
The CC\_UGRIZW1 selection has cuts similar to the Fisher\_UGRIZW1 (see Table \ref{tab:selection}) selection except the cut on the Fisher discriminant, which is replaced by
\begin{equation}
        \begin{array}{l}
                g-r > 2.0 \times (u-r)-3.00,\\
                g-r < 1.2 \times (r-i),\\
                i-z > -2.4 \times (r-i)+0.60, \; \textnormal{and}\\
                r-W1 > 2.0 \times (g-r)+0.35.\\
        \end{array}
        \label{eq:cc_ugrizw1}
\end{equation}
The CC\_GRIW1 selection has cuts similar to the Fisher\_GRIW1 (see Table \ref{tab:selection}) selection except the cut on the Fisher discriminant, which is replaced by
\begin{equation}
        \begin{array}{l}
                g-r < 1.0, \\
                r-i > 0.5, \; \textnormal{and}\\
                r-W1 > 2.5 \times (g-r)+0.25.\\
        \end{array}
        \label{eq:cc_griw1}
\end{equation}

%=========================
% ELGsel/Selection properties
%=========================
\subsection{Selection properties \label{sec:sel_properties}}

We recall that the redshift requirements for a 180 deg$^{-2}$ eBOSS/ELG target selection are
(1) an efficiency of 70\% in the $\redshiftmin \le z \le \redshiftmax$ range (i.e. at least 70\% of the targets have a measurable $z_{\rm spec}$ with $\redshiftmin \le z \le \redshiftmax$);
(2) a redshift failure rate $\lesssim$1\% in $\redshiftmin \le z \le \redshiftmax$.

\begin{figure}
        \includegraphics[width=0.95\columnwidth]{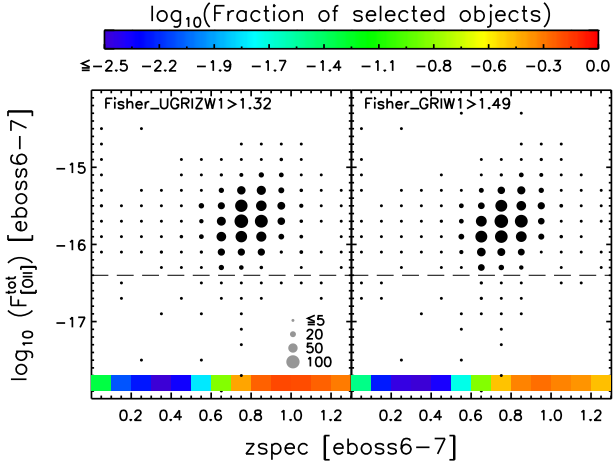}\\   
        \caption{Dependence of the Fisher\_UGRIZW1 and Fisher\_GRIW1 selections on $z_{\rm spec}$ and $F_{[\rm O\textsc{ii}]}^{\rm tot}$ for our observed \texttt{eboss6-7} plates.
The dot size scales with the number of objects entering the bin.
For each redshift bin (using $z_{\rm phot}$), we colour-coded at log$_{10}$($F_{[\rm O\textsc{ii}]}^{\rm tot})=-17.8$ the ratio of the number of selected photometric objects to the number of photometric objects passing all the cuts in Table \ref{tab:selection} except the cut on the Fisher discriminant.
The horizontal dashed line represents the approximate $F_{[\rm O\textsc{ii}]}^{\rm tot}$ of objects with $S/N = 3$.}
        \label{fig:Fisher_OIIzspec_eBOSS}
\end{figure}

Figure \ref{fig:Fisher_OIIzspec_eBOSS} represents the Fisher\_UGRIZW1 and Fisher\_GRIW1 selections in the [O\textsc{ii}] flux versus redshift diagram using our \texttt{eboss6-7} test plates measurements.
For each redshift bin (using $z_{\rm phot}$), we colour-code at log$_{10}$($F_{[\rm O\textsc{ii}]}^{\rm tot})=-17.8$ the ratio of the number of selected photometric objects to the number of photometric objects passing all the cuts in Table \ref{tab:selection}, except the cut on the Fisher discriminant.
We do not display the Fisher\_GRIW1$^{OII}$  and Fisher\_GRZW1 selections because their \texttt{eboss6-7} observations are biased.
As expected from the preliminary analysis of Figure \ref{fig:Fisher_OIIzspec}, the two Fisher selections are efficient in selecting $\redshiftmin \le z \le \redshiftmax$ galaxies with a significant [O\textsc{ii}] flux.
More precisely, we see that the cut on the Fisher discriminant is very efficient in rejecting $z_{\rm spec}<0.6$ objects, as it rejects $>99\%$ of those.

% Figure: eBOSS - Fisher vs. OII-zspec

We now study the magnitude and colours distributions for our tested selections, then the redshift distributions and the [O\textsc{ii}] emission; we finally present the overall statistics for all the tested selections.

%
% Colors
%
\subsubsection{Magnitudes and colours \label{sec:sel_col}}
\begin{figure}
        \includegraphics[width=0.95\columnwidth]{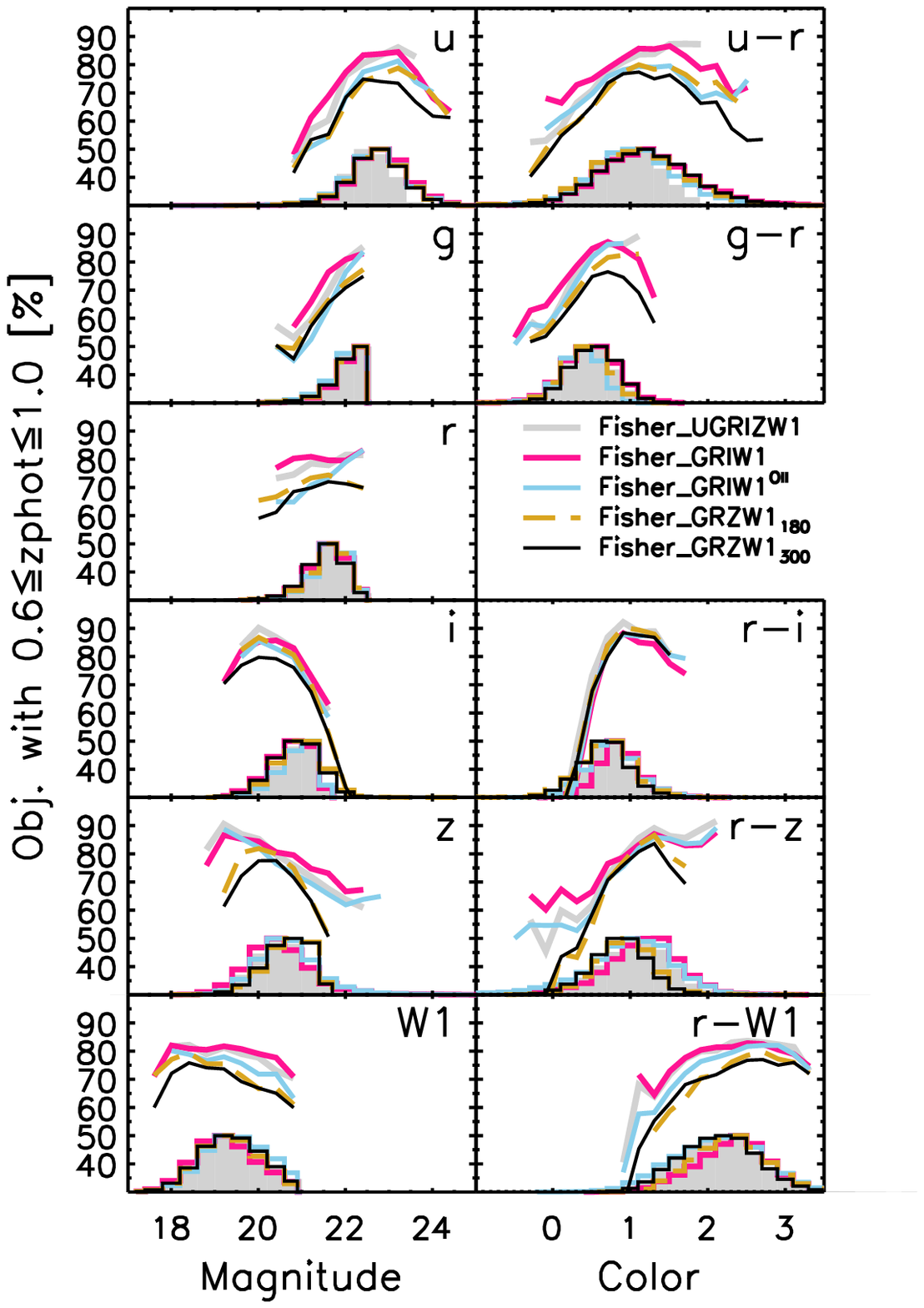}\\
        \caption{Photometric properties of the five Fisher and the CC\_UGRI selections. For each magnitude and colour, we report the percentage of selected objects having $\redshiftmin \le z_{\rm phot} \le \redshiftmax$, along with the magnitude normalised distribution of the selection.
For the Fisher\_GRZW1 selections, the $z$ band corresponds to the DECaLS-like photometry obtained from the CFHTLS-Wide $z$-band photometry.}
        \label{fig:selmagcol}
\end{figure}

Figure \ref{fig:selmagcol} summarises the magnitude and colour distributions of the selections, along with the percentage of selected objects with $\redshiftmin \le z_{\rm phot} \le \redshiftmax$.
Overall, the five Fisher selections have broadly similar magnitudes distributions, except for the $u$ band and the DECaLS/$z$-band, where we apply different magnitude cuts.
When compared to the four other Fisher selections, the Fisher\_UGRIZW1 selection almost has no objects with red $u-r>2$ colours, which are likely to be more passive galaxies.
One reason for that is the presence of an upper $u$-band magnitude limit in the selection criteria, which requires a minimum flux in the $u$ band.
For all colours, the Fisher\_GRIW1$^{OII}$ selection has a distribution that is bluer than the Fisher\_GRIW1 selection one, confirming the link between blue colours and [O\textsc{ii}] emission.
When compared to the Fisher\_GRZW1$_{180}$ selection, the Fisher\_GRZW1$_{300}$ selection shows small differences; for instance, it has slightly redder $g-r$ colours and slightly bluer $r-W1$ colours, consistent with a lower redshift (see Figure \ref{fig:grW1}).
Additionally, all our Fisher selections show common trends in the percentage of selected objects with $\redshiftmin \le z_{\rm phot} \le \redshiftmax$.
For instance, the selected objects with faint $u$ or $g$ magnitudes are, for the large majority, in our desired redshift range; however, the depth of the images and the requirement to obtain an usable spectrum with a 1h observation prevent us from pushing the selection at fainter magnitudes.

% Figure: mag/col histograms

\begin{figure}
        \includegraphics[width=0.95\columnwidth]{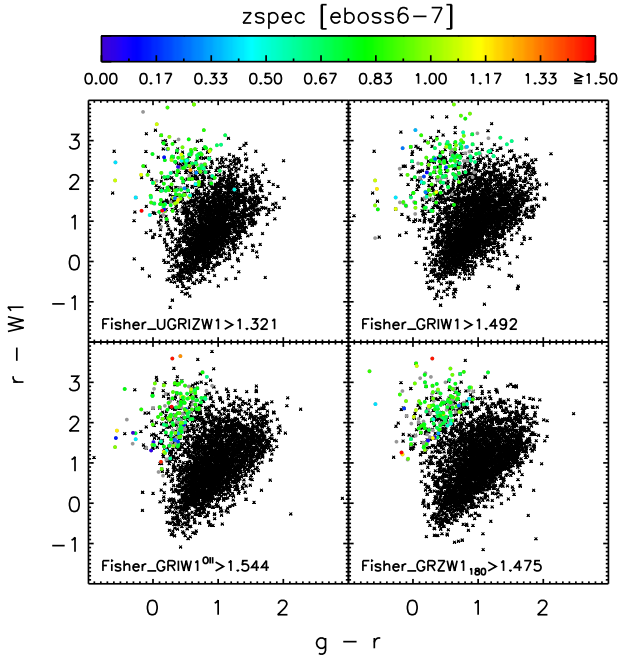}\\       
        \caption{$r-W1$ vs. $g-r$ colour-colour plot of the four defined Fisher discriminants.
For each selection, crosses represent photometric objects in the \texttt{eboss6-7} area passing all the cuts in Table \ref{tab:selection}, except the cut on the Fisher discriminant. Filled dots represent those passing the cut on the Fisher disriminant with the colour coding $z_{\rm spec}$ when observed with the \texttt{eboss6-7} test plates (grey colour when no reliable $z_{\rm spec}$ could be estimated from the spectrum or if the object has not been observed).
For clarity, we display only one out of ten objects.}
        \label{fig:Fisher_grW1}
\end{figure}

In Section \ref{sec:w1_improvement}, we studied the location of the $\redshiftmin \le z_{\rm phot} \le \redshiftmax$ ELGs in the $r-W1$ vs. $g-r$ colour-colour diagram.
We look in Figure \ref{fig:Fisher_grW1} at the same diagram for our tested Fisher selections: overall, the selected galaxies are indeed located in the expected region.
We can see some small differences in the loci occupied by the different selections.
For instance, the approximate cut in this diagram has a steeper slope for the Fisher\_UGRIZW1 and Fisher\_GRIW1$^{OII}$ selections than for the Fisher\_GRIW1 and Fisher\_GRZW1$_{180}$ selections, and this steeper slope implies the selection of more galaxies at $(g-r,r-W1) \sim (0,1)$ that have higher redshift and are strongly star-forming, while the flatter slope implies the selection of more galaxies at $(g-r,r-W1) \sim (1,2)$, with less star formation.

% Figure: grW1 for our field

%
% Redshift
%
\subsubsection{Redshift  \label{sec:sel_z}}
\begin{figure}
        \includegraphics[width=0.95\columnwidth]{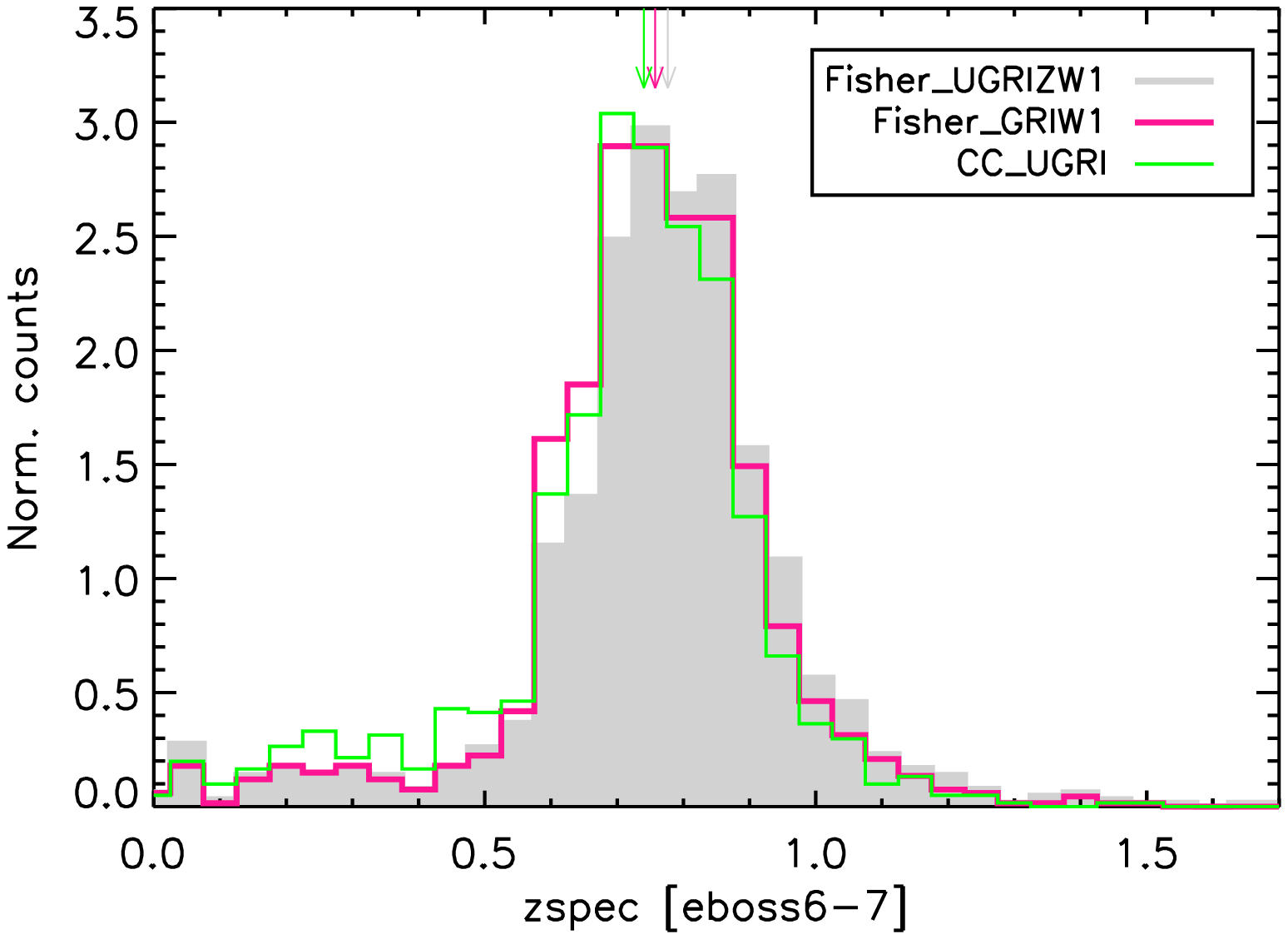}\\
        \includegraphics[width=0.95\columnwidth]{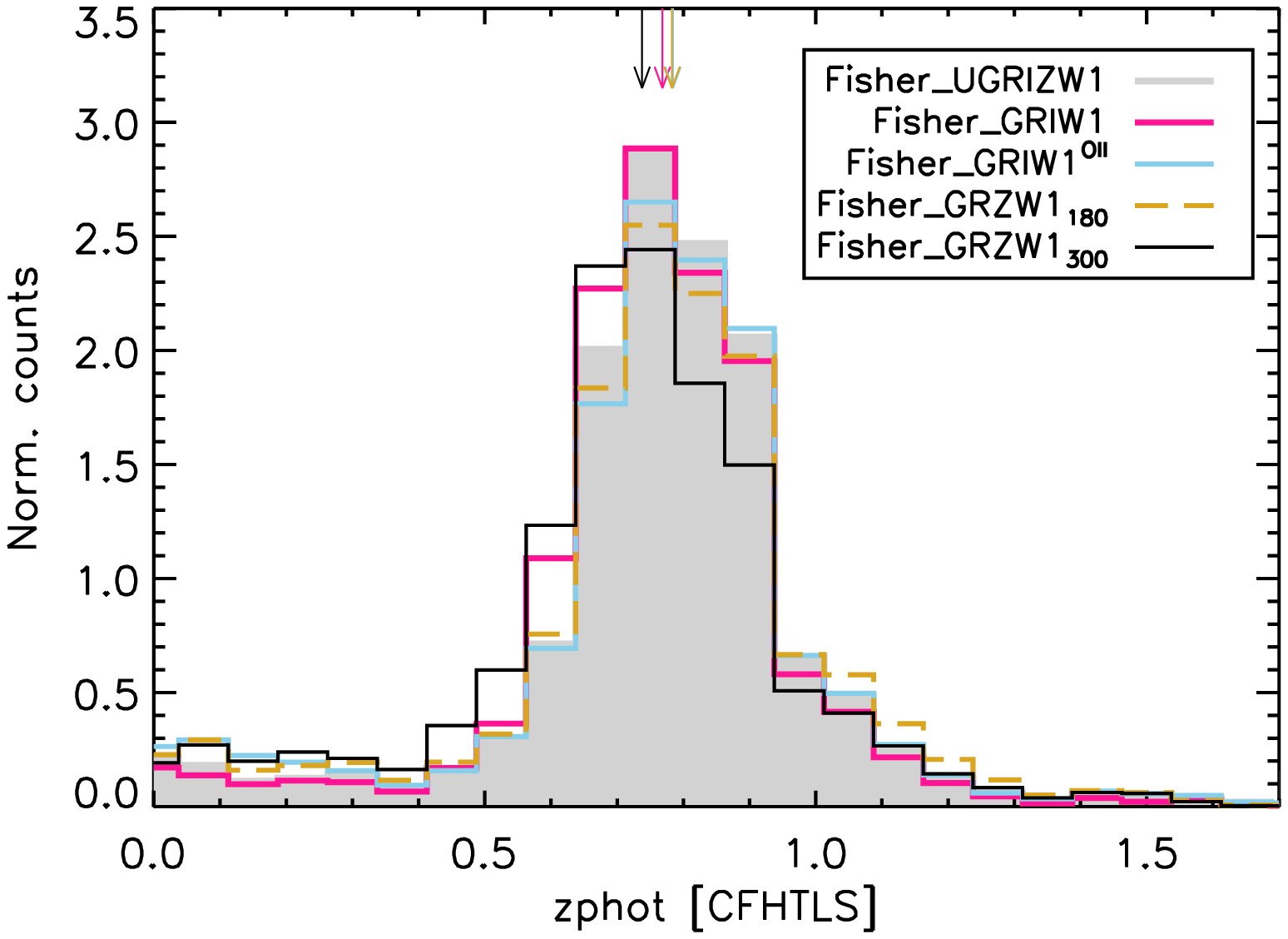}\\
        \includegraphics[width=0.95\columnwidth]{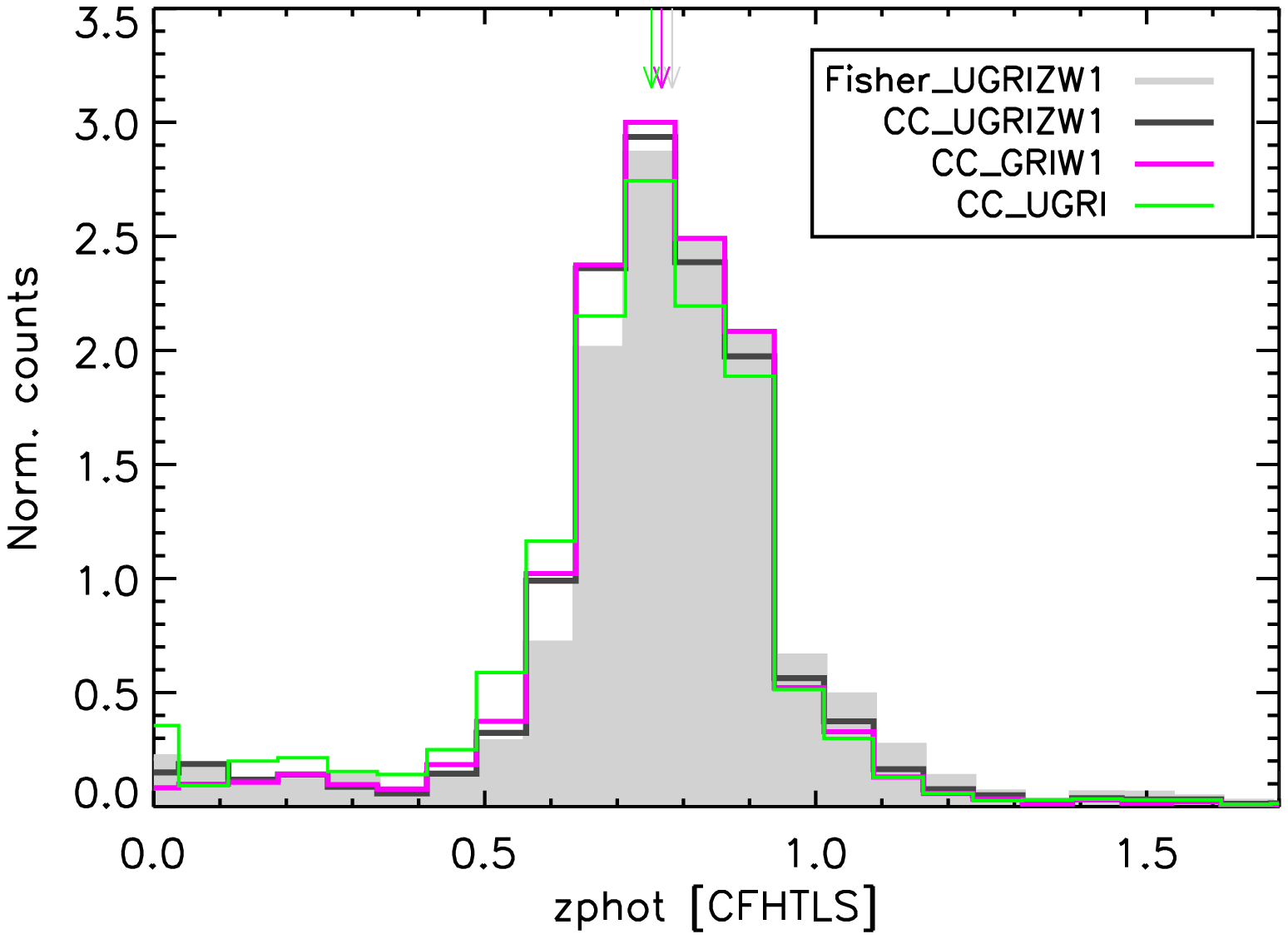}\\
        \caption{
Redshift distributions for the five Fisher and the CC\_UGRI selections.
The median redshift value is indicated by an arrow on the top x-axis.
\textit{Top panel}: \texttt{eboss6-7} $z_{\rm spec}$.
\textit{Middle panel}: CFHTLS $z_{\rm phot}$ for the Fisher selections.
\textit{Bottom panel}: CFHTLS $z_{\rm phot}$ for the colour-colour selections. We report the Fisher\_UGRIZW1 selection to facilitate the comparison.}
        \label{fig:selredshift}
\end{figure}

We now study the redshift distributions for the Fisher and colour-colour selections.
The top panel of the Figure \ref{fig:selredshift} represents the $z_{\rm spec}$ distribution from \texttt{eboss6-7} observations for the three selections that have unbiased sampling.
The middle (bottom, respectively) panel represents the CFHTLS/$z_{\rm phot}$ distribution for the Fisher selections (colour-colour selections, respectively).
We observe that the \texttt{eboss6-7} $z_{\rm spec}$ distributions for the Fisher\_UGRIZW1 and Fisher\_GRIW1 selections are close, i.e. peaked between $\redshiftmin \le z_{\rm phot} \le \redshiftmax$ with little contamination from $z_{\rm spec}<0.6$ objects.
The Fisher\_UGRIZW1 selection has a slightly higher median $z_{\rm spec}$ than the Fisher\_GRIW1 selection.
The CC\_UGRI selection has a distribution that is slightly shifted to lower redshifts with more contamination from $z_{\rm spec}<0.6$ objects.

We verify in the middle and bottom panels that the $z_{\rm phot}$ distributions faithfully mimic the $z_{\rm spec}$ distributions for the three selections of the top panel, though with a slight shift towards higher values of redshift at $\sim$1, as expected from Appendix \ref{sec:zphot_properties}.
Furthermore, we see that the Fisher\_GRIW1$^{OII}$ selection has more contamination from $z_{\rm phot}<0.6$ objects than the Fisher\_GRIW1 selection.
The Fisher\_GRZW1$_{180}$ selection has slightly more galaxies at $z>1$, while the Fisher\_GRZW1$_{300}$ selection has a distribution peaking at a lower redshift and has more $z_{\rm phot}<0.6$ contamination than the Fisher\_GRZW1$_{180}$ selection.

The colour-colour selections including the $W1$ band have $z_{\rm phot}$ distributions comparable to the Fisher selections.
We see that including the $W1$ band reduces the contamination from $z_{\rm phot}<0.6$ objects.

% Figure: redshift histograms

%
% OII flux
%
\subsubsection{[O\textsc{ii}] flux \label{sec:sel_oii}}
\begin{figure}
        \includegraphics[width=0.95\columnwidth]{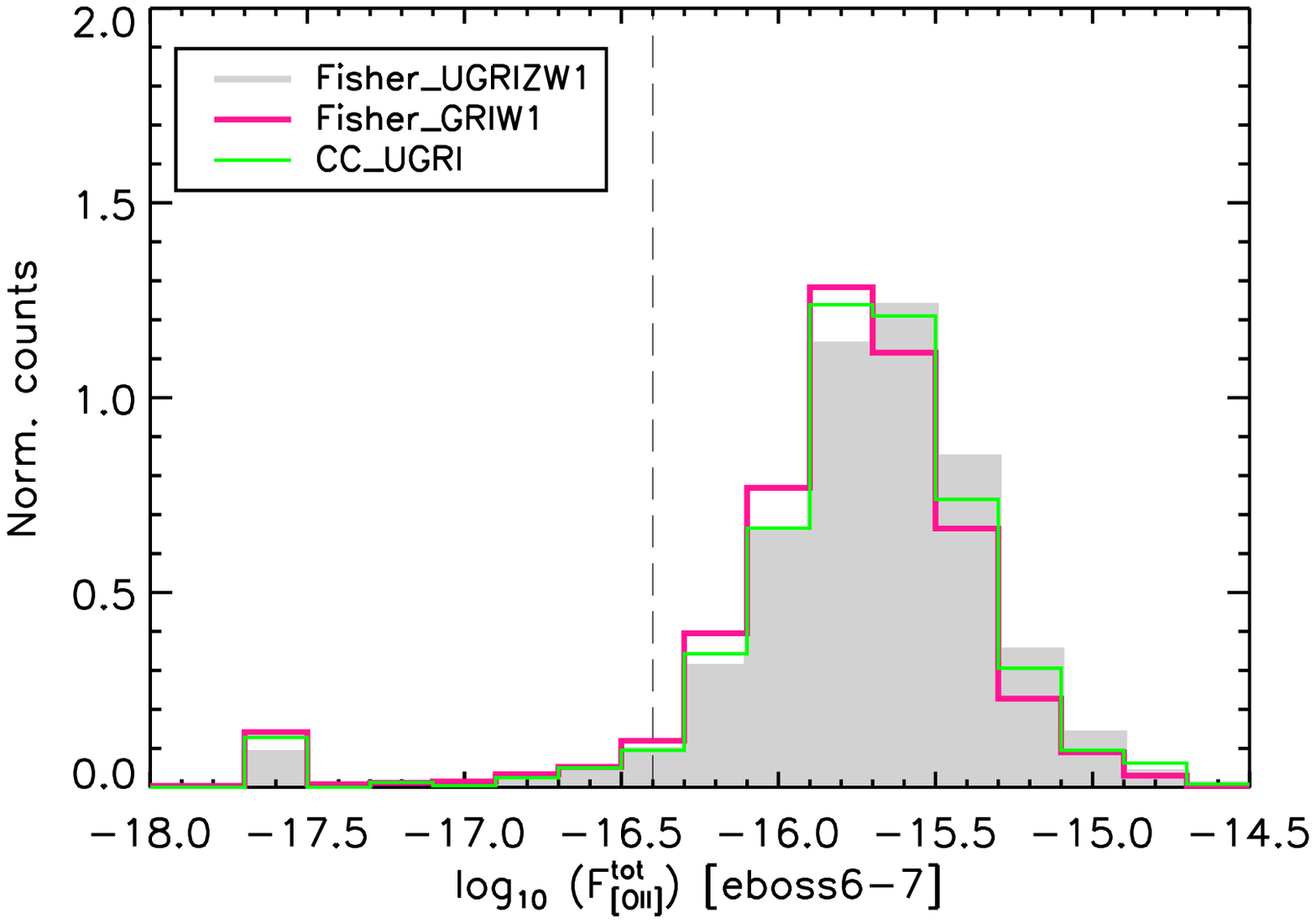}\\
        \includegraphics[width=0.95\columnwidth]{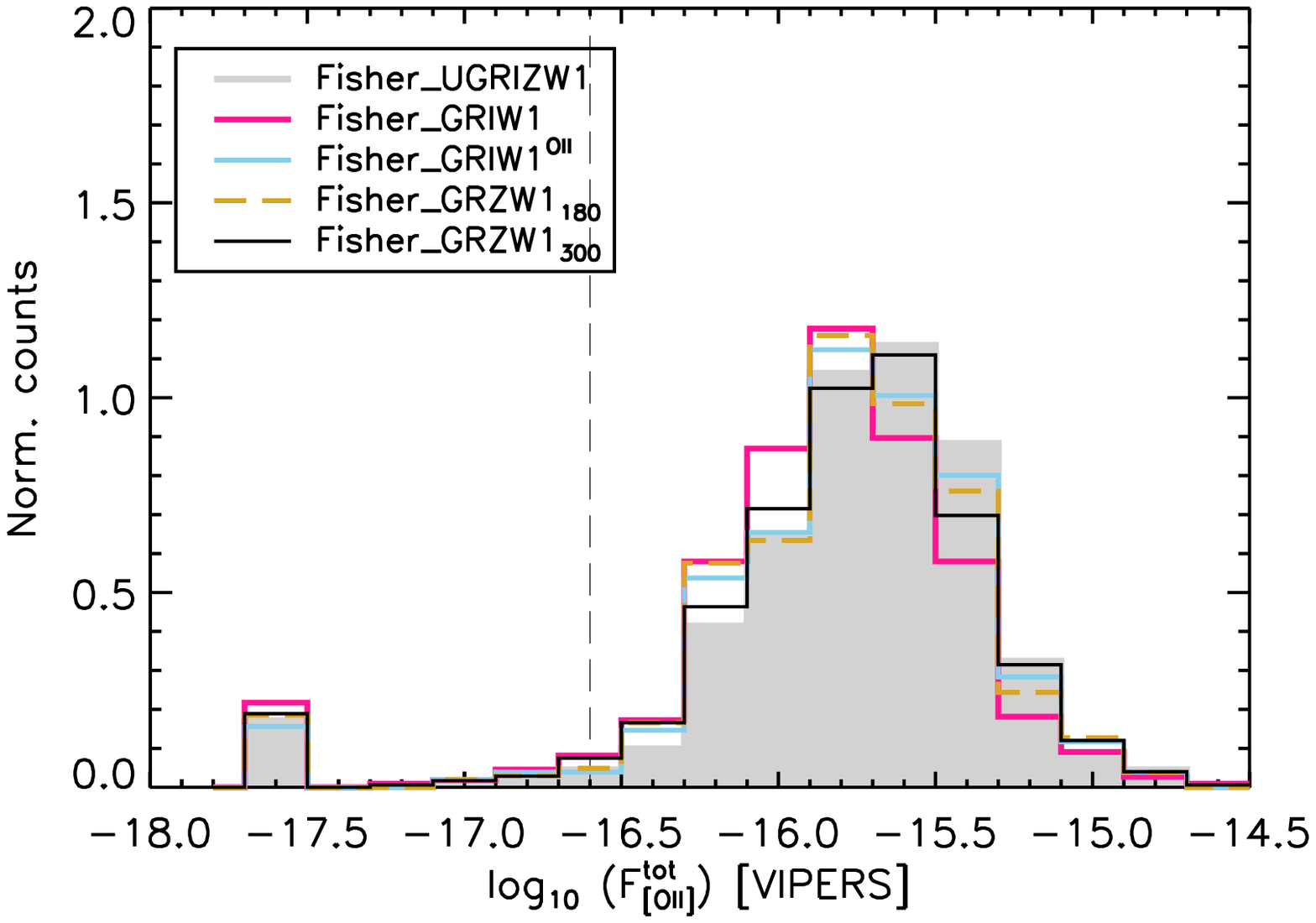}\\
        \includegraphics[width=0.95\columnwidth]{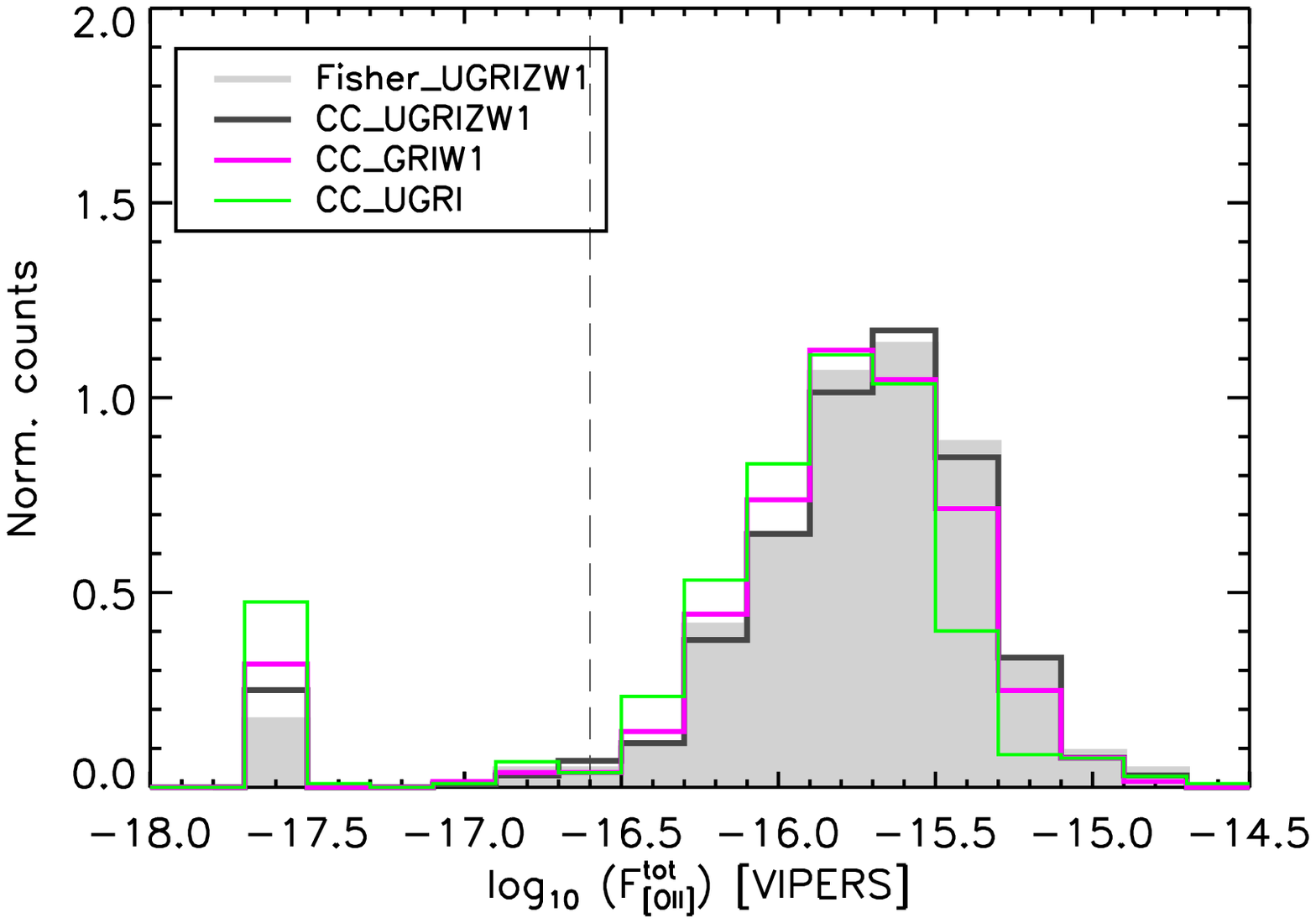}\\
        \caption{
Total [O\textsc{ii}] flux distributions for the five Fisher and the CC\_UGRI selections.
The vertical dashed line represents the approximate $F_{[\rm O\textsc{ii}]}^{\rm tot}$ of objects with $S/N = 3$.
Objects with unreliable [O\textsc{ii}] flux measurement are represented at $f_{[\rm O\textsc{ii}]}^{\rm tot} = -17.6$.
We consider here only those galaxies with a reliable $z_{\rm spec}$ measurement.
\textit{Top panel}: \texttt{eboss6-7} galaxies.
\textit{Middle panel}: VIPERS galaxies for the Fisher selections.
\textit{Bottom panel}: VIPERS galaxies for the colour-colour selections. We report the Fisher\_UGRIZW1 selection to facilitate the comparison.}
        \label{fig:FOII}
\end{figure}

\begin{figure}
        \includegraphics[width=0.95\columnwidth]{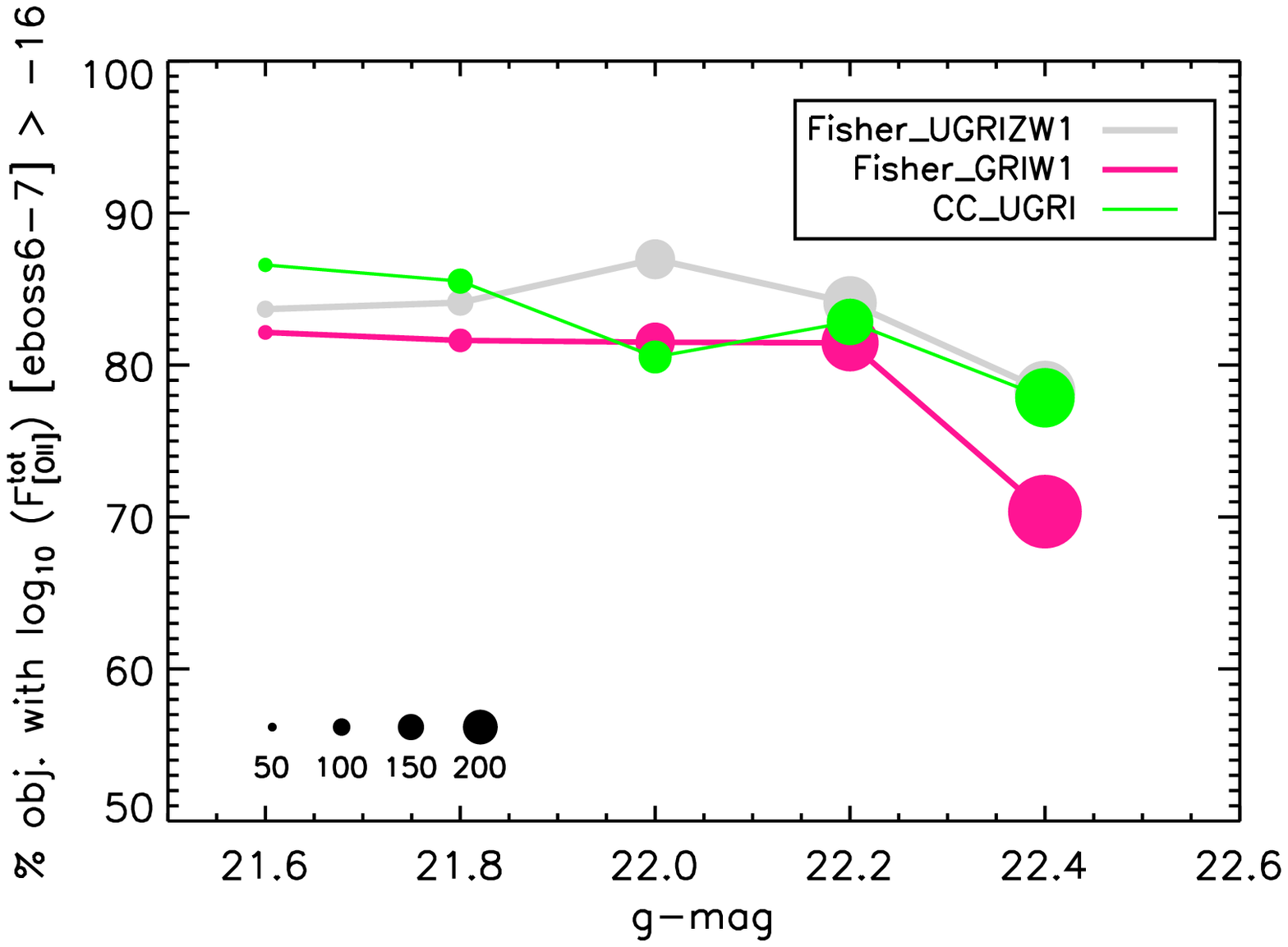}\\
        \includegraphics[width=0.95\columnwidth]{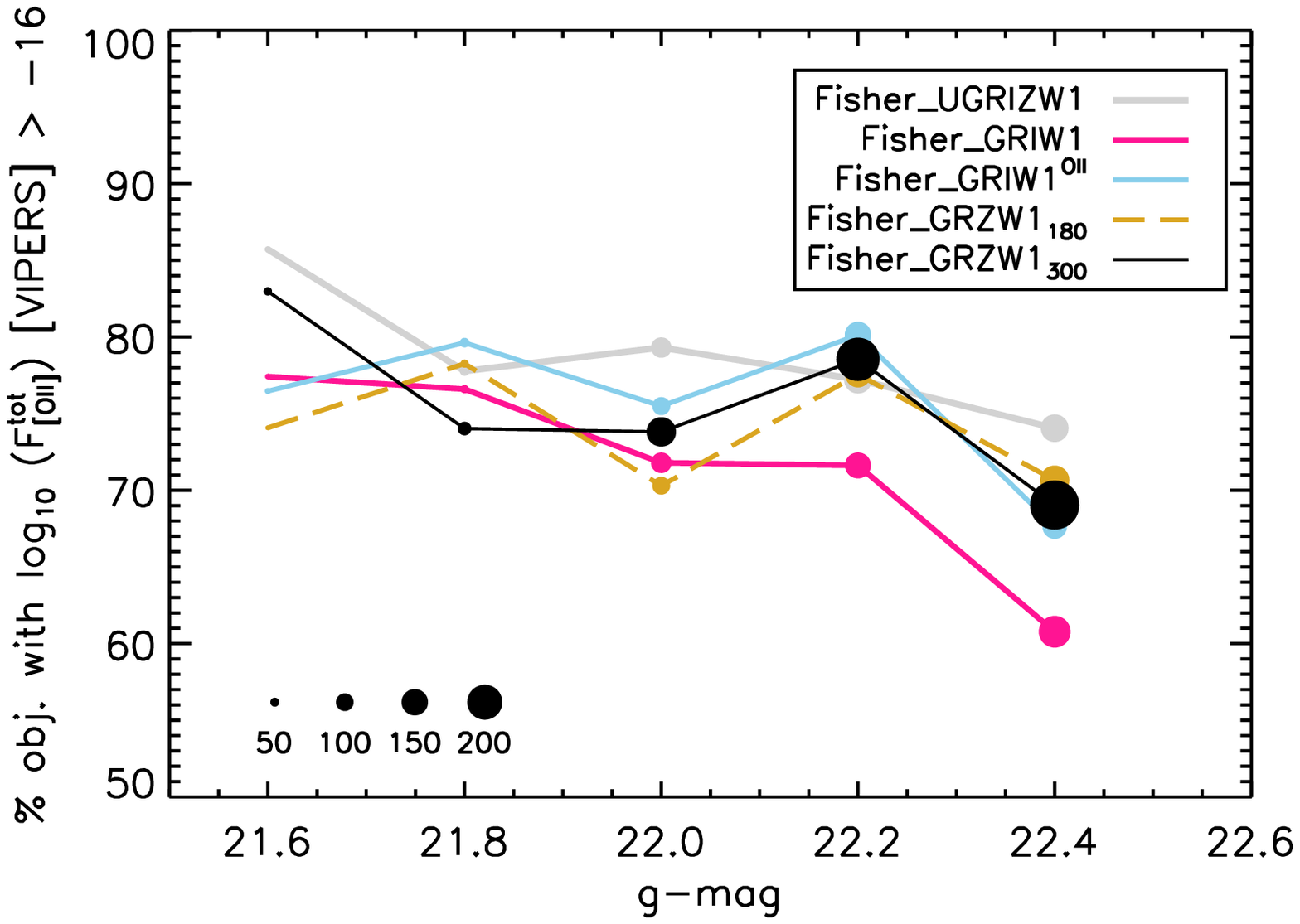}\\
        \includegraphics[width=0.95\columnwidth]{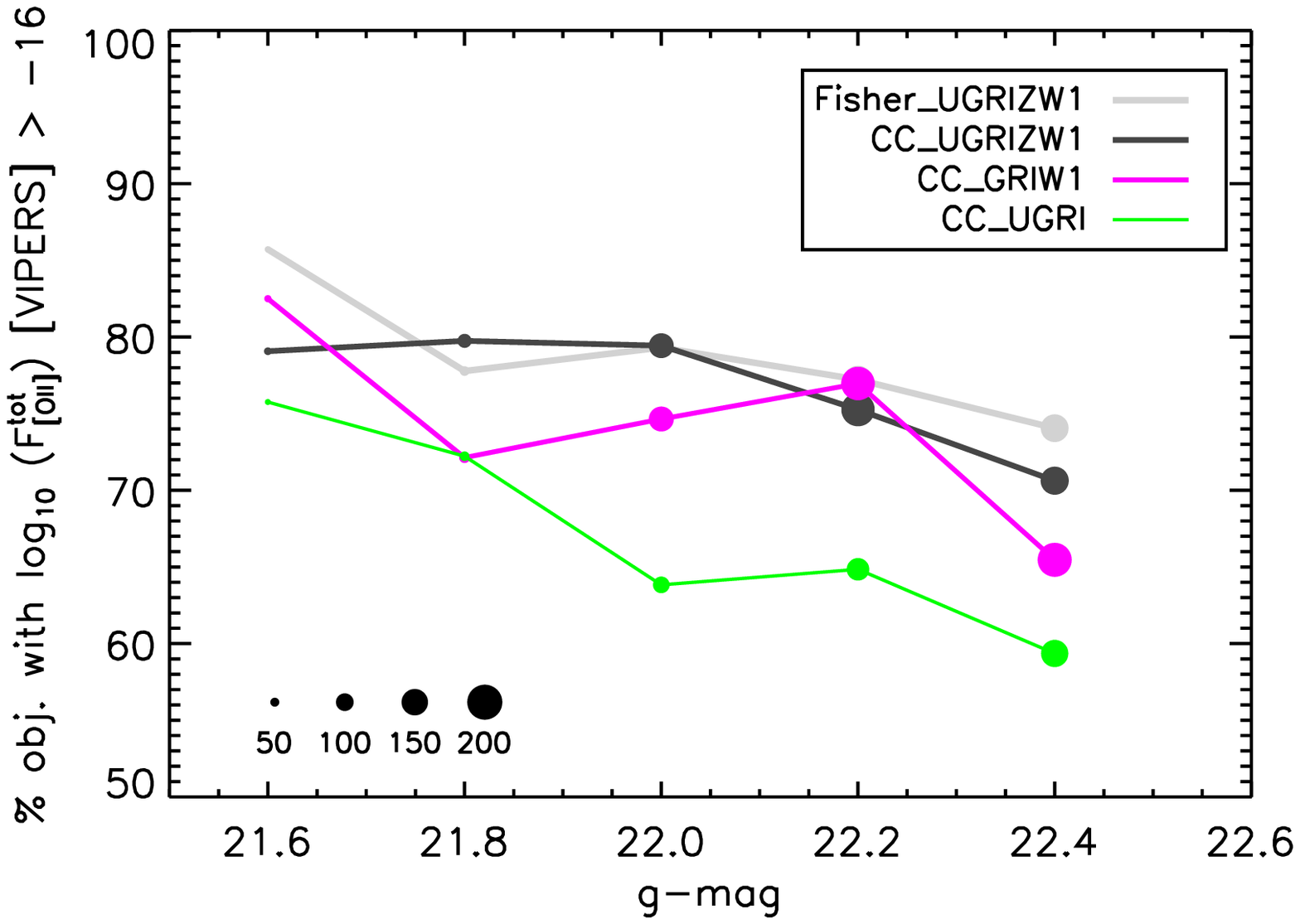}\\
        \caption{
Total [O\textsc{ii}] flux distributions vs. the $g$-band magnitude for the five Fisher and the CC\_UGRI selections.
We consider here only galaxies with a reliable $z_{\rm spec}$ measurement.
The dot size scales with the number of objects entering the bin.
\textit{Top panel}: \texttt{eboss6-7} galaxies.
\textit{Middle panel}: VIPERS galaxies for the Fisher selections.
\textit{Bottom panel}: VIPERS galaxies for the colour-colour selections. We report the Fisher\_UGRIZW1 selection to facilitate the comparison.}
        \label{fig:gFOII}
\end{figure}

 In Figure \ref{fig:FOII} we display the total [O\textsc{ii}] flux distributions, and in Figure \ref{fig:gFOII} the percentage of objects with $f_{[\rm O\textsc{ii}]}^{\rm tot} \ge 10^{-16}$ erg.cm$^{-2}$.s$^{-1}$.\AA$^{-1}$ as a function of the $g$-band magnitude.
Total fluxes above $10^{-16}$ erg.cm$^{-2}$.s$^{-1}$.\AA$^{-1}$ are detected at more than 5$\sigma$ in both the VIPERS and eBOSS. This threshold allows us to directly compare the percentages using the VIPERS and the eBOSS, overcoming the fact the VIPERS spectra have a slightly higher signal-to-noise ratio than the eBOSS spectra.\\

The Fisher\_UGRIZW1 selection emits more [O\textsc{ii}] than does the Fisher\_GRIW1 selection. This is consistently seen from \texttt{eboss6-7} and VIPERS data.
When looking at the VIPERS data, we see that, as expected from the Fisher training, the Fisher\_GRIW1$^{OII}$ selection is more [O\textsc{ii}]-emitting than the Fisher\_GRIW1 selection.
The Fisher\_GRZW1 selection is intermediate, and the [O\textsc{ii}] emission is similar for both tested densities (180 and 300 deg$^{-2}$).
The VIPERS data also show that the colour-colour selection has slightly less [O\textsc{ii}] emission than the Fisher\_UGRIZW1 selection.
Noticeably, the CC\_UGRI selection is significantly less [O\textsc{ii}]-emitting than the other selections.

Finally, we see the trend for all selections and \texttt{eboss6-7} and VIPERS data towards galaxies that are fainter in the $g$ band to have less [O\textsc{ii}] emission on average.
This result agrees with those from the \citet{comparat15a} study of the [O\textsc{ii}] luminosity function.

% Figure: OII flux histograms

% Figure: OII flux vs. g-mag

%
% Summary
%
\subsubsection{Statistics and summary  \label{sec:sel_summary}}
%
% Global statistics
%
% Tables 3&4
We give details on the properties of our tested Fisher selections in Table \ref{tab:selstats} and of the colour-colour selections in Table \ref{tab:selstats_colcol}. We present the statistics for redshift, [O\textsc{ii}] flux, and overall efficiency.
More precisely,
% photometric data
lines (L1-L6) present information computed with the photometric data over a $\sim$50 deg$^2$ area within the CFHTLS/W1 field (density, overlap with LRGs, photometric redshifts statistics);
% VIPERS
lines (L7-L8) report the numbers of $0.6 < z_{\rm spec} < 1.0$ VIPERS galaxies passing the selection, along with the percentage of those having $f_{[\rm O\textsc{ii}]}^{\rm tot}>10^{-16}$ erg.cm$^{-1}$.s$^{-2}$.
% eboss6-7
 Lines (L9-L14) present spectroscopic information for the \texttt{eboss6-7} plates observations covering an area of 8.82 deg$^2$ (number of galaxies, percentage of observed galaxies, percentage of galaxies with non reliable $z_{\rm spec}$ measurement, $z_{\rm spec}$ statistics). Finally, lines (L15-L16) present spectroscopic information for the \texttt{eboss6-7} galaxies with a reliable $z_{\rm spec}$ measurement with $\redshiftmin \le z \le \redshiftmax$, which are those objects that would be used for a BAO measurement (mean $z_{\rm spec}$, percentage of galaxies with $f_{[\rm O\textsc{ii}]}^{\rm tot}>10^{-16}$ erg.cm$^{-1}$.s$^{-2}$, expected percentage of galaxies with catastrophic $z_{\rm spec}$ estimation).
We now summarise the selection properties based on the above analysis of Figures \ref{fig:Fisher_OIIzspec_eBOSS}-\ref{fig:gFOII} and statistics from Tables \ref{tab:selstats}-\ref{tab:selstats_colcol}.\\

% Table: global stats
\begin{table*}
        \begin{tabular}{llccccc}
                \hline
                \hline
                 && Fisher\_UGRIZW1 & Fisher\_GRIW1 & Fisher\_GRIW1$^{OII}$  & Fisher\_GRZW1$_{180}$ & Fisher\_GRZW1$_{300}$\\
                \hline
                &\textbf{Photometric data ($\sim$50 deg$^2$)} &&&&&\\
                (L1) & Density (deg$^{-2}$)& 180 & 182 & 181 & 183 & 301\\
                (L2) & eBOSS/LRG overlap (deg$^{-2}$) & 2.0 & 3.5 & 2.4 & 2.7 & 3.4\\
                (L3) & median($z_{\rm phot}$) & 0.78 & 0.77 & 0.78 & 0.78 & 0.74\\
                (L4) & $z_{\rm phot}$ peak width$^{\dagger}$ & 0.12 & 0.13 & 0.12 & 0.13 & 0.14\\
                (L5) & \% with $\redshiftmin \le z_{\rm phot} \le \redshiftmax$ & 79\% & 80\% & 75\% & 73\% & 71\%\\
                (L6) & mean($z_{\rm phot}$) ($\redshiftmin \le z_{\rm phot} \le \redshiftmax$) & 0.79 & 0.78 & 0.79 & 0.79 & 0.77\\
                &&&&&&\\
                &\textbf{VIPERS ($0.6 \le z_{\rm spec} \le 1.0$)} &&&&&\\
                (L7) & N selected galaxies & 555 & 552 & 512 & 513 & 874\\
                (L8) & $f_{[\rm O\textsc{ii}]}^{\rm tot}>10^{-16}$ erg.cm$^{-1}$.s$^{-2}$ & 78\% & 69\% & 76\% & 74\% & 75\%\\
                &&&&&&\\
                &\textbf{Plates \texttt{eboss6-7} (8.82 deg$^2$)} &&&&&\\
                (L9) & N selected galaxies & 1609 & 1685 & 1684 & 1669 & 2763\\
                (L10) & Targeted & 92\% & 93\% & 89\% & 83\% & 68\%\\
                (L11) & Unreliable $z_{\rm spec}$ & 12\% & 15\% & (15\%) & (13\%) & (13\%)\\
                (L12) & Median($z_{\rm spec}^{\rm reliable}$) & 0.78 & 0.76 & (0.78) & (0.78) & (0.76)\\
                (L13) & $z_{\rm spec}^{\rm reliable}$ peak width$^{\dagger}$ & 0.12 & 0.12 & (0.12) & (0.12) & (0.13)\\
                (L14) & Efficiency ($\redshiftmin < z_{\rm spec}^{\rm reliable} < \redshiftmax$) & 71\% &71\% & (67\%) & (71\%) & (71\%)\\
                &&&&&&\\
                &\textbf{Plates \texttt{eboss6-7}},&&&&&\\
                &\textbf{\hspace{30pt} $\redshiftmin < z_{\rm spec}^{\rm reliable} < \redshiftmax$ only} &&&&&\\
                (L15) & Mean($z_{\rm spec}^{\rm reliable}$) & 0.79 & 0.77 & (0.79) & (0.78) & (0.77)\\
                (L16) & $f_{[\rm O\textsc{ii}]}^{\rm tot}>10^{-16}$ erg.cm$^{-1}$.s$^{-2}$ & 85\% & 81\% & (84\%) & (81\%) & (82\%)\\
                (L17) & Expected $z_{\rm spec}$ failure  & 0.5\% & 0.9\% & (0.7\%) & (0.8\%) & (0.8\%)\\
                \hline
        \end{tabular}
        \caption{
Summary of the Fisher selection properties.
% photometric data
Lines (L1-L6): information computed with the photometric data over a $\sim$50 deg$^2$ area within the CFHTLS/W1 field (density, overlap with LRGs identified using the cuts defined in Prakash et al. (2015), photometric redshifts statistics).
% VIPERS
Lines (L7-L8): number of $0.6 \le z_{\rm spec} \le 1.0$ VIPERS galaxies passing the selection, and percentage of those having $f_{[\rm O\textsc{ii}]}^{\rm tot}>10^{-16}$ erg.cm$^{-1}$.s$^{-2}$.
% eboss6-7
Lines (L9-L14): spectroscopic information for the \texttt{eboss6-7} plates observations covering an area of 8.82 deg$^2$ (number of galaxies, percentage of observed galaxies, percentage of galaxies with unreliable $z_{\rm spec}$ measurement, $z_{\rm spec}$ statistics).
% eboss6-7 0.6<z<1.0
Lines (L15-L16) present spectroscopic information for the \texttt{eboss6-7} galaxies with a reliable $z_{\rm spec}$ measurement with $\redshiftmin \le z \le \redshiftmax$ (mean $z_{\rm spec}$, percentage of galaxies with $f_{[\rm O\textsc{ii}]}^{\rm tot}>10^{-16}$ erg.cm$^{-1}$.s$^{-2}$, expected percentage of galaxies with catastrophic $z_{\rm spec}$ estimation).
For lines (L10-L16), we report in brackets the quantities derived from our spectroscopic observations for the Fisher\_GRIW1$^{OII}$ and Fisher\_GRZW1 selections: those quantities are biased because they are obtained from a non-random subsample constituted of objects passing the Fisher\_UGRIZW1 or Fisher\_GRIW1 selections.
$^{\dagger}$: the width is estimated through the fitting of Gaussian.}
        \label{tab:selstats}
\end{table*}

% Table: global stats for color-color cuts selections
\begin{table*}
        \begin{centering}
        \begin{tabular}{llccc}
                \hline
                \hline
                && CC\_UGRI & CC\_UgrizW1 & CC\_griW1\\
                \hline
                &\textbf{Photometric data ($\sim$50 deg$^2$)} &&&\\
                (L1) & Density (deg$^{-2}$)& 183 & 179 & 183\\
                (L2) & eBOSS/LRG overlap (deg$^{-2}$) & 2.4 &1.8 & 2.6\\
                (L3) & median($z_{\rm phot}$) & 0.75 & 0.77 & 0.77\\
                (L4) & $z_{\rm phot}$ peak width & 0.13 &0.12 & 0.12\\
                (L5) & \% with $\redshiftmin \le z_{\rm phot} \le \redshiftmax$ & 76\% & 81\% & 83\%\\
                (L6) & mean($z_{\rm phot}$) ($\redshiftmin \le z_{\rm phot} \le \redshiftmax$) & 0.78 & 0.78 & 0.78\\
                &&&\\
                &\textbf{VIPERS ($0.6 \le z_{\rm spec} \le 1.0$)} &&\\
                (L7) & N selected galaxies & 536 & 666 & 668\\
                (L8) & $f_{[\rm O\textsc{ii}]}^{\rm tot}>10^{-16}$ erg.cm$^{-1}$.s$^{-2}$ & 64\% & 76\% & 73\%\\
                &&&\\
                &\textbf{Plates \texttt{eboss6-7} (8.82 deg$^2$)} &&\\
                (L9) & N selected galaxies & 1604 & 1670 & 1692\\
                (L10) & Targeted & 96\% & (81\%) & (87\%)\\
                (L11) & Unreliable $z_{\rm spec}$ & 26\% & (10\%) & (13\%)\\
                (L12) & Median($z_{\rm spec}^{\rm reliable}$) & 0.74 & (0.77) & (0.76)\\
                (L13) & $z_{\rm spec}^{\rm reliable}$ peak width & 0.12 & (0.12) & (0.12)\\
                (L14) & Efficiency ($\redshiftmin < z_{\rm spec}^{\rm reliable} < \redshiftmax$) & 59\% & (76\%) & (73\%)\\
                &&&\\
                &\textbf{Plates \texttt{eboss6-7}}&&\\
                &\textbf{\hspace{30pt} $\redshiftmin < z_{\rm spec}^{\rm reliable} < \redshiftmax$ only} &&\\
                (L15) & Mean($z_{\rm spec}^{\rm reliable}$) & 0.77 & (0.78) & (0.78)\\
                (L16) & $f_{[\rm O\textsc{ii}]}^{\rm tot}>10^{-16}$ erg.cm$^{-1}$.s$^{-2}$ & 78\% & (86\%) & (85\%)\\
                (L17) & Expected $z_{\rm spec}$ failure & 1.1\% & (0.5\%) & (0.7\%)\\
                \hline
        \end{tabular}
        \caption{Summary of the colour-colour selections properties. Lines are similar to Table \ref{tab:selstats}}
        \label{tab:selstats_colcol}
        \end{centering}
\end{table*}

% Fisher\_UGRIZW1
\textbf{\textit{Fisher\_UGRIZW1 selection.}}
This selection meets the initial eBOSS/ELG target selection redshift criteria.
It has an efficiency of 71\% and an expected $z_{\rm spec}$ failure rate of 0.5\%.
We observe that it has a narrow $z_{\rm spec}$ distribution with a typical width of 0.12. The $z_{\rm phot}$ distribution is very similar.
The median $z_{\rm spec}$ of the selection is 0.78.
As expected from the preliminary study (Figure \ref{fig:Fisher_OIIzspec}), the Fisher\_UGRIZW1 selection is efficient in selecting [O\textsc{ii}] emitters, and this can be seen in particular in Figure \ref{fig:gFOII} and also in the lower value of objects with an unreliable $z_{\rm spec}$ or with the small overlap with LRGs (12\% and 2.0 deg$^{-2}$, respectively; see Table \ref{tab:selstats}).\\

% Fisher\_GRIW1
\textbf{\textit{Fisher\_GRIW1 and Fisher\_GRIW1$^{OII}$ selections.}}
The Fisher\_GRIW1 selection also meets the initial eBOSS/ELG target selection redshift criteria, the efficiency and expected $z_{\rm spec}$ failure rate in $\redshiftmin \le z_{\rm spec} \le \redshiftmax$ being close to the requirements (71\% and 0.9\%, respectively).
The shape of the redshift distribution is close to the one of the Fisher\_UGRIZW1 selection, but shifted to a slightly lower value (0.76 vs. 0.78 for $z_{\rm spec}$).
The Fisher\_GRIW1 selection is also a little more efficient at removing low-redshift objects, as expected from Figure \ref{fig:Fisher_OIIzspec}, where we see a strong correlation between $z_{\rm spec}$ and the Fisher discriminant.
These features are also visible in the $z_{\rm phot}$ distribution.
An important characteristic of the Fisher\_GRIW1 selection is that it tends to select fewer [O\textsc{ii}] emitters than the Fisher\_UGRIZW1 selection.
Figure \ref{fig:gFOII} consistently supports this observation, using both the eBOSS and the VIPERS measurements; in addition, this can also be seen in Table \ref{tab:selstats}, where the number of LRGs per square degree is almost twice higher than for the Fisher\_UGRIZW1 selection, or in Figure \ref{fig:selmagcol}, where the selected objects have redder $u-r$ colours.
The Fisher\_GRIW1$^{OII}$ selection succeeds to select more [O\textsc{ii}] emitters (middle panels of Figures \ref{fig:FOII} and \ref{fig:gFOII}, and overlap with LRGs in Table \ref{tab:selstats}), but at the cost of being less efficient at removing low-redshift objects (middle panel of Figure \ref{fig:selredshift}).
For example, it can be seen in Figures \ref{fig:selmagcol} and \ref{fig:Fisher_grW1} that the Fisher\_GRIW1$^{OII}$ selection has colours bluer than in the Fisher\_GRIW1 selection.
Unfortunately, even if we probe 89\% of the Fisher\_GRIW1$^{OII}$ selection with the \texttt{eboss6-7} test plates, we cannot infer robust statistics with them because $\sim$5\% of the untargeted objects are a biased subsample. Typically, low-redshift objects belonging to the Fisher\_GRIW1$^{OII}$ selection but not to the Fisher\_UGRIZW1 and Fisher\_GRIW1 selections will not be targeted.\\

% Fisher\_GRZW1
\textbf{\textit{Fisher\_GRZW1$_{180}$ and Fisher\_GRZW1$_{300}$ selections.}}
The Fisher\_GRZW1$_{180}$ selection offers an interesting alternative.
The photometric redshift distribution is similar to the one of the Fisher\_UGRIZW1 selection, though it  includes a hardly higher number of $z_{\rm phot} \le 0.2$ galaxies (see Table \ref{tab:selstats} and middle panel of Figure \ref{fig:selredshift}).
In terms of [O\textsc{ii}] emission, the VIPERS data indicate that this selection lies midway between the Fisher\_UGRIZW1 and the Fisher\_GRIW1 selections.
If we increase the target density to 300 deg$^{-2}$, the $z_{\rm phot}$ distribution broadens and is shifted to lower values (0.74 vs. 0.78), but the Fisher\_GRZW1$_{300}$ selection has similar [O\textsc{ii}] emission properties to a target density of 180 deg$^{-2}$.\\

% Colour-colour selections
\textbf{\textit{Colour-colour selections.}}
With the \texttt{eboss6-7} data, we see that the CC\_UGRI selection has a lower median redshift (0.74) and low efficiency (59\%), because this selection emits less [O\textsc{ii}], implying more galaxies without exploitable redshift because of the lack of significant emission lines.
Thus, the WISE/$W1$ data help in increasing the selection redshift and [O\textsc{ii}] flux, hence in increasing the selection efficiency.
The other two tested selections (CC\_UGRIZW1 and CC\_GRIW1) have characteristics that are broadly similar to the corresponding Fisher selections.
More precisely, on the one hand, they have slightly more objects with $\redshiftmin \le z_{\rm phot} \le \redshiftmax$, but on the other, their median redshift is slightly lower, and they have slightly less [O\textsc{ii}] emission.\\

%=========================
% ELGsel/Adjusting the selection density
%=========================
\subsection{Adjusting the selection density \label{sec:sel_sgc}}
Our tests in Section \ref{sec:sel_properties} were done on a single location in a rather small area compared to the aim of 1500 deg$^2$ of the eBOSS/ELG survey. Here we investigate 1) the mean object density over large SGC areas and 2) the way the selection efficiency varies if we change the selection density.
We refer to Paper III for a complete analysis over the full SGC.

The lower cut on the Fisher discriminant, $X_{FI,\rm min}$, is set so that our selections have an object density of 180 deg$^{-2}$ over a $\sim$50 deg$^2$ area included in the CFHTLS-Wide W1 field, approximately centred at R.A.=34 and DEC.=-6.5.
Thanks to the recent development of catalogue tools for the Paper III analysis, it is now feasible to apply our selections using SDSS, WISE, and SCUSS photometry over larger SDSS footprints.
We computed the object density for two typical SGC areas of $\sim$700 deg$^2$ each: one that we label LowDec ($-35<\textnormal{R.A.} <40$ and $-5<\textnormal{DEC.}<5$) and one labelled HiDec ($0<\textnormal{R.A.} <30$ and $5<\textnormal{DEC.}<30$).

The Fisher\_UGRIZW1 selection has a mean object density of 183 deg$^{-2}$ (166 deg$^{-2}$, respectively) over the LowDec (HiDec, respectively) area, whereas the Fisher\_GRIW1 and the Fisher\_GRIW1$^{OII}$ selections have a mean object density of 183-187 deg$^{-2}$ over both areas.
The Fisher\_UGRIZW1 thus seems to have a less homogeneous density over the SGC than the Fisher\_GRIW1 and Fisher\_GRIW1$^{OII}$ selections.\\

\begin{figure}
        \includegraphics[width=0.95\columnwidth]{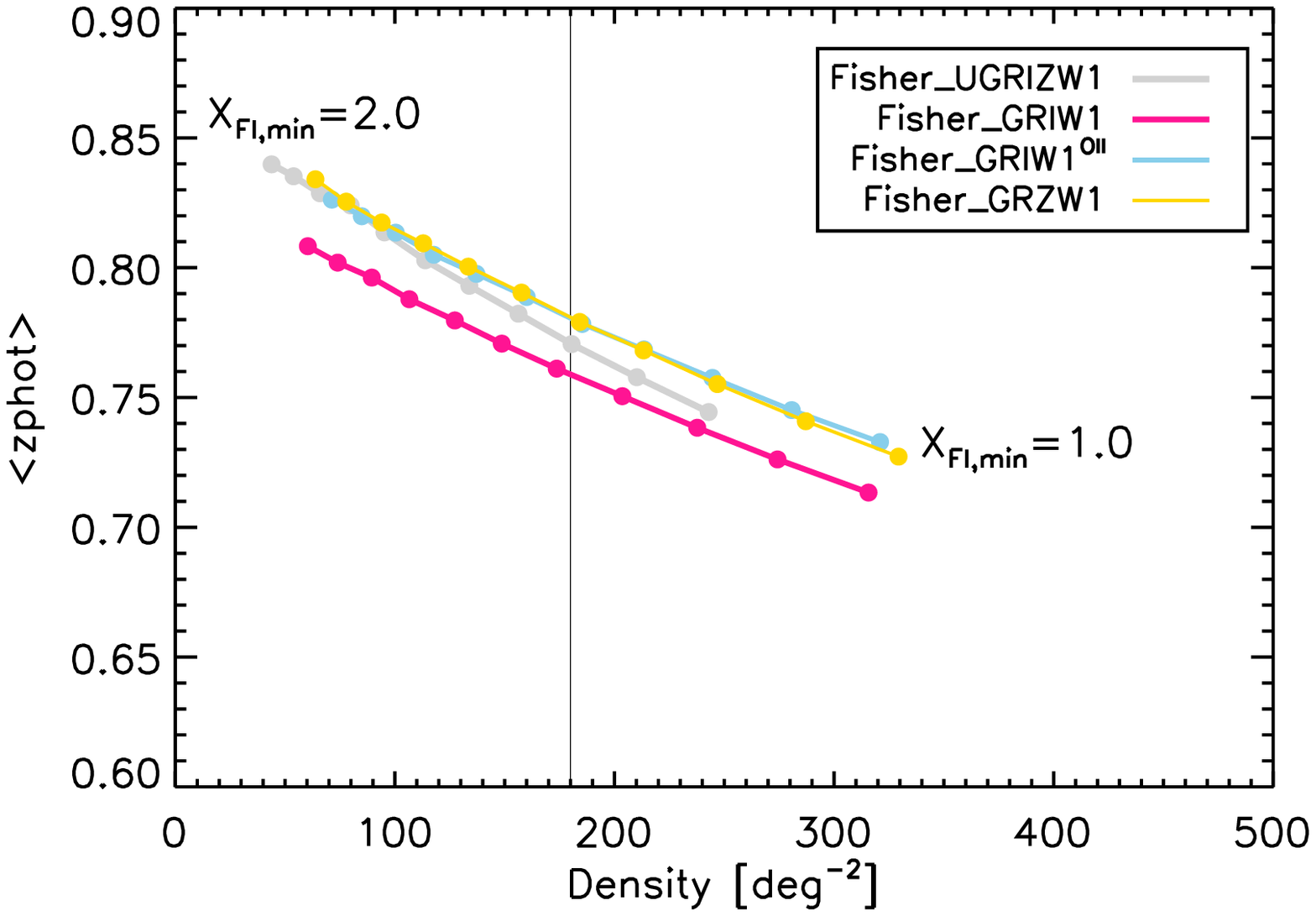}\\
        \includegraphics[width=0.95\columnwidth]{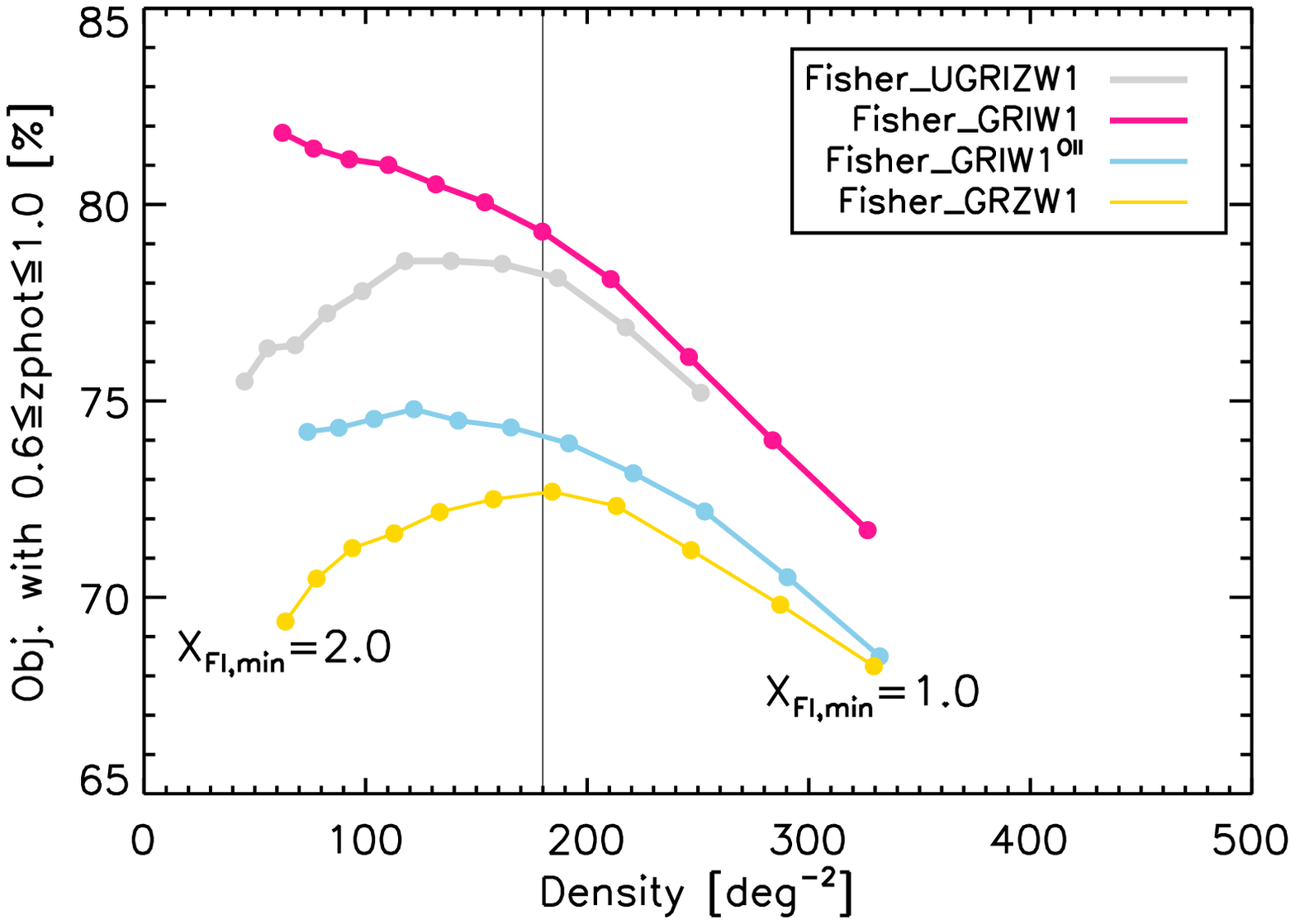}\\
        \caption{Selection dependence on $X_{FI,\rm min}$, the Fisher discriminant threshold cut.
$X_{FI,\rm min}$ varies from 2.0 (low densities, left) to 1.0 (high densities, right) with a step of 0.1.
The sky region considered here is our $\sim$50 deg$^2$ test area within the CFHTLS W1 field.
\textit{Top panel}: mean $z_{\rm phot}$ of the selection.
\textit{Bottom panel}: percentage of galaxies with $\redshiftmin \le z_{\rm phot} \le \redshiftmax$ estimated with the CFHTLS $z_{\rm phot}$. The thin vertical line illustrates the 180 deg$^{-2}$ density used to set the Fisher discriminant threshold cut in Section  \ref{sec:sel_properties}.
}
        \label{fig:effdens}
\end{figure}

We note that the Fisher\_UGRIZW1 selection object density can be increased by lowering $X_{FI,\rm min}$, the threshold cut on the Fisher discriminant.
To illustrate this flexibility in our method, we do the following exercise using the CFHTLS $z_{\rm phot}$ for our $\sim$50 deg$^2$ test area.
We look at the variations of the mean $z_{\rm phot}$ and of the percentage of galaxies with $\redshiftmin \le z_{\rm phot} \le \redshiftmax$ as a function of the selection density, when varying the Fisher discriminant threshold cut $X_{FI,\rm min}$ from 2.0 to 1.0 in steps of 0.1.
We note that the efficiencies computed with the $z_{\rm phot}$ are higher than the ones computed with the $z_{\rm spec}$ because there is no requirement to have a measurable redshift from the spectrum; however, we are here interested in the relative variation with the threshold cut on the Fisher discriminant.
The results are displayed in Figure \ref{fig:effdens}, where we see that the four Fisher discriminants have similar behaviours.
If we decrease the threshold cut on the Fisher discriminant (from left to right), the density increases and the mean $z_{\rm phot}$ decreases, meaning that we select more galaxies but they have a lower redshift.
The percentage of galaxies with $\redshiftmin \le z_{\rm phot} \le \redshiftmax$ decreases when the density increases, because there are more selected galaxies at $z_{\rm phot} < \redshiftmin$.
When the density decreases, we see two different types of behaviour.
For the Fisher\_GRIW1 selection, the percentage increases: this is because the mean redshift also increases, but staying at $\lesssim 0.8$  implies that more galaxies are included in the $\redshiftmin \le z_{\rm phot} \le \redshiftmax$ range.
However, we see a different behaviour for the three other selections, where the percentage flattens or starts to reverse when going to low densities.
This is explained by the fact that the selections start to include galaxies at $z_{\rm phot} > \redshiftmax.$ The reversal at low densities for the Fisher\_UGRIZW1 selection is due to the higher redshift of the selection and, for the Fisher\_GRZW1 selection, is due to a higher redshift and a broader redshift distribution.
For low densities, the Fisher\_GRIW1$^{OII}$ selection has a slightly lower redshift than the Fisher\_UGRIZW1 selection and a slightly narrower distribution than the Fisher\_GRZW1 selection, which explains that we only observe a flattening and not a reversal.

Additionally, for the four Fisher discriminant selections, this percentage is fairly constant, within a few percentage points, for densities between $\sim$150 deg$^{-2}$ and $\sim$250 deg$^{-2}$.
Overall, this means that increasing the selection density while lowering the threshold cut on the Fisher discriminant should still provide satisfactory results.
We note that the percentage of galaxies with $\redshiftmin \le z_{\rm phot} \le \redshiftmax$ is higher than our computed efficiency with $z_{\rm spec}$, because it does not require the additional criterion to have a reliable $z_{\rm spec}$ measurement.
To quantitatively illustrate the impact of increasing the target density, we report in Table \ref{tab:selstats2} the properties for the Fisher\_UGRIZW1 selection when the cut on the Fisher discriminant is set to have a target density of 210 deg$^{-2}$ ($1.209<X_{FI}$) over our $\sim$50 deg$^2$ test area.

% Figure: efficiency vs. density

% Table: UGRIZW1 210 /deg2
\begin{table}
        \begin{tabular}{lc}
                \hline
                \hline
                 & Fisher\_UGRIZW1\\
                \hline
                \textbf{Photometric data ($\sim$50 deg$^2$)} &\\
                Density (deg$^{-2}$)& 212\\
                eBOSS/LRG overlap (deg$^{-2}$) & 2.1\\
                median($z_{\rm phot}$) & 0.77\\
                $z_{\rm phot}$ peak width & 0.12\\
                Percentage with $\redshiftmin \le z_{\rm phot} \le \redshiftmax$ & 78\%\\
                mean($z_{\rm phot}$) ($\redshiftmin \le z_{\rm phot} \le \redshiftmax$) & 0.78\\
                &\\
                \textbf{VIPERS ($0.6 \le z_{\rm spec} \le 1.0$)} &\\
                N selected galaxies & 671\\
                log$_{10}(f_{[\rm O\textsc{ii}]}^{\rm tot})>-16.0$ & 78\%\\
                &\\
                \textbf{Plates \texttt{eboss6-7} (8.82 deg$^2$)} &\\
                N selected galaxies & 1902\\
                Targeted & 86\%\\
                Non-reliable $z_{\rm spec}$ & (12\%)\\
                median($z_{\rm spec}^{\rm reliable}$) & (0.77)\\
                $z_{\rm spec}^{\rm reliable}$ peak width$^{\dagger\dagger}$ & (0.12)\\
                Efficiency ($\redshiftmin \le z_{\rm spec}^{\rm reliable} \le \redshiftmax$) & (69\%)\\
                &\\
                \textbf{Plates \texttt{eboss6-7}, $\redshiftmin \le z_{\rm spec}^{\rm reliable} \le \redshiftmax$ only} &\\
                mean($z_{\rm spec}^{\rm reliable}$) & (0.78)\\
                log$_{10}(f_{[\rm O\textsc{ii}]}^{\rm tot})>-16.0$ & (81\%)\\
                Expected $z_{\rm spec}$ failure  & (0.7\%)\\
                \hline
        \end{tabular}
        \caption{Table illustrating the Fisher\_UGRIZW1 selection properties for a target density of 210 deg$^{-2}$ ($1.209<X_{FI}$) over our $\sim$50 deg$^2$ test area.
Reported quantities are simliar to those in Table \ref{tab:selstats}.}
        \label{tab:selstats2}
\end{table}

%============================================================
% FISHER SELECTIONS STACKED PROPERTIES
%============================================================
\section{Fisher\_UGRIZW1 and Fisher\_GRIW1 selection stacked properties \label{sec:stack}}

We have shown in the previous section that the Fisher\_UGRIZW1 and Fisher\_GRIW1 selections successfully select galaxies in the $\redshiftmin \le z_{\rm spec} \le \redshiftmax$ range with [O\textsc{ii}] emission, thus permitting $\sim$70\% of the selection to be in the desired redshift range with a reliable $z_{\rm spec}$ measurement in the 1h observation with the BOSS spectrograph. Although allowing a reliable redshift measurement,
 the typical individual spectra are noisy (see Figure \ref{fig:exspectra}), which prevents us from visualising or measuring the typical features of the selected galaxies.
Stacking the data allows us to significantly increase the signal-to-noise ratio in the data, thereby enabling this visualisation or measurement of typical features of the selections that would not be visible or measurable in the individual data.

In this section, we take advantage of the unbiased, almost complete coverage of the Fisher\_UGRIZW1 and Fisher\_GRIW1 selections with the ten \texttt{eboss6-7} test plates to study some global physical properties of those two selections through the use of stacked data.
We stack the \texttt{eboss6-7} spectra in Section \ref{sec:stack_spectra} and the CFHTLS-Wide images in Section \ref{sec:stack_images}.

%=========================
% ELGsel/Stacked spectra
%=========================
\subsection{Stacked spectra \label{sec:stack_spectra}}
\begin{figure}
        \includegraphics[width=0.95\columnwidth]{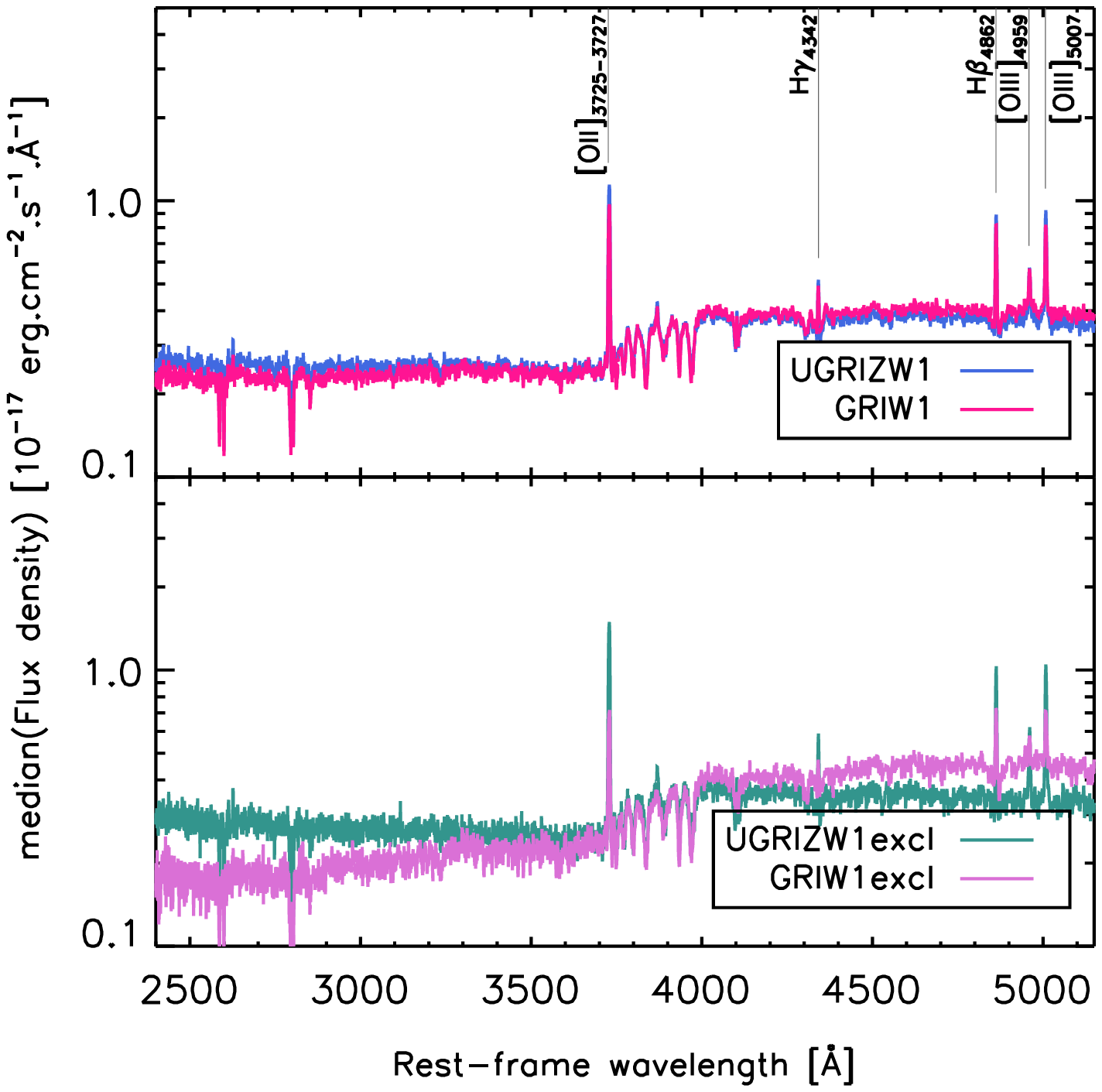}\\
        \caption{Stacked spectra from the \texttt{eboss6-7} plates for the Fisher\_UGRIZW1 and Fisher\_GRIW1 selections.
\textit{Top panel}: the stacking (average $S/N$ of $\sim$5) is done using all galaxies passing the selections ($\sim$1100 galaxies per stack).
\textit{Bottom panel}: the stacking (average $S/N$ of $\sim$3) is done using only galaxies not belonging to the intersection between Fisher\_UGRIZW1 and Fisher\_GRIW1 selections ($\sim$300 galaxies per stack), to enhance the differences.}
        \label{fig:selstack}
\end{figure}

To illustrate the typical spectral features of the $\redshiftmin \le z_{\rm spec} \le \redshiftmax$ galaxies passing  the Fisher\_UGRIZW1 and Fisher\_GRIW1 selections, we display the median stacked spectra of the observed galaxies passing the Fisher\_UGRIZW1 and Fisher\_GRIW1 selections and having $\redshiftmin \le z_{\rm spec} \le \redshiftmax$ (about 1100 galaxies per stack) in the top panel of Figure \ref{fig:selstack}.
We display the rest-frame wavelength range falling in the BOSS spectrograph from $z_{\rm spec} = \redshiftmin$ to $z_{\rm spec} = \redshiftmax$. The stacked spectra have a $S/N$ of $\sim$5 in this range.
To enhance the differences between the two selections in the
bottom panel of Figure \ref{fig:selstack}, we display the median stacked spectra, removing objects that at the same time pass the two selections. The stack is thus done with $\sim$300 galaxies and has a $S/N$ of $\sim$3.\\

Firstly, the clear [O\textsc{ii}] emission line in both stacked spectra confirms the ELG nature of the selections.
Though both selections have comparable stacked spectra at first order, they nevertheless present small differences: the Fisher\_UGRIZW1 stacked spectrum has stronger emission lines, a smaller 4000~\AA~break, and more emission in the near ultra-violet.
All point to a more star-forming nature for the Fisher\_UGRIZW1 selection compared to the Fisher\_GRIW1 selection, in agreement with the analysis done in the Section \ref{sec:sel_properties}.
Furthermore, as expected, those differences are increased in the bottom panel of Figure \ref{fig:selstack}.
We recall that this stacking is only used for a qualitative visualisation of the typical spectral features of the $\redshiftmin \le z_{\rm spec} \le \redshiftmax$ galaxies passing the Fisher\_UGRIZW1 and Fisher\_GRIW1 selections.
A more thorough analysis of those stacked spectra is presented in Paper I.\\

% Figure: stacked spectra

%=========================
% ELGsel/Stacked CFHTLS image
%=========================
\subsection{Stacked CFHTLS-Wide image \label{sec:stack_images}}
We took advantage of the \texttt{eboss6-7} plates being within the CFHTLS-wide survey W1 field to do a simple morphological analysis for the two Fisher\_UGRIZW1 and Fisher\_GRIW1 selections.
The CFHTLS-Wide images are about three magnitudes deeper than the SDSS and have better resolution and seeing (pixel scale of 0.187\arcsec.pix$^{-1}$ and seeing of 0.7\arcsec-0.8\arcsec).
Galaxy surface brightness distribution can be modelled with a \citet{sersic68} profile $I(r) = I_e \times \exp\{-\kappa [(r/r_e)^{1/n_{\rm ser}}-1]\}$, where $I(r)$ is the surface brightness at $r$, and $I_e$ is the surface brightness at the effective radius $r_e$, which is the radius which encloses half of the emitted light.
The S\'{e}rsic index $n_{\rm ser}$ translates the shape of the profile, with a higher value corresponding to a profile more peaked at the centre and with larger wings: $n_{\rm ser}=4$ corresponds to a \citet{de-vaucouleurs48} profile, which is typical of early-type galaxies, while $n_{\rm ser}=1$ corresponds to an exponential profile, typical of late-type galaxies.
\citet{wuyts11a} have shown that, for $0.1 < z_{\rm spec} < 2.5$, typical passive galaxies have $n_{\rm ser} \sim 4$, while typical star-forming galaxies have $n_{\rm ser} \sim 1$-2.

We used the CFHTLS $i$-band images and restricted ourselves to the \texttt{eboss6-7} galaxies passing the Fisher\_UGRIZW1 and Fisher\_GRIW1 selections with $0.7 \le z_{\rm spec} \le 0.8$ to mitigate the effect of redshift on the angular size of the galaxies. In this redshift range, which corresponds to the peak of the $z_{\rm spec}$ distribution for the two considered selections, the $i$ band probes the rest frame 4100~\AA--4400~\AA.
We obtain similar results if we use the CFHTLS $r$-band images, which probe the 3400~\AA--3700~\AA~rest-frame at $0.7 \le z_{\rm spec} \le 0.8$.
For each stamp used in the median stacked image, we masked neighbouring objects beforehand, subtracted the sky, and scaled the galaxy fluxes to a normalised absolute magnitude.
For both selections, we created two median stacked images, using the \texttt{eboss6-7} galaxies with $0.7 \le z_{\rm spec} \le 0.8$: 1): (1) We used all the observed objects passing the selection ($\sim$370 galaxies); 2) we used only objects passing one selection but not the other ($\sim$90 galaxies).
We fitted the surface brightness distribution with the \textsc{Galfit} software \citep[v3.0.5:][]{peng10a}.
During the fit, we set the axis ratio to 1 and the position angle to 0; besides this, we used spectroscopic stars ($18 \le i_{\rm AB} \le 21$) in the \texttt{eboss6-7} plates area to create a point-spread function (PSF) stamp.\\

\begin{figure*}
        \begin{tabular}{lr}
        \includegraphics[width=0.95\columnwidth]{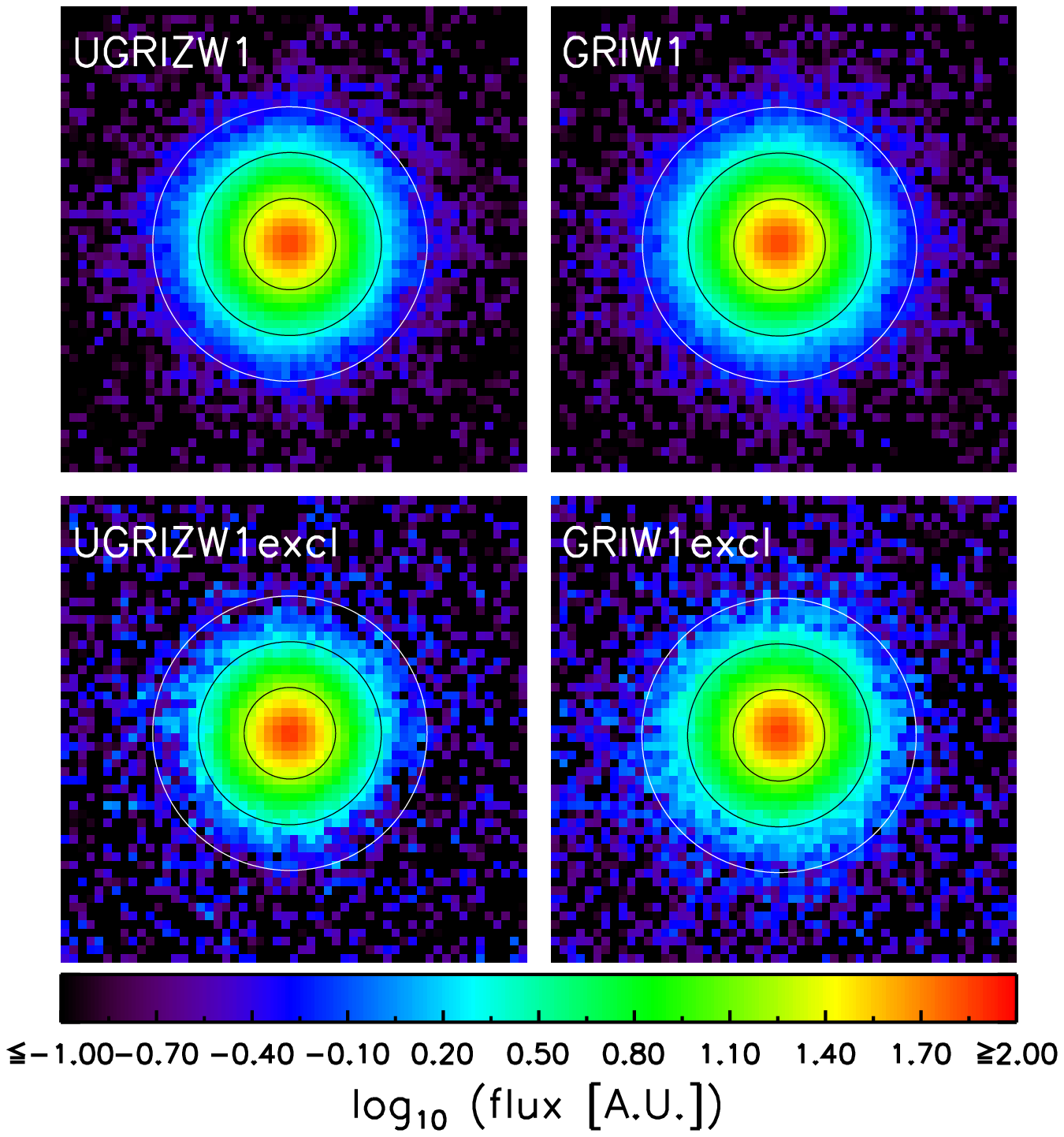} &
        \includegraphics[width=0.95\columnwidth]{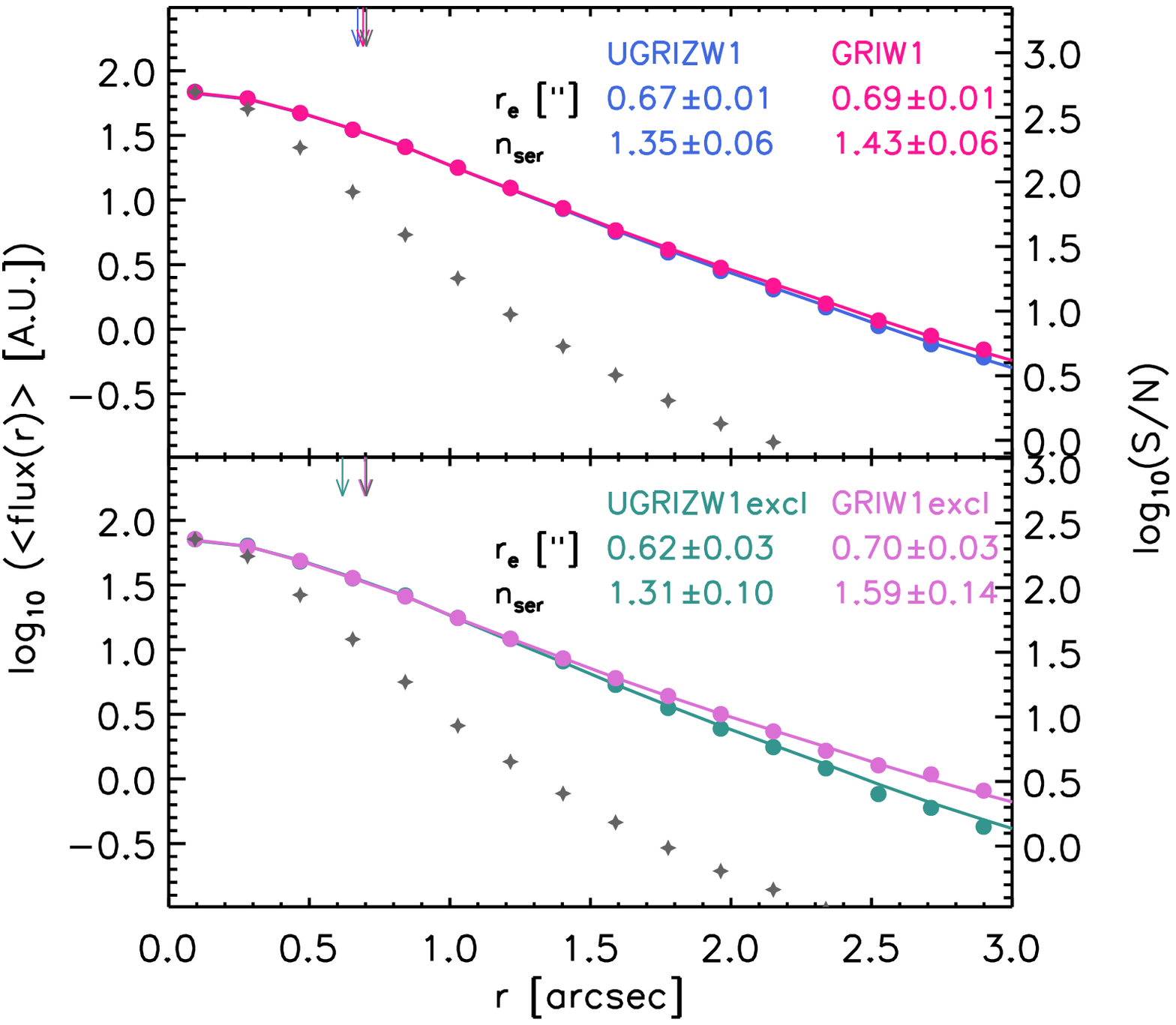}\\
        \end{tabular}
        \caption{CFHTLS-Wide $i$-band stacked images and structural parameters for the \texttt{eboss6-7} plates galaxies with $0.7 \le z_{\rm spec} \le 0.8$ passing the Fisher\_UGRIZW1 and Fisher\_GRIW1 selections.
We labelled UGRIZW1 and GRIW1 when the stacking was done using all galaxies with $0.7 \le z_{\rm spec} \le 0.8$ and passing the selections ($\sim$370 galaxies per stack).
We labelled UGRIZW1excl and GRIW1excl when the stacking was done using only galaxies with $0.7 \le z_{\rm spec} \le 0.8$ and not belonging to the intersection between Fisher\_UGRIZW1 and Fisher\_GRIW1 selections ($\sim$90 galaxies per stack), to enhance the differences.
\textit{Left panel}: stacked images. Each displayed image has a 10\arcsec~ width and the circles have a 1\arcsec, 2\arcsec and 3\arcsec~ radius.
\textit{Right panel}: radial profile computed from the stacked images.
Dots and diamonds represent the data, solid lines represent the fitted profile, corresponding to the reported values.
The grey stars indicate the PSF profile.
The arrows on the top x-axis represent the fitted $r_e$ and the PSF FWHM.}
        \label{fig:selstackCFHTLS}
\end{figure*}

We present the stacked images and their radial profile in the Figure \ref{fig:selstackCFHTLS}.
We also report in this figure the estimated $r_e$ and $n_{\rm ser}$, along with their uncertainty computed via a thousand bootstrap realisations for each case.
The stacked galaxies have small sizes, though clearly resolved: the surface brightness profile extends significantly farther than the PSF full-width-half-maximum.
Regarding the surface brightness profile shape, we observe that both selections have a S\'{e}rsic index of $\sim$1.3-1.4, typical of star-forming galaxies (top panel).
Interestingly, we see that the Fisher\_GRIW1 selection galaxies have a slightly higher S\'{e}rsic index, which is consistent with the trend seen in previous sections toward the Fisher\_GRIW1 selection forming fewer stars than the Fisher\_UGRIZW1 selection. This trend is more significant on the stacked images using galaxies belonging to only one of the selections (bottom panel).
In addition, the Fisher\_GRIW1 selection galaxies also have sizes slightly larger than the Fisher\_UGRIZW1 selection galaxies ones.
Finally, we note that the Fisher\_UGRIZW1 selection galaxies tend to have relatively more flux in a 2\arcsec~ aperture -- corresponding to the BOSS spectrograph fibre diameter -- than the Fisher\_GRIW1 selection, though this effect is minor.
For the fitted parameters corresponding to the bottom panel of Figure \ref{fig:selstackCFHTLS}, 90\% of the total flux is included in a 2\arcsec~ aperture for the Fisher\_UGRIZW1 selection galaxies, versus 84\% for the Fisher\_GRIW1 selection galaxies.\\

% Figure: stacked image profile

%============================================================
% CONCLUSIONS
%============================================================
\section{Conclusions \label{sec:conclusions}}
We have studied possible $z \sim 0.8$ ELG selection schemes in preparation of the eBOSS/ELG survey.
The initial eBOSS/ELG requirements are to select 180 deg$^{-2}$ SDSS galaxies, 70\% of which have a reliable $z_{\rm spec}$ measurement in a $\sim$1h exposure observation with the BOSS spectrograph, $\redshiftmin \le z_{\rm spec} \le \redshiftmax,$ and a catastrophic failure rate $\lesssim$1\% in this redshift range.
Our selection schemes are based on the Fisher discriminant approach, which consists in computing the Fisher discriminant, a linear combination of colours defined from a spectroscopic training sample, and a simple selection with cuts on magnitudes and on this Fisher discriminant.
This type of selection is simple and has the advantage of being flexible, since the density can be adjusted by modifying the Fisher discriminant threshold.

We studied the use of different photometric surveys: SCUSS/$u$+SDSS/$grz$+WISE/$W1$, SDSS/$gri$+WISE/$W1$, and SDSS/$gr$+DECaLS/$z$+WISE/$W1$.
We quantified the properties of our selections in terms of redshift, [O\textsc{ii}] emission, and efficiency, using dedicated eBOSS/ELG test plates and public photometric and spectroscopic data.
We did a parallel analysis of colour-colour selections and showed, on the one hand, that the $W1$-band is crucial in improving the efficiency and, on the other hand, that the Fisher selections are competitive with colour-colour selections.

The Fisher\_UGRIZW1 selection meets the eBOSS/ELG redshift requirements. It has a median redshift of 0.78, and 68\% of the selection has $\redshiftmin \le z_{\rm spec} \le \redshiftmax$; among those 68\% galaxies, $\sim$80\% have a significant [O\textsc{ii}]  emission ($f_{[\rm O\textsc{ii}]}^{\rm tot} \ge 10^{-16}$ erg.cm$^{-2}$.s$^{-1}$.\AA$^{-1}$), and the catastrophic $z_{\rm spec}$ measurement is expected to be 0.6\% with the current instrumental setup and pipeline.
The Fisher\_GRIW1 selection also meets the eBOSS/ELG redshift requirements with a 69\% efficiency in the $\redshiftmin \le z_{\rm spec} \le \redshiftmax$ range and an expected catastrophic $z_{\rm spec}$ measurement of 1.1\% in this redshift range.
This selection has, on average, less [O\textsc{ii}]  emission than the Fisher\_UGRIZW1 selection.
Training the Fisher method with $z_{\rm spec}$ and [O\textsc{ii}]  flux (Fisher\_UGRIZW1$^{OII}$ selection) allows us to increase the [O\textsc{ii}]  emission of the selection, but at the cost of a slightly lower mean redshift.
The Fisher\_GRZW selection using the DECaLS/$z$-band seems to provide an acceptable alternative if set to a 180 deg$^{-2}$ target density.
Finally, we show that the density can be increased while keeping a reasonably high number of galaxies with $\redshiftmin \le z_{\rm phot} \le \redshiftmax$.

In addition, we also studied the properties of the stacked spectra and stacked CFHTLS-Wide images for the Fisher\_UGRIZW1 and Fisher\_GRIW1 selections.
Those stacked data present typical features of star-forming galaxies and also indicate that the Fisher\_UGRIZW1 selection tends to favour more star-forming galaxies than the Fisher\_GRIW1 selection.

For the two most efficient selections, the Fisher\_UGRIZW1 and Fisher\_GRIW1 selections, the homogeneity over the SGC along with the possible systematic dependence on various quantities, is studied in Paper III.
Paper III also presents the catalogue release over the SGC for those two selections.

To conclude, this Fisher discriminant approach can be used more generally if one desires to select a galaxy population with desired properties with multi-band photometry.
The method is simple and offers flexibility, the only requirement being the use of a spectroscopic training sample.
Future massive spectroscopic surveys, such as
DESI,
4MOST\footnote{\href{https://www.4most.eu/}{https://www.4most.eu/}}, or 
the Prime Focus Spectrograph \citep[PFS;][]{sugai12}, will provide large samples that are particularly well-suited to the Fisher discriminant approach.

%============================================================
% ACKNOWLEDGEMENTS
%============================================================
\begin{acknowledgements}
% Funding
AR acknowledges funding from the P2IO LabEx (ANR-10-LABX-0038) in the framework "Investissements d\textquoteright Avenir" (ANR-11-IDEX-0003-01) managed by the French National Research Agency (ANR).
JC acknowledges financial support from MINECO (Spain) under project number AYA2012-31101.
JPK and TD acknowledge support from the ERC advanced grant LIDA.
% SDSS III + IV
This study is based on data from SDSS-III (full text acknowledgement: \href{http:// www.sdss3.org/collaboration/boiler-plate.php}{http:// www.sdss3.org/collaboration/boiler-plate.php}) and is done in the context of SDSS-IV (full text acknowledgement: \href{http://www.sdss.org/collaboration/\#OfficialSDSSAcknowledgment}{http://www.sdss.org/collaboration/\#OfficialSDSSAcknowledgment}).
% SCUSS
The SCUSS is funded by the Main Direction Program of Knowledge Innovation of Chinese Academy of Sciences (No. KJCX2-EW-T06). It is also an international cooperative project between National Astronomical Observatories, Chinese Academy of Sciences and Steward Observatory, University of Arizona, USA. Technical support and observational assistances of the Bok telescope are provided by Steward Observatory. The project is managed by the National Astronomical Observatory of China and Shanghai Astronomical Observatory.
% WISE
This publication makes use of data products from the Wide-field Infrared Survey Explorer, which is a joint project of the University of California, Los Angeles, the Jet Propulsion Laboratory/California Institute of Technology, and NEOWISE, which is a project of the Jet Propulsion Laboratory/California Institute of Technology. WISE and NEOWISE are funded by the National Aeronautics and Space Administration.
% CFHTLS
This work is based on observations obtained with MegaPrime/MegaCam, a joint project of CFHT and CEA/ DAPNIA, at the Canada--France--Hawaii Telescope (CFHT), which is operated by the National Research Council (NRC) of Canada, the Institut National des Sciences de l'Univers of the Centre National de la Recherche Scientifique (CNRS) of France, and the University of Hawaii. This research used the facilities of the Canadian Astronomy Data Centre operated by the National Research Council of Canada with the support of the Canadian Space Agency.
% VVDS
This research uses data from the VIMOS VLT Deep Survey, obtained from the VVDS database operated by Cesam, Laboratoire d'Astrophysique de Marseille, France. 
% VIPERS
This paper uses data from the VIMOS Public Extragalactic Redshift Survey (VIPERS). VIPERS was performed with the ESO Very Large Telescope under the "Large Programme" 182.A- 0886. The participating institutions and funding agencies are listed at \href{http://vipers.inaf.it}{http://vipers.inaf.it}.
% referee
We thank the anonymous referee for his report, which helped us improve the clarity of the paper.
\end{acknowledgements}

%============================================================
% BIBLIOGRAPHY
%============================================================
%\bibliography{raichoor_master}
\bibliography{ms.bib}

\begin{thebibliography}{50}
\expandafter\ifx\csname natexlab\endcsname\relax\def\natexlab#1{#1}\fi

\bibitem[{{Alam} {et~al.}(2015){Alam}, {Albareti}, {Allende Prieto}, {Anders},
  {Anderson}, {Anderton}, {Andrews}, {Armengaud}, {Aubourg}, {Bailey}, \&
  et~al.}]{alam15}
{Alam}, S., {Albareti}, F.~D., {Allende Prieto}, C., {et~al.} 2015, \apjs, 219,
  12

\bibitem[{{Anderson} {et~al.}(2012){Anderson}, {Aubourg}, {Bailey}, {Bizyaev},
  {Blanton}, {Bolton}, {Brinkmann}, {Brownstein}, {Burden}, {Cuesta}, {da
  Costa}, {Dawson}, {de Putter}, {Eisenstein}, {Gunn}, {Guo}, {Hamilton},
  {Harding}, {Ho}, {Honscheid}, {Kazin}, {Kirkby}, {Kneib}, {Labatie},
  {Loomis}, {Lupton}, {Malanushenko}, {Malanushenko}, {Mandelbaum}, {Manera},
  {Maraston}, {McBride}, {Mehta}, {Mena}, {Montesano}, {Muna}, {Nichol},
  {Nuza}, {Olmstead}, {Oravetz}, {Padmanabhan}, {Palanque-Delabrouille}, {Pan},
  {Parejko}, {P{\^a}ris}, {Percival}, {Petitjean}, {Prada}, {Reid}, {Roe},
  {Ross}, {Ross}, {Samushia}, {S{\'a}nchez}, {Schlegel}, {Schneider},
  {Sc{\'o}ccola}, {Seo}, {Sheldon}, {Simmons}, {Skibba}, {Strauss}, {Swanson},
  {Thomas}, {Tinker}, {Tojeiro}, {Maga{\~n}a}, {Verde}, {Wagner}, {Wake},
  {Weaver}, {Weinberg}, {White}, {Xu}, {Y{\`e}che}, {Zehavi}, \&
  {Zhao}}]{anderson12}
{Anderson}, L., {Aubourg}, E., {Bailey}, S., {et~al.} 2012, \mnras, 427, 3435

\bibitem[{{Bertin} \& {Arnouts}(1996)}]{bertin96}
{Bertin}, E. \& {Arnouts}, S. 1996, \aaps, 117, 393

\bibitem[{{Bruzual} \& {Charlot}(2003)}]{bruzual03}
{Bruzual}, G. \& {Charlot}, S. 2003, \mnras, 344, 1000

\bibitem[{{Calzetti} {et~al.}(2000){Calzetti}, {Armus}, {Bohlin}, {Kinney},
  {Koornneef}, \& {Storchi-Bergmann}}]{calzetti00}
{Calzetti}, D., {Armus}, L., {Bohlin}, R.~C., {et~al.} 2000, \apj, 533, 682

\bibitem[{{Comparat} {et~al.}(2015{\natexlab{a}}){Comparat}, {Delubac},
  {Jouvel}, {Raichoor}, {Kneib}, {Yeche}, {Abdalla}, {Le Cras}, {Maraston},
  {Wilkinson}, {Zhu}, {Jullo}, {Prada}, {Schlegel}, {Xu}, {Zou}, {Bautista},
  {Bizyaev}, {Bolton}, {Brownstein}, {Dawson}, {Gaulme}, {Kinemuchi},
  {Malanushenko}, {Malanushenko}, {Mariappan}, {Newman}, {Oravetz}, {Pan},
  {Percival}, {Prakash}, {Schneider}, {Simmons}, {Allam}, {Banerji},
  {Benoit-L{\'e}vy}, {Bertin}, {Brooks}, {Capozzi}, {Carnero Rosell}, {Carrasco
  Kind}, {Carretero}, {Castander}, {Cunha}, {da Costa}, {Desai}, {Doel},
  {Eifler}, {Estrada}, {Flaugher}, {Fosalba}, {Frieman}, {Gaztanaga}, {Gerdes},
  {Gruen}, {Gruendl}, {Gutierrez}, {Honscheid}, {James}, {Kuehn}, {Kuropatkin},
  {Lahav}, {Lima}, {Maia}, {March}, {Marshall}, {Miquel}, {Plazas}, {Reil},
  {Roe}, {Romer}, {Roodman}, {Rykoff}, {Sako}, {Sanchez}, {Scarpine},
  {Sevilla-Noarbe}, {Soares-Santos}, {Sobreira}, {Suchyta}, {Swanson}, {Tarle},
  {Thaler}, {Thomas}, {Walker}, \& {Zhang}}]{comparat15}
{Comparat}, J., {Delubac}, T., {Jouvel}, S., {et~al.} 2015{\natexlab{a}}, \aap,
  submitted [arxiv:1509.05045]

\bibitem[{{Comparat} {et~al.}(2015{\natexlab{b}}){Comparat}, {Richard},
  {Kneib}, {Ilbert}, {Gonzalez-Perez}, {Tresse}, {Zoubian}, {Arnouts},
  {Brownstein}, {Baugh}, {Delubac}, {Ealet}, {Escoffier}, {Ge}, {Jullo},
  {Lacey}, {Ross}, {Schlegel}, {Schneider}, {Steele}, {Tasca}, {Yeche},
  {Lesser}, {Jiang}, {Jing}, {Fan}, {Fan}, {Ma}, {Nie}, {Wang}, {Wu}, {Zhang},
  {Zhou}, {Zhou}, \& {Zou}}]{comparat15a}
{Comparat}, J., {Richard}, J., {Kneib}, J.-P., {et~al.} 2015{\natexlab{b}},
  \aap, 575, A40

\bibitem[{{Coupon} {et~al.}(2009){Coupon}, {Ilbert}, {Kilbinger}, {McCracken},
  {Mellier}, {Arnouts}, {Bertin}, {Hudelot}, {Schultheis}, {Le F{\`e}vre}, {Le
  Brun}, {Guzzo}, {Bardelli}, {Zucca}, {Bolzonella}, {Garilli}, {Zamorani},
  {Zanichelli}, {Tresse}, \& {Aussel}}]{coupon09}
{Coupon}, J., {Ilbert}, O., {Kilbinger}, M., {et~al.} 2009, \aap, 500, 981

\bibitem[{{Dawson} {et~al.}(2013){Dawson}, {Schlegel}, {Ahn}, {Anderson},
  {Aubourg}, {Bailey}, {Barkhouser}, {Bautista}, {Beifiori}, {Berlind},
  {Bhardwaj}, {Bizyaev}, {Blake}, {Blanton}, {Blomqvist}, {Bolton}, {Borde},
  {Bovy}, {Brandt}, {Brewington}, {Brinkmann}, {Brown}, {Brownstein}, {Bundy},
  {Busca}, {Carithers}, {Carnero}, {Carr}, {Chen}, {Comparat}, {Connolly},
  {Cope}, {Croft}, {Cuesta}, {da Costa}, {Davenport}, {Delubac}, {de Putter},
  {Dhital}, {Ealet}, {Ebelke}, {Eisenstein}, {Escoffier}, {Fan}, {Filiz Ak},
  {Finley}, {Font-Ribera}, {G{\'e}nova-Santos}, {Gunn}, {Guo}, {Haggard},
  {Hall}, {Hamilton}, {Harris}, {Harris}, {Ho}, {Hogg}, {Holder}, {Honscheid},
  {Huehnerhoff}, {Jordan}, {Jordan}, {Kauffmann}, {Kazin}, {Kirkby}, {Klaene},
  {Kneib}, {Le Goff}, {Lee}, {Long}, {Loomis}, {Lundgren}, {Lupton}, {Maia},
  {Makler}, {Malanushenko}, {Malanushenko}, {Mandelbaum}, {Manera}, {Maraston},
  {Margala}, {Masters}, {McBride}, {McDonald}, {McGreer}, {McMahon}, {Mena},
  {Miralda-Escud{\'e}}, {Montero-Dorta}, {Montesano}, {Muna}, {Myers},
  {Naugle}, {Nichol}, {Noterdaeme}, {Nuza}, {Olmstead}, {Oravetz}, {Oravetz},
  {Owen}, {Padmanabhan}, {Palanque-Delabrouille}, {Pan}, {Parejko},
  {P{\^a}ris}, {Percival}, {P{\'e}rez-Fournon}, {P{\'e}rez-R{\`a}fols},
  {Petitjean}, {Pfaffenberger}, {Pforr}, {Pieri}, {Prada}, {Price-Whelan},
  {Raddick}, {Rebolo}, {Rich}, {Richards}, {Rockosi}, {Roe}, {Ross}, {Ross},
  {Rossi}, {Rubi{\~n}o-Martin}, {Samushia}, {S{\'a}nchez}, {Sayres}, {Schmidt},
  {Schneider}, {Sc{\'o}ccola}, {Seo}, {Shelden}, {Sheldon}, {Shen}, {Shu},
  {Slosar}, {Smee}, {Snedden}, {Stauffer}, {Steele}, {Strauss}, {Streblyanska},
  {Suzuki}, {Swanson}, {Tal}, {Tanaka}, {Thomas}, {Tinker}, {Tojeiro},
  {Tremonti}, {Vargas Maga{\~n}a}, {Verde}, {Viel}, {Wake}, {Watson}, {Weaver},
  {Weinberg}, {Weiner}, {West}, {White}, {Wood-Vasey}, {Yeche}, {Zehavi},
  {Zhao}, \& {Zheng}}]{dawson13}
{Dawson}, K.~S., {Schlegel}, D.~J., {Ahn}, C.~P., {et~al.} 2013, \aj, 145, 10

\bibitem[{{de Vaucouleurs}(1948)}]{de-vaucouleurs48}
{de Vaucouleurs}, G. 1948, Annales d'Astrophysique, 11, 247

\bibitem[{{Delubac} {et~al.}(2014){Delubac}, {Bautista}, {Busca}, {Rich},
  {Kirkby}, {Bailey}, {Font-Ribera}, {Slosar}, {Lee}, {Pieri}, {Hamilton},
  {Aubourg}, {Blomqvist}, {Bovy}, {Brinkmann}, {Carithers}, {Dawson},
  {Eisenstein}, {Kneib}, {Le Goff}, {Margala}, {Miralda-Escud{\'e}}, {Myers},
  {Nichol}, {Noterdaeme}, {O'Connell}, {Olmstead}, {Palanque-Delabrouille},
  {P{\^a}ris}, {Petitjean}, {Ross}, {Rossi}, {Schlegel}, {Schneider},
  {Weinberg}, {Y{\`e}che}, \& {York}}]{delubac14}
{Delubac}, T., {Bautista}, J.~E., {Busca}, N.~G., {et~al.} 2014, ArXiv e-prints

\bibitem[{{Drinkwater} {et~al.}(2010){Drinkwater}, {Jurek}, {Blake}, {Woods},
  {Pimbblet}, {Glazebrook}, {Sharp}, {Pracy}, {Brough}, {Colless}, {Couch},
  {Croom}, {Davis}, {Forbes}, {Forster}, {Gilbank}, {Gladders}, {Jelliffe},
  {Jones}, {Li}, {Madore}, {Martin}, {Poole}, {Small}, {Wisnioski}, {Wyder}, \&
  {Yee}}]{drinkwater10}
{Drinkwater}, M.~J., {Jurek}, R.~J., {Blake}, C., {et~al.} 2010, \mnras, 401,
  1429

\bibitem[{{Eisenstein} {et~al.}(2001){Eisenstein}, {Annis}, {Gunn}, {Szalay},
  {Connolly}, {Nichol}, {Bahcall}, {Bernardi}, {Burles}, {Castander},
  {Fukugita}, {Hogg}, {Ivezi{\'c}}, {Knapp}, {Lupton}, {Narayanan}, {Postman},
  {Reichart}, {Richmond}, {Schneider}, {Schlegel}, {Strauss}, {SubbaRao},
  {Tucker}, {Vanden Berk}, {Vogeley}, {Weinberg}, \& {Yanny}}]{eisenstein01}
{Eisenstein}, D.~J., {Annis}, J., {Gunn}, J.~E., {et~al.} 2001, \aj, 122, 2267

\bibitem[{{Eisenstein} {et~al.}(2005){Eisenstein}, {Zehavi}, {Hogg},
  {Scoccimarro}, {Blanton}, {Nichol}, {Scranton}, {Seo}, {Tegmark}, {Zheng},
  {Anderson}, {Annis}, {Bahcall}, {Brinkmann}, {Burles}, {Castander},
  {Connolly}, {Csabai}, {Doi}, {Fukugita}, {Frieman}, {Glazebrook}, {Gunn},
  {Hendry}, {Hennessy}, {Ivezi{\'c}}, {Kent}, {Knapp}, {Lin}, {Loh}, {Lupton},
  {Margon}, {McKay}, {Meiksin}, {Munn}, {Pope}, {Richmond}, {Schlegel},
  {Schneider}, {Shimasaku}, {Stoughton}, {Strauss}, {SubbaRao}, {Szalay},
  {Szapudi}, {Tucker}, {Yanny}, \& {York}}]{eisenstein05}
{Eisenstein}, D.~J., {Zehavi}, I., {Hogg}, D.~W., {et~al.} 2005, \apj, 633, 560

\bibitem[{{Fisher}(1936)}]{fisher36}
{Fisher}, R.~A. 1936, Annals of Eugenics, 7, 179

\bibitem[{{Font-Ribera} {et~al.}(2014){Font-Ribera}, {Kirkby}, {Busca},
  {Miralda-Escud{\'e}}, {Ross}, {Slosar}, {Rich}, {Aubourg}, {Bailey},
  {Bhardwaj}, {Bautista}, {Beutler}, {Bizyaev}, {Blomqvist}, {Brewington},
  {Brinkmann}, {Brownstein}, {Carithers}, {Dawson}, {Delubac}, {Ebelke},
  {Eisenstein}, {Ge}, {Kinemuchi}, {Lee}, {Malanushenko}, {Malanushenko},
  {Marchante}, {Margala}, {Muna}, {Myers}, {Noterdaeme}, {Oravetz},
  {Palanque-Delabrouille}, {P{\^a}ris}, {Petitjean}, {Pieri}, {Rossi},
  {Schneider}, {Simmons}, {Viel}, {Yeche}, \& {York}}]{font-ribera14}
{Font-Ribera}, A., {Kirkby}, D., {Busca}, N., {et~al.} 2014, \jcap, 5, 27

\bibitem[{{Franzetti} {et~al.}(2014){Franzetti}, {Garilli}, {Guzzo},
  {Marchetti}, \& {Scodeggio}}]{franzetti14}
{Franzetti}, P., {Garilli}, B., {Guzzo}, L., {Marchetti}, A., \& {Scodeggio},
  M. 2014, \aap, submitted [arXiv:1410.0465]

\bibitem[{{Garilli} {et~al.}(2014){Garilli}, {Guzzo}, {Scodeggio},
  {Bolzonella}, {Abbas}, {Adami}, {Arnouts}, {Bel}, {Bottini}, {Branchini},
  {Cappi}, {Coupon}, {Cucciati}, {Davidzon}, {De Lucia}, {de la Torre},
  {Franzetti}, {Fritz}, {Fumana}, {Granett}, {Ilbert}, {Iovino}, {Krywult}, {Le
  Brun}, {Le F{\`e}vre}, {Maccagni}, {Ma{\l}ek}, {Marulli}, {McCracken},
  {Paioro}, {Polletta}, {Pollo}, {Schlagenhaufer}, {Tasca}, {Tojeiro},
  {Vergani}, {Zamorani}, {Zanichelli}, {Burden}, {Di Porto}, {Marchetti},
  {Marinoni}, {Mellier}, {Moscardini}, {Nichol}, {Peacock}, {Percival},
  {Phleps}, \& {Wolk}}]{garilli14}
{Garilli}, B., {Guzzo}, L., {Scodeggio}, M., {et~al.} 2014, \aap, 562, A23

\bibitem[{{Guzzo} {et~al.}(2014){Guzzo}, {Scodeggio}, {Garilli}, {Granett},
  {Fritz}, {Abbas}, {Adami}, {Arnouts}, {Bel}, {Bolzonella}, {Bottini},
  {Branchini}, {Cappi}, {Coupon}, {Cucciati}, {Davidzon}, {De Lucia}, {de la
  Torre}, {Franzetti}, {Fumana}, {Hudelot}, {Ilbert}, {Iovino}, {Krywult}, {Le
  Brun}, {Le F{\`e}vre}, {Maccagni}, {Ma{\l}ek}, {Marulli}, {McCracken},
  {Paioro}, {Peacock}, {Polletta}, {Pollo}, {Schlagenhaufer}, {Tasca},
  {Tojeiro}, {Vergani}, {Zamorani}, {Zanichelli}, {Burden}, {Di Porto},
  {Marchetti}, {Marinoni}, {Mellier}, {Moscardini}, {Nichol}, {Percival},
  {Phleps}, \& {Wolk}}]{guzzo14}
{Guzzo}, L., {Scodeggio}, M., {Garilli}, B., {et~al.} 2014, \aap, 566, A108

\bibitem[{{Gwyn}(2012)}]{gwyn12}
{Gwyn}, S.~D.~J. 2012, \aj, 143, 38

\bibitem[{{Hildebrandt} {et~al.}(2012){Hildebrandt}, {Erben}, {Kuijken}, {van
  Waerbeke}, {Heymans}, {Coupon}, {Benjamin}, {Bonnett}, {Fu}, {Hoekstra},
  {Kitching}, {Mellier}, {Miller}, {Velander}, {Hudson}, {Rowe}, {Schrabback},
  {Semboloni}, \& {Ben{\'{\i}}tez}}]{hildebrandt12}
{Hildebrandt}, H., {Erben}, T., {Kuijken}, K., {et~al.} 2012, \mnras, 421, 2355

\bibitem[{{Hopkins} \& {Beacom}(2006)}]{hopkins06}
{Hopkins}, A.~M. \& {Beacom}, J.~F. 2006, \apj, 651, 142

\bibitem[{{Ilbert} {et~al.}(2006){Ilbert}, {Arnouts}, {McCracken},
  {Bolzonella}, {Bertin}, {Le F{\`e}vre}, {Mellier}, {Zamorani}, {Pell{\`o}},
  {Iovino}, {Tresse}, {Le Brun}, {Bottini}, {Garilli}, {Maccagni}, {Picat},
  {Scaramella}, {Scodeggio}, {Vettolani}, {Zanichelli}, {Adami}, {Bardelli},
  {Cappi}, {Charlot}, {Ciliegi}, {Contini}, {Cucciati}, {Foucaud}, {Franzetti},
  {Gavignaud}, {Guzzo}, {Marano}, {Marinoni}, {Mazure}, {Meneux}, {Merighi},
  {Paltani}, {Pollo}, {Pozzetti}, {Radovich}, {Zucca}, {Bondi}, {Bongiorno},
  {Busarello}, {de La Torre}, {Gregorini}, {Lamareille}, {Mathez}, {Merluzzi},
  {Ripepi}, {Rizzo}, \& {Vergani}}]{ilbert06a}
{Ilbert}, O., {Arnouts}, S., {McCracken}, H.~J., {et~al.} 2006, \aap, 457, 841

\bibitem[{{Ilbert} {et~al.}(2009){Ilbert}, {Capak}, {Salvato}, {Aussel},
  {McCracken}, {Sanders}, {Scoville}, {Kartaltepe}, {Arnouts}, {Le Floc'h},
  {Mobasher}, {Taniguchi}, {Lamareille}, {Leauthaud}, {Sasaki}, {Thompson},
  {Zamojski}, {Zamorani}, {Bardelli}, {Bolzonella}, {Bongiorno}, {Brusa},
  {Caputi}, {Carollo}, {Contini}, {Cook}, {Coppa}, {Cucciati}, {de la Torre},
  {de Ravel}, {Franzetti}, {Garilli}, {Hasinger}, {Iovino}, {Kampczyk},
  {Kneib}, {Knobel}, {Kovac}, {Le Borgne}, {Le Brun}, {F{\`e}vre}, {Lilly},
  {Looper}, {Maier}, {Mainieri}, {Mellier}, {Mignoli}, {Murayama}, {Pell{\`o}},
  {Peng}, {P{\'e}rez-Montero}, {Renzini}, {Ricciardelli}, {Schiminovich},
  {Scodeggio}, {Shioya}, {Silverman}, {Surace}, {Tanaka}, {Tasca}, {Tresse},
  {Vergani}, \& {Zucca}}]{ilbert09}
{Ilbert}, O., {Capak}, P., {Salvato}, M., {et~al.} 2009, \apj, 690, 1236

\bibitem[{{Ilbert} {et~al.}(2013){Ilbert}, {McCracken}, {Le F{\`e}vre},
  {Capak}, {Dunlop}, {Karim}, {Renzini}, {Caputi}, {Boissier}, {Arnouts},
  {Aussel}, {Comparat}, {Guo}, {Hudelot}, {Kartaltepe}, {Kneib}, {Krogager},
  {Le Floc'h}, {Lilly}, {Mellier}, {Milvang-Jensen}, {Moutard}, {Onodera},
  {Richard}, {Salvato}, {Sanders}, {Scoville}, {Silverman}, {Taniguchi},
  {Tasca}, {Thomas}, {Toft}, {Tresse}, {Vergani}, {Wolk}, \& {Zirm}}]{ilbert13}
{Ilbert}, O., {McCracken}, H.~J., {Le F{\`e}vre}, O., {et~al.} 2013, \aap, 556,
  A55

\bibitem[{{Ilbert} {et~al.}(2005){Ilbert}, {Tresse}, {Zucca}, {Bardelli},
  {Arnouts}, {Zamorani}, {Pozzetti}, {Bottini}, {Garilli}, {Le Brun}, {Le
  F{\`e}vre}, {Maccagni}, {Picat}, {Scaramella}, {Scodeggio}, {Vettolani},
  {Zanichelli}, {Adami}, {Arnaboldi}, {Bolzonella}, {Cappi}, {Charlot},
  {Contini}, {Foucaud}, {Franzetti}, {Gavignaud}, {Guzzo}, {Iovino},
  {McCracken}, {Marano}, {Marinoni}, {Mathez}, {Mazure}, {Meneux}, {Merighi},
  {Paltani}, {Pello}, {Pollo}, {Radovich}, {Bondi}, {Bongiorno}, {Busarello},
  {Ciliegi}, {Lamareille}, {Mellier}, {Merluzzi}, {Ripepi}, \&
  {Rizzo}}]{ilbert05}
{Ilbert}, O., {Tresse}, L., {Zucca}, E., {et~al.} 2005, \aap, 439, 863

\bibitem[{{Jouvel} {et~al.}(2015){Jouvel}, {Delubac}, {Comparat}, {Carnero},
  {Camacho}, {Abdalla}, {Kneib}, {Merson}, {Lima}, {Sobreira}, {da Costa},
  {Prada}, {Zhu}, {Benoit-Levy}, {De La Macora}, {Kuropatkin}, {Lin}, {Abbott},
  {Allam}, {Banerji}, {Bertin}, {Brooks}, {Capozzi}, {Carrasco Kind},
  {Carretero}, {Castander}, {Cunha}, {Desai}, {Doel}, {Eifler}, {Estrada},
  {Fausti Neto}, {Flaugher}, {Fosalba}, {Frieman}, {Gaztanaga}, {Gerdes},
  {Gruen}, {Gruendl}, {Gutierrez}, {Honscheid}, {James}, {Kuehn}, {Lahav},
  {Li}, {Maia}, {March}, {Marshall}, {Miquel}, {Percival}, {Plazas}, {Reil},
  {Romer}, {Roodman}, {Rykoff}, {Sako}, {Sanchez}, {Santiago}, {Scarpine},
  {Sevilla-Noarbe}, {Soares Santos}, {Suchyta}, {Tarle}, {Thaler}, {Thomas},
  {Walker}, \& {Zhang}}]{jouvel15}
{Jouvel}, S., {Delubac}, T., {Comparat}, J., {et~al.} 2015, \mnras, submitted
  [arXiv:1509.07121]

\bibitem[{{Kauffmann} {et~al.}(2003){Kauffmann}, {Heckman}, {White}, {Charlot},
  {Tremonti}, {Peng}, {Seibert}, {Brinkmann}, {Nichol}, {SubbaRao}, \&
  {York}}]{kauffmann03a}
{Kauffmann}, G., {Heckman}, T.~M., {White}, S.~D.~M., {et~al.} 2003, \mnras,
  341, 54

\bibitem[{{Lang} {et~al.}(2014){Lang}, {Hogg}, \& {Schlegel}}]{lang14a}
{Lang}, D., {Hogg}, D.~W., \& {Schlegel}, D.~J. 2014, \aj, submitted
  [arXiv:1410.7397]

\bibitem[{{Le F{\`e}vre} {et~al.}(2013){Le F{\`e}vre}, {Cassata}, {Cucciati},
  {Garilli}, {Ilbert}, {Le Brun}, {Maccagni}, {Moreau}, {Scodeggio}, {Tresse},
  {Zamorani}, {Adami}, {Arnouts}, {Bardelli}, {Bolzonella}, {Bondi},
  {Bongiorno}, {Bottini}, {Cappi}, {Charlot}, {Ciliegi}, {Contini}, {de la
  Torre}, {Foucaud}, {Franzetti}, {Gavignaud}, {Guzzo}, {Iovino}, {Lemaux},
  {L{\'o}pez-Sanjuan}, {McCracken}, {Marano}, {Marinoni}, {Mazure}, {Mellier},
  {Merighi}, {Merluzzi}, {Paltani}, {Pell{\`o}}, {Pollo}, {Pozzetti},
  {Scaramella}, {Tasca}, {Vergani}, {Vettolani}, {Zanichelli}, \&
  {Zucca}}]{le-fevre13}
{Le F{\`e}vre}, O., {Cassata}, P., {Cucciati}, O., {et~al.} 2013, \aap, 559,
  A14

\bibitem[{{Le F{\`e}vre} {et~al.}(2003){Le F{\`e}vre}, {Saisse}, {Mancini},
  {Brau-Nogue}, {Caputi}, {Castinel}, {D'Odorico}, {Garilli}, {Kissler-Patig},
  {Lucuix}, {Mancini}, {Pauget}, {Sciarretta}, {Scodeggio}, {Tresse}, \&
  {Vettolani}}]{le-fevre03}
{Le F{\`e}vre}, O., {Saisse}, M., {Mancini}, D., {et~al.} 2003, Society of
  Photo-Optical Instrumentation Engineers (SPIE) Conference Series, 4841, 1670

\bibitem[{{Le F{\`e}vre} {et~al.}(2005){Le F{\`e}vre}, {Vettolani}, {Garilli},
  {Tresse}, {Bottini}, {Le Brun}, {Maccagni}, {Picat}, {Scaramella},
  {Scodeggio}, {Zanichelli}, {Adami}, {Arnaboldi}, {Arnouts}, {Bardelli},
  {Bolzonella}, {Cappi}, {Charlot}, {Ciliegi}, {Contini}, {Foucaud},
  {Franzetti}, {Gavignaud}, {Guzzo}, {Ilbert}, {Iovino}, {McCracken}, {Marano},
  {Marinoni}, {Mathez}, {Mazure}, {Meneux}, {Merighi}, {Paltani}, {Pell{\`o}},
  {Pollo}, {Pozzetti}, {Radovich}, {Zamorani}, {Zucca}, {Bondi}, {Bongiorno},
  {Busarello}, {Lamareille}, {Mellier}, {Merluzzi}, {Ripepi}, \&
  {Rizzo}}]{le-fevre05}
{Le F{\`e}vre}, O., {Vettolani}, G., {Garilli}, B., {et~al.} 2005, \aap, 439,
  845

\bibitem[{{Lilly} {et~al.}(2009){Lilly}, {Le Brun}, {Maier}, {Mainieri},
  {Mignoli}, {Scodeggio}, {Zamorani}, {Carollo}, {Contini}, {Kneib}, {Le
  F{\`e}vre}, {Renzini}, {Bardelli}, {Bolzonella}, {Bongiorno}, {Caputi},
  {Coppa}, {Cucciati}, {de la Torre}, {de Ravel}, {Franzetti}, {Garilli},
  {Iovino}, {Kampczyk}, {Kovac}, {Knobel}, {Lamareille}, {Le Borgne}, {Pello},
  {Peng}, {P{\'e}rez-Montero}, {Ricciardelli}, {Silverman}, {Tanaka}, {Tasca},
  {Tresse}, {Vergani}, {Zucca}, {Ilbert}, {Salvato}, {Oesch}, {Abbas},
  {Bottini}, {Capak}, {Cappi}, {Cassata}, {Cimatti}, {Elvis}, {Fumana},
  {Guzzo}, {Hasinger}, {Koekemoer}, {Leauthaud}, {Maccagni}, {Marinoni},
  {McCracken}, {Memeo}, {Meneux}, {Porciani}, {Pozzetti}, {Sanders},
  {Scaramella}, {Scarlata}, {Scoville}, {Shopbell}, \& {Taniguchi}}]{lilly09}
{Lilly}, S.~J., {Le Brun}, V., {Maier}, C., {et~al.} 2009, \apjs, 184, 218

\bibitem[{{Lilly} {et~al.}(1996){Lilly}, {Le Fevre}, {Hammer}, \&
  {Crampton}}]{lilly96}
{Lilly}, S.~J., {Le Fevre}, O., {Hammer}, F., \& {Crampton}, D. 1996, \apjl,
  460, L1

\bibitem[{{Madau} {et~al.}(1998){Madau}, {Pozzetti}, \& {Dickinson}}]{madau98}
{Madau}, P., {Pozzetti}, L., \& {Dickinson}, M. 1998, \apj, 498, 106

\bibitem[{{Newman} {et~al.}(2013){Newman}, {Cooper}, {Davis}, {Faber}, {Coil},
  {Guhathakurta}, {Koo}, {Phillips}, {Conroy}, {Dutton}, {Finkbeiner}, {Gerke},
  {Rosario}, {Weiner}, {Willmer}, {Yan}, {Harker}, {Kassin}, {Konidaris},
  {Lai}, {Madgwick}, {Noeske}, {Wirth}, {Connolly}, {Kaiser}, {Kirby},
  {Lemaux}, {Lin}, {Lotz}, {Luppino}, {Marinoni}, {Matthews}, {Metevier}, \&
  {Schiavon}}]{newman13a}
{Newman}, J.~A., {Cooper}, M.~C., {Davis}, M., {et~al.} 2013, \apjs, 208, 5

\bibitem[{{Peng} {et~al.}(2010){Peng}, {Ho}, {Impey}, \& {Rix}}]{peng10a}
{Peng}, C.~Y., {Ho}, L.~C., {Impey}, C.~D., \& {Rix}, H.-W. 2010, \aj, 139,
  2097

\bibitem[{{Percival} {et~al.}(2010){Percival}, {Reid}, {Eisenstein}, {Bahcall},
  {Budavari}, {Frieman}, {Fukugita}, {Gunn}, {Ivezi{\'c}}, {Knapp}, {Kron},
  {Loveday}, {Lupton}, {McKay}, {Meiksin}, {Nichol}, {Pope}, {Schlegel},
  {Schneider}, {Spergel}, {Stoughton}, {Strauss}, {Szalay}, {Tegmark},
  {Vogeley}, {Weinberg}, {York}, \& {Zehavi}}]{percival10}
{Percival}, W.~J., {Reid}, B.~A., {Eisenstein}, D.~J., {et~al.} 2010, \mnras,
  401, 2148

\bibitem[{{Schlegel} {et~al.}(1998){Schlegel}, {Finkbeiner}, \&
  {Davis}}]{schlegel98}
{Schlegel}, D.~J., {Finkbeiner}, D.~P., \& {Davis}, M. 1998, \apj, 500, 525

\bibitem[{{Seo} {et~al.}(2012){Seo}, {Ho}, {White}, {Cuesta}, {Ross}, {Saito},
  {Reid}, {Padmanabhan}, {Percival}, {de Putter}, {Schlegel}, {Eisenstein},
  {Xu}, {Schneider}, {Skibba}, {Verde}, {Nichol}, {Bizyaev}, {Brewington},
  {Brinkmann}, {Nicolaci da Costa}, {Gott}, {Malanushenko}, {Malanushenko},
  {Oravetz}, {Palanque-Delabrouille}, {Pan}, {Prada}, {Ross}, {Simmons}, {de
  Simoni}, {Shelden}, {Snedden}, \& {Zehavi}}]{seo12}
{Seo}, H.-J., {Ho}, S., {White}, M., {et~al.} 2012, \apj, 761, 13

\bibitem[{{Sersic}(1968)}]{sersic68}
{Sersic}, J.~L. 1968, {Atlas de galaxias australes (Observatorio Astronomico,
  Cordoba, Argentina)}

\bibitem[{{Smee} {et~al.}(2013){Smee}, {Gunn}, {Uomoto}, {Roe}, {Schlegel},
  {Rockosi}, {Carr}, {Leger}, {Dawson}, {Olmstead}, {Brinkmann}, {Owen},
  {Barkhouser}, {Honscheid}, {Harding}, {Long}, {Lupton}, {Loomis}, {Anderson},
  {Annis}, {Bernardi}, {Bhardwaj}, {Bizyaev}, {Bolton}, {Brewington}, {Briggs},
  {Burles}, {Burns}, {Castander}, {Connolly}, {Davenport}, {Ebelke}, {Epps},
  {Feldman}, {Friedman}, {Frieman}, {Heckman}, {Hull}, {Knapp}, {Lawrence},
  {Loveday}, {Mannery}, {Malanushenko}, {Malanushenko}, {Merrelli}, {Muna},
  {Newman}, {Nichol}, {Oravetz}, {Pan}, {Pope}, {Ricketts}, {Shelden},
  {Sandford}, {Siegmund}, {Simmons}, {Smith}, {Snedden}, {Schneider},
  {SubbaRao}, {Tremonti}, {Waddell}, \& {York}}]{smee13}
{Smee}, S.~A., {Gunn}, J.~E., {Uomoto}, A., {et~al.} 2013, \aj, 146, 32

\bibitem[{{Strauss} {et~al.}(2002){Strauss}, {Weinberg}, {Lupton}, {Narayanan},
  {Annis}, {Bernardi}, {Blanton}, {Burles}, {Connolly}, {Dalcanton}, {Doi},
  {Eisenstein}, {Frieman}, {Fukugita}, {Gunn}, {Ivezi{\'c}}, {Kent}, {Kim},
  {Knapp}, {Kron}, {Munn}, {Newberg}, {Nichol}, {Okamura}, {Quinn}, {Richmond},
  {Schlegel}, {Shimasaku}, {SubbaRao}, {Szalay}, {Vanden Berk}, {Vogeley},
  {Yanny}, {Yasuda}, {York}, \& {Zehavi}}]{strauss02}
{Strauss}, M.~A., {Weinberg}, D.~H., {Lupton}, R.~H., {et~al.} 2002, \aj, 124,
  1810

\bibitem[{{Sugai} {et~al.}(2012){Sugai}, {Karoji}, {Takato}, {Tamura},
  {Shimono}, {Ohyama}, {Ueda}, {Ling}, {Vital de Arruda}, {Barkhouser},
  {Bennett}, {Bickerton}, {Braun}, {Bruno}, {Carr}, {Batista de Carvalho
  Oliveira}, {Chang}, {Chen}, {Dekany}, {Pereira Dominici}, {Ellis}, {Fisher},
  {Gunn}, {Heckman}, {Ho}, {Hu}, {Jaquet}, {Karr}, {Kimura}, {Le F{\`e}vre},
  {Le Mignant}, {Loomis}, {Lupton}, {Madec}, {Marrara}, {Martin}, {Murayama},
  {Cesar de Oliveira}, {Mendes de Oliveira}, {Souza de Oliveira}, {Orndorff},
  {de Paiva Vila{\c c}a}, {Macanhan}, {Prieto}, {Bispo dos Santos}, {Seiffert},
  {Smee}, {Smith}, {Sodr{\'e}}, {Spergel}, {Surace}, {Vives}, {Wang}, \&
  {Yan}}]{sugai12}
{Sugai}, H., {Karoji}, H., {Takato}, N., {et~al.} 2012, in Society of
  Photo-Optical Instrumentation Engineers (SPIE) Conference Series, Vol. 8446,
  Society of Photo-Optical Instrumentation Engineers (SPIE) Conference Series,
  0

\bibitem[{{The Dark Energy Survey Collaboration}(2005)}]{des05}
{The Dark Energy Survey Collaboration}. 2005, arXiv:astro-ph/0510346

\bibitem[{{van Dokkum} {et~al.}(2010){van Dokkum}, {Whitaker}, {Brammer},
  {Franx}, {Kriek}, {Labb{\'e}}, {Marchesini}, {Quadri}, {Bezanson},
  {Illingworth}, {Muzzin}, {Rudnick}, {Tal}, \& {Wake}}]{van-dokkum10}
{van Dokkum}, P.~G., {Whitaker}, K.~E., {Brammer}, G., {et~al.} 2010, \apj,
  709, 1018

\bibitem[{{Wright} {et~al.}(2010){Wright}, {Eisenhardt}, {Mainzer}, {Ressler},
  {Cutri}, {Jarrett}, {Kirkpatrick}, {Padgett}, {McMillan}, {Skrutskie},
  {Stanford}, {Cohen}, {Walker}, {Mather}, {Leisawitz}, {Gautier}, {McLean},
  {Benford}, {Lonsdale}, {Blain}, {Mendez}, {Irace}, {Duval}, {Liu}, {Royer},
  {Heinrichsen}, {Howard}, {Shannon}, {Kendall}, {Walsh}, {Larsen}, {Cardon},
  {Schick}, {Schwalm}, {Abid}, {Fabinsky}, {Naes}, \& {Tsai}}]{wright10}
{Wright}, E.~L., {Eisenhardt}, P.~R.~M., {Mainzer}, A.~K., {et~al.} 2010, \aj,
  140, 1868

\bibitem[{{Wuyts} {et~al.}(2011){Wuyts}, {F{\"o}rster Schreiber}, {van der
  Wel}, {Magnelli}, {Guo}, {Genzel}, {Lutz}, {Aussel}, {Barro}, {Berta},
  {Cava}, {Graci{\'a}-Carpio}, {Hathi}, {Huang}, {Kocevski}, {Koekemoer},
  {Lee}, {Le Floc'h}, {McGrath}, {Nordon}, {Popesso}, {Pozzi}, {Riguccini},
  {Rodighiero}, {Saintonge}, \& {Tacconi}}]{wuyts11a}
{Wuyts}, S., {F{\"o}rster Schreiber}, N.~M., {van der Wel}, A., {et~al.} 2011,
  \apj, 742, 96

\bibitem[{{York} {et~al.}(2000){York}, {Adelman}, {Anderson}, {Anderson},
  {Annis}, {Bahcall}, {Bakken}, {Barkhouser}, {Bastian}, {Berman}, {Boroski},
  {Bracker}, {Briegel}, {Briggs}, {Brinkmann}, {Brunner}, {Burles}, {Carey},
  {Carr}, {Castander}, {Chen}, {Colestock}, {Connolly}, {Crocker}, {Csabai},
  {Czarapata}, {Davis}, {Doi}, {Dombeck}, {Eisenstein}, {Ellman}, {Elms},
  {Evans}, {Fan}, {Federwitz}, {Fiscelli}, {Friedman}, {Frieman}, {Fukugita},
  {Gillespie}, {Gunn}, {Gurbani}, {de Haas}, {Haldeman}, {Harris}, {Hayes},
  {Heckman}, {Hennessy}, {Hindsley}, {Holm}, {Holmgren}, {Huang}, {Hull},
  {Husby}, {Ichikawa}, {Ichikawa}, {Ivezi{\'c}}, {Kent}, {Kim}, {Kinney},
  {Klaene}, {Kleinman}, {Kleinman}, {Knapp}, {Korienek}, {Kron}, {Kunszt},
  {Lamb}, {Lee}, {Leger}, {Limmongkol}, {Lindenmeyer}, {Long}, {Loomis},
  {Loveday}, {Lucinio}, {Lupton}, {MacKinnon}, {Mannery}, {Mantsch}, {Margon},
  {McGehee}, {McKay}, {Meiksin}, {Merelli}, {Monet}, {Munn}, {Narayanan},
  {Nash}, {Neilsen}, {Neswold}, {Newberg}, {Nichol}, {Nicinski}, {Nonino},
  {Okada}, {Okamura}, {Ostriker}, {Owen}, {Pauls}, {Peoples}, {Peterson},
  {Petravick}, {Pier}, {Pope}, {Pordes}, {Prosapio}, {Rechenmacher}, {Quinn},
  {Richards}, {Richmond}, {Rivetta}, {Rockosi}, {Ruthmansdorfer}, {Sandford},
  {Schlegel}, {Schneider}, {Sekiguchi}, {Sergey}, {Shimasaku}, {Siegmund},
  {Smee}, {Smith}, {Snedden}, {Stone}, {Stoughton}, {Strauss}, {Stubbs},
  {SubbaRao}, {Szalay}, {Szapudi}, {Szokoly}, {Thakar}, {Tremonti}, {Tucker},
  {Uomoto}, {Vanden Berk}, {Vogeley}, {Waddell}, {Wang}, {Watanabe},
  {Weinberg}, {Yanny}, \& {Yasuda}}]{york00}
{York}, D.~G., {Adelman}, J., {Anderson}, Jr., J.~E., {et~al.} 2000, \aj, 120,
  1579

\bibitem[{{Zou} {et~al.}(2015){Zou}, {Jiang}, {Zhou}, {Wu}, {Ma}, {Fan}, {Fan},
  {He}, {Jing}, {Lesser}, {Li}, {Nie}, {Shen}, {Wang}, {Zhang}, \&
  {Zhou}}]{zou15}
{Zou}, H., {Jiang}, Z., {Zhou}, X., {et~al.} 2015, \aj, 150, 104

\end{thebibliography}

\appendix

%=========================
% CFHTLS ZPHOT
%=========================
\section{CFHTLS photometric redshift reliability \label{sec:zphot_properties}}
In Section \ref{sec:sel_properties}, we make use of the CFHTLS-Wide photometric redshifts \citep[T0007 release\footnote{\href{http://terapix.iap.fr/rubrique.php?id\_rubrique=267}{http://terapix.iap.fr/rubrique.php?id\_rubrique=267}};][]{ilbert06a,coupon09} to estimate the redshift distribution of our selection schemes.
Those photometric redshifts have been proven to be of very good quality up to $i<22.5$ (bias below 1\%, scatter of $\sim$0.04, and less than 4\% outliers).
Nevertheless, those statistics have been computed for magnitude-limited samples, whereas the galaxies under study in this paper are mainly star-forming galaxies.
Owing to the lack of features (weak 4000~\AA~ break, power law spectrum), this class of galaxies is well-known for having slightly less accurate photometric redshift, which results in a higher outlier rate \citep[e.g.][]{ilbert06a,hildebrandt12}.

\begin{figure}
        \includegraphics[width=0.95\columnwidth]{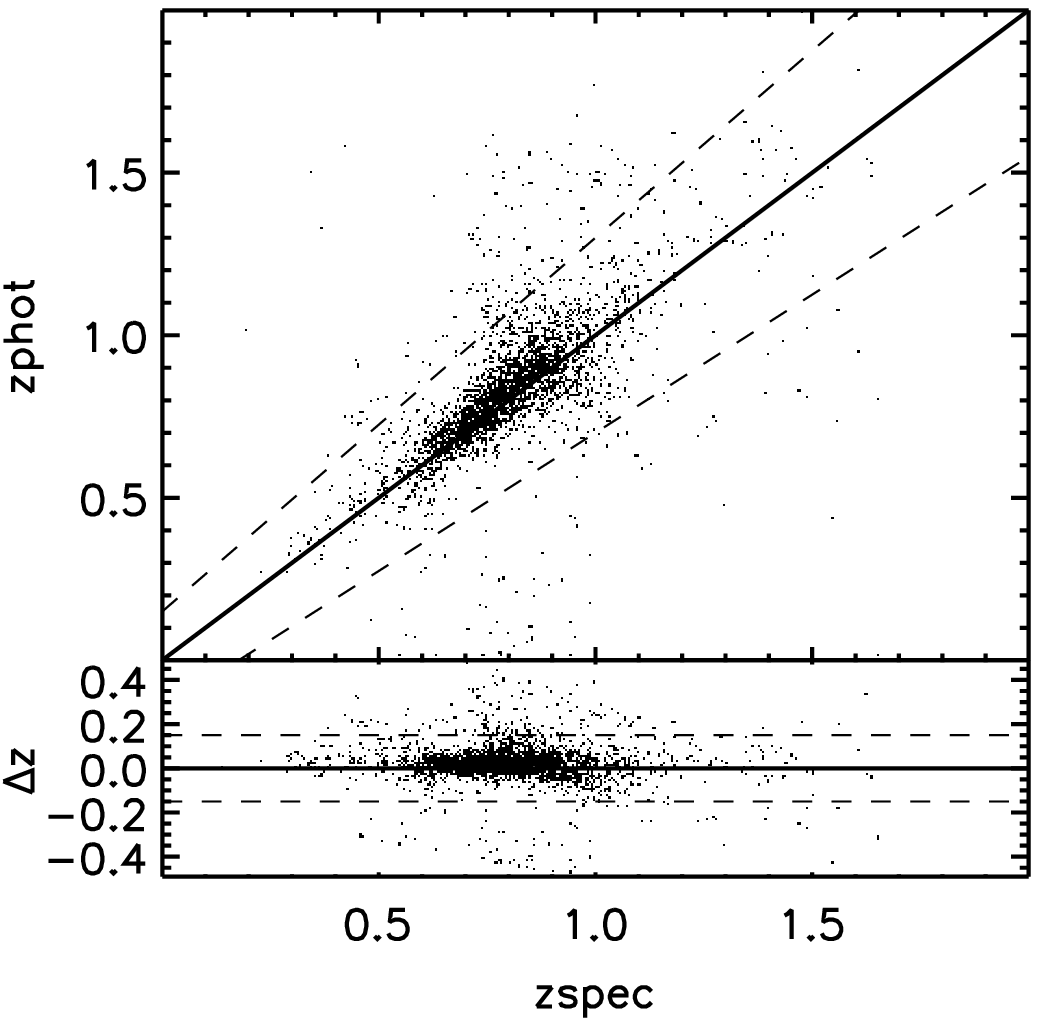}\\
        \includegraphics[width=0.95\columnwidth]{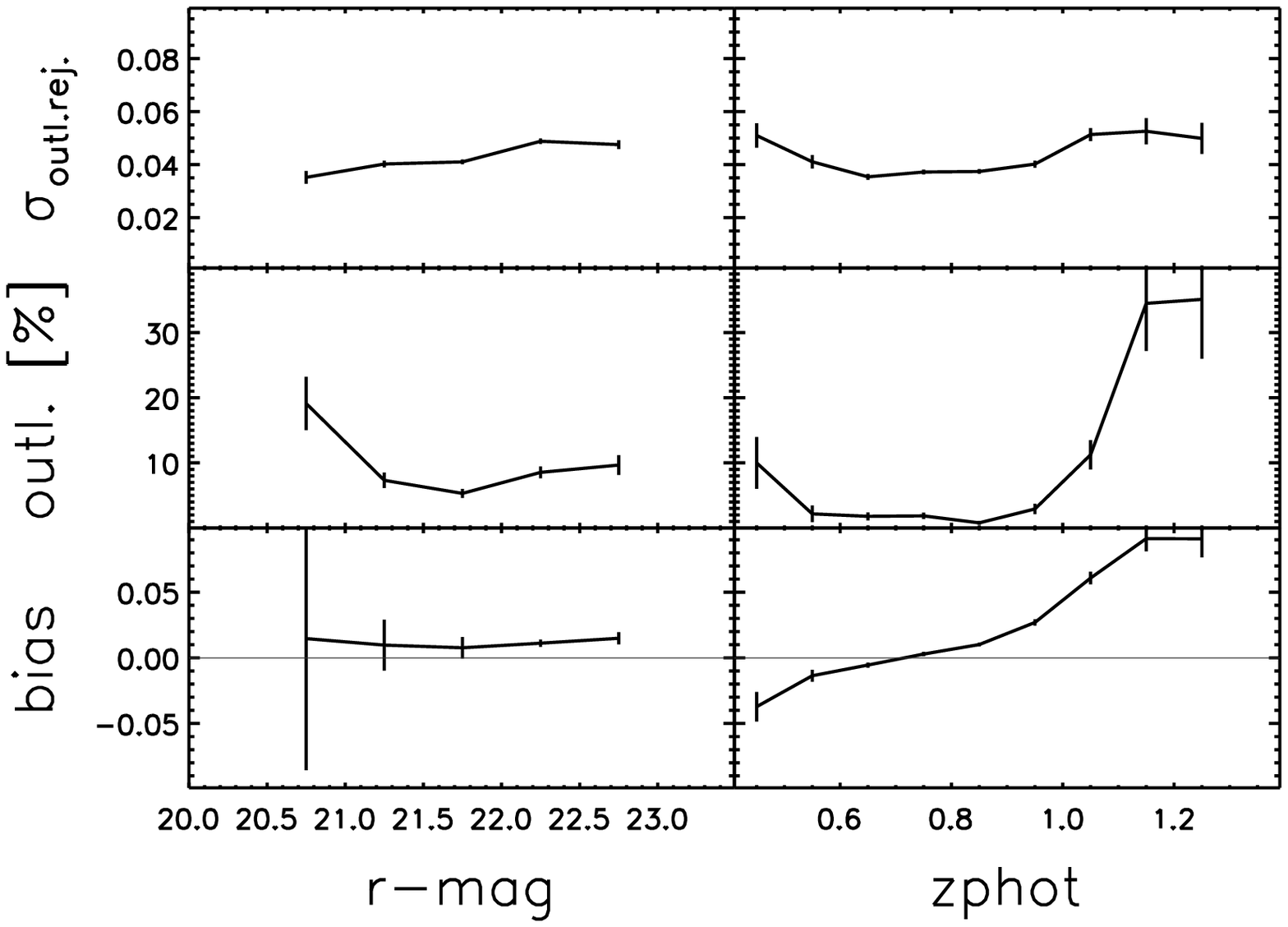}\\
        \caption{Reliability of the CFHTLS photometric redshifts for $z \sim 0.8$ ELGs, using the \texttt{eboss6-7} observed galaxies with a secure $z_{\rm spec}$ and $L_{[\rm O\textsc{ii}]}^{\rm tot} > 10^{41}$ erg.s$^{-1}$.
\textit{Top panel}: $z_{\rm phot}$ vs. $z_{\rm spec}$.
\textit{Bottom panel}: $z_{\rm phot}$ statistics, as a function of magnitude (\textit{left}) and redshift (\textit{right}). We only report quantities for the bins where we have more than 50 galaxies, and error bars are calculated assuming a Poissonian distribution.}
        \label{fig:zphot_properties}
\end{figure}

Using the \texttt{eboss6-7} test plates, we demonstrate in Figure \ref{fig:zphot_properties} the reliability of those photometric redshifts for ELGs up to redshift $\sim$1.
We selected \texttt{eboss6-7} galaxies with a secure spectroscopic redshift and a [O\textsc{ii}] total luminosity $L_{[\rm O\textsc{ii}]}^{\rm tot}$ greater than 10$^{41}$ erg.s$^{-1}$ (3712 galaxies with $z_{\rm spec} = 0.80 \pm 0.17$).
For each object in our spectroscopic sample, we calculated $\Delta z = \frac{z_{\rm phot}-z_{\rm spec}}{1+z_{\rm spec}}$ and classify it as an outlier if $| \Delta z | > 0.15$.
For each binned subsample, we report \textit{bias}: the median value of $\Delta z$; \textit{outl.}: the percentage of outliers; and $\sigma_{\rm outl.rej.}$: the standard deviation of $\Delta z$ when outliers have been excluded.
These quantities are used to facilitate comparison with other works. As mentioned in \citet{hildebrandt12}, the outlier definition is arbitrary.
We observe that the photometric redshifts are slightly biased high ($bias \sim +0.01$, for all magnitudes), the bias becoming significant for $z_{\rm phot} \gtrsim 1.1$.
The scatter is reasonable ($\sigma_{\rm outl.rej.} \sim 0.04$), as is the outlier rate (5-10\%).
Those results qualitatively agree with previous studies \citep{ilbert06a,hildebrandt12}.

% Figure: zphot

%\newpage

\end{document}